\documentclass[aps, pra,reprint,10pt,superscriptaddress,amssymb,amsfonts]{revtex4-1}
\usepackage{amsmath,amssymb,amsthm,amscd,latexsym,epsfig,bm}

\usepackage[colorlinks,bookmarks=false,citecolor=blue,linkcolor=red,urlcolor=blue]{hyperref}
\usepackage{verbatim}

\newcommand{\zz}{\mathbb{Z}_2}
\newcommand{\z}{\mathbb{Z}}

\def\br{\boldsymbol{r}}

\def\bx{\boldsymbol{x}}
\def\by{\boldsymbol{y}}

\begin{document}

\title{Topological phases protected by point group symmetry}

\author{Hao Song}
\altaffiliation[Current address: ]{Departamento de F\'{\i}sica Te\'{o}rica I, Universidad Complutense, 28040 Madrid, Spain}
\affiliation{Department of Physics, University of Colorado, Boulder, Colorado 80309, USA}
\affiliation{Center for Theory of Quantum Matter, University of Colorado, Boulder, Colorado 80309, USA}
\author{Sheng-Jie Huang}
\affiliation{Department of Physics, University of Colorado, Boulder, Colorado 80309, USA}
\affiliation{Center for Theory of Quantum Matter, University of Colorado, Boulder, Colorado 80309, USA}
\author{Liang Fu}
\affiliation{Department of Physics, Massachusetts Institute of Technology, Cambridge, Massachusetts 02139, USA}
\author{Michael Hermele}
\affiliation{Department of Physics, University of Colorado, Boulder, Colorado 80309, USA}
\affiliation{Center for Theory of Quantum Matter, University of Colorado, Boulder, Colorado 80309, USA}
%\thanks{}
\date{\today}

\begin{abstract}
We consider symmetry protected topological (SPT) phases with crystalline point group symmetry, dubbed point group SPT (pgSPT) phases.
We show that such phases can be understood in terms of lower-dimensional topological phases with on-site symmetry, and can be constructed as stacks and arrays of these lower-dimensional states.  This provides the basis for a general framework to classify and characterize bosonic and fermionic pgSPT phases, that can be applied for arbitrary crystalline point group symmetry and in arbitrary spatial dimension.  We develop and illustrate this framework by means of a few examples, focusing on three-dimensional states.  We classify bosonic pgSPT phases and fermionic topological crystalline superconductors with $\zz^P$ (reflection) symmetry, electronic topological crystalline insulators (TCIs) with ${\rm U}(1) \times \zz^P$ symmetry, and bosonic pgSPT phases with $C_{2v}$ symmetry, which is generated by two perpendicular mirror reflections.  We also study surface properties, with a focus on gapped, topologically ordered surface states.  For electronic TCIs we find a $\z_8 \times \zz$ classification, where the $\z_8$ corresponds to known states obtained from non-interacting electrons, and the $\zz$ corresponds to a ``strongly correlated'' TCI that requires strong interactions in the bulk.  Our approach may also point the way toward a general theory of symmetry enriched topological (SET) phases with crystalline point group symmetry.
\end{abstract}

\maketitle

\section{Introduction}

Topological band insulators host fascinating and rich properties on their surfaces \cite{hasan10,hasan11,qi11}. Spurred on by these phenomena, it has been recognized over the past few years that topological insulators are  one example in a large family of symmetry protected topological (SPT) phases \cite{kitaev09, ryu10, gu09tensor,pollmann10,fidkowski11,turner11,chen11a,schuch11,chen13symmetry,levin12braiding}.  Such states are now well-understood for free fermion systems with internal symmetry \cite{kitaev09, ryu10}, and much attention has now turned to ``strongly correlated'' SPT phases, those which require strong interactions in the bulk.  Motivation to understand strongly correlated SPT phases comes in part from materials such as SmB$_6$, where strongly interacting $f$-electrons are proposed to play a crucial role in forming a topological insulator ground state \cite{dzero16topological}.

In order to find SPT phases in real systems, it is important to consider realistic symmetries.  Most existing theory of strongly correlated SPT phases focuses on internal, or on-site, symmetries, such as ${\rm U}(1)$ charge conservation, $\zz^T$ time reversal and ${\rm SO}(3)$ spin rotation.  Of course, crystalline point group and space group symmetries are often important in solids, and a good deal is now understood about free-fermion SPT phases protected by such symmetries (see \cite{ando15topological} and references therein), including some results on these states when interactions are included \cite{isobe15,yqi15anomalous,yoshida15correlation,morimoto15breakdown,chsieh16global}.  

Much less is understood about strongly correlated SPT phases protected by crystalline symmetries.  While there has been some progress for one- and two-dimensional states  \cite{gu09tensor,pollmann10,chen11a,schuch11,chen13symmetry,chsieh14symmetry,chsieh14CPT,cho15,yoshida15bosonic,ware15topological,lapa16interaction}, and a few works pertaining to three dimensions \cite{chen13symmetry,you14symmetry,kapustin15fermionic,hermele15,cheng15translational}, there is currently no generally applicable framework to classify and characterize crystalline SPT phases.  Many of the powerful approaches used to study SPT phases with internal symmetry, such as group cohomology \cite{chen13symmetry} or gauging of symmetry \cite{levin12braiding}, cannot straightforwardly be generalized to crystalline symmetries.  This is thus a  significant gap in the theoretical understanding of SPT phases, which we fill in this paper for crystalline point group symmetry.

\begin{table*}
\label{tab}
\begin{tabular}{c|c|c|c}
Bosonic/fermionic & Protecting symmetry & Classification & Section of paper \\
\hline
Bosonic & $\zz^P$ & $\zz \times \zz$ & \ref{sec:3dz2p} \\
\hline
Bosonic & $\zz^P$ with translation & $(\zz)^3$ & \ref{sec:3dz2p} \\
\hline
Fermionic & ${\rm U}(1) \times \zz^P$ & $\z_8 \times \zz$ & \ref{sec:tci} \\
\hline
Fermionic & $\zz^P$, $\sigma^2 = 1$ & $\z_{16}$ & \ref{sec:tcsc} \\
\hline
Fermionic & $\zz^P$, $\sigma^2 = (-1)^F$ & Trivial & \ref{sec:tcsc} \\
\hline
Bosonic & $C_{2v}$ & $(\zz)^4$ & \ref{sec:other-point-groups}
\end{tabular}
\caption{Summary of the classifications obtained for point group SPT phases in three dimensions.  The first column indicates whether we are considering bosonic or fermionic systems,  the second column gives the protecting symmetry, and the third column gives the classification.  $\zz^P$ denotes reflection symmetry, $\sigma$ the reflection operation, and $(-1)^F$ the fermion parity operator. $C_{2v}$ is the three-dimensional point group generated by two perpendicular mirror reflections.  Translation refers to discrete translation symmetry normal to the mirror planes.  The last column shows the section of the paper where each classification is obtained.}
\end{table*}

We consider SPT phases protected by crystalline point group symmetry, which we dub point group SPT (pgSPT) phases.  More precisely, we consider crystalline symmetry groups leaving at least one point fixed \footnote{The symmetries we focus on are more properly called site symmetries; in crystallography, the point group is defined as the quotient of the space group by its translation subgroup. We abuse terminology slightly and refer to site symmetries as point group symmetries, as this is a more evocative and commonly understood term, and because site symmetry groups are always themselves crystallographic point groups.}.
We show that any pgSPT state in spatial dimension $d$ can be adiabatically connected, preserving symmetry, to a system composed of lower-dimensional topological states with on-site symmetry.  This dimensional reduction allows us to classify bosonic and fermionic pgSPT phases in any spatial dimension, to  study symmetry-preserving surfaces, and to explicitly construct pgSPT phases as stacks and arrays of lower-dimensional states.  

We illustrate our approach via a number of physically interesting examples, devoting particular attention to the case of mirror reflection symmetry (referred to as $\zz^P$) in three dimensions ($3d$).  We consider both bosonic and fermionic pgSPT phases protected by $\zz^P$, obtaining classifications (summarized in Table~\ref{tab}) and studying surface properties.  Remarkably, all the states we find can be constructed as stacks of two-dimensional topological phases.

Among fermionic pgSPT phases, an especially physically relevant case is that of electronic topological crystalline insulators (TCIs) with charge conservation and reflection symmetry [${\rm U}(1) \times \zz^P$], which have been predicted and observed in the SnTe material class \cite{thsieh12topological,tanaka12experimental,dziawa12topological,syxu12observation}.  At the free-fermion level, these systems obey a $\z$ classification, which breaks down to $\z_8$ for interacting electrons \cite{isobe15}.  We show that the full classification of such states is $\z_8 \times \zz$. The root state generating the additional $\zz$ factor requires strong interactions in the bulk, and can be understood as a topological paramagnet, where the spin sector is in a bosonic pgSPT phase.  This state is analogous to topological paramagnets found in the classification of interacting topological insulators protected by internal symmetry \cite{cwang14classification}.

Our approach can be applied to any point group, and we illustrate this for $3d$ bosonic pgSPT phases with $C_{2v}$ symmetry, which is generated by two perpendicular mirror reflections.  We find a $(\zz)^4$ classification, where the states can be understood in terms of $2d$ topological phases on the mirror planes, and in terms of $1d$ SPT phases located on the line where the mirror planes intersect.  Extensions to other point groups, including for fermionic systems, are left for future work.

In crystalline solids, point group symmetry always occurs as a subgroup of a larger space group including translational symmetry.  We emphasize that, in general, 
our dimensional reduction argument cannot be applied so as to respect translational symmetry. Instead, the strategy is to focus on point subgroups of the full space group, and treat each one separately, while ignoring the rest of the symmetry.  For each point subgroup we can obtain a classification of pgSPT phases, and by considering relations among different subgroups imposed by the full space group symmetry, we can obtain a partial classification of SPT phases invariant under the full space group. However, while we know of no concrete examples, our approach could miss SPT phases with non-trivial interplay between translation and point group symmetries, and should not be considered a full classification of space group SPT phases.

We expect the ideas developed here to be applicable beyond the domain of SPT phases.  In particular, the essence of our approach can be applied to symmetry enriched topological (SET) phases with crystalline point group symmetry.  SET phases are those that remain non-trivial even if all symmetries are broken explicitly, for instance due to the presence of fractional excitations with non-trivial braiding statistics (\emph{i.e.} anyons, in two dimensions).  Despite some progress \cite{wen02,essin13,barkeshli14,yqi15detecting,zaletel15}, so far there there is no general theoretical framework to classify and characterize SET phases with crystalline symmetries; we believe that, combined with other ideas, the approach developed here could form the basis for such a framework.   This possible extension of our results is discussed further in Sec.~\ref{sec:discussion}.

Our main focus is on $3d$ pgSPT phases, so we now discuss some prior work in three dimensions.  In particular, we note the work of Isobe and Fu\cite{isobe15}, who showed that interactions reduce the classification of TCIs with ${\rm U}(1) \times \zz^P$ symmetry from $\z$ to $\z_8$.  They imposed a spatially-varying Dirac mass term that produces, at the surface, an array of well-separated one-dimensional conductors on axes of reflection symmetry.  They then pointed out that these one-dimensional conductors are identical to edges of $2d$ electronic topological phases protected by internal ${\rm U}(1) \times \zz$ symmetry, and drew attention to the connection between these two apparently different kinds of topological phases.  Indeed, these observations are an instance of the general connection between $d$-dimensional pgSPT phases and lower-dimensional topological phases that we obtain.  By exposing this general connection, without relying on a non-interacting description as a starting point or focusing only on edge and surface theories, we are able to go beyond Ref.~\onlinecite{isobe15} to classify general pgSPT phases.

A few works obtained some prior results on strongly correlated pgSPT phases in three dimensions.  Y.-Z. You and C. Xu studied $3d$  SPT phases protected by spatial inversion symmetry, also combined with internal symmetries, using a non-linear sigma model approach \cite{you14symmetry}.  
Hermele and X.~Chen identified some $3d$ bosonic SPT phases protected by a combination of ${\rm U}(1)$ and crystalline symmetries, by developing a method to test for anomalies in candidate surface theories \cite{hermele15}.  
Finally, Kapustin \emph{et. al.} used the cobordism approach developed in~\cite{kapustin14symmetry,kapustin14bosonic} to study fermionic SPT phases \cite{kapustin15fermionic}.  While their focus was on internal symmetries, results agreeing with ours were also quoted for fermionic topological superconductors protected by $\zz^P$.

We now give an outline of the remainder of the paper.  We illustrate our approach in Sec.~\ref{sec:3dz2p}, where we discuss $3d$ bosonic pgSPT phases protected by $\zz^P$ symmetry.  The approach is based on reduction to a $2d$ state on the mirror plane, where the $\zz^P$ reflection symmetry acts effectively as an on-site $\zz$ symmetry.  The reduction procedure is described in Sec.~\ref{sec:reduction}, and is then used in Sec.~\ref{sec:3dz2p-classification} to obtain a $\zz \times \zz$ classification of pgSPT phases.  Section~\ref{sec:3d-translation} discusses the role of translation symmetry normal to the mirror plane, which expands the classification to $(\zz)^3$.  Sections~\ref{sec:3dz2p-classification} and~\ref{sec:3d-translation} also show that the pgSPT phases we find can be understood as stacks of two-dimensional topological phases.  For the $\zz \times \zz$ classification obtained with $\zz^P$ symmetry alone, there are two root states; one of these can be understood as a stack of non-trivial $2d$ SPT phases with on-site $\zz$ symmetry (the $\zz$ root state), while the other can be understood as a stack of bosonic $E_8$ states \cite{kitaev11KITP} with alternating chirality (the $E_8$ root state).

Surface properties of these states are considered in Sec.~\ref{sec:surface}, focusing on gapped, topologically ordered surfaces.  The $\zz$ root state admits a surface with toric code topological order and anomalous reflection symmetry fractionalization, while the $E_8$ root state admits a reflection-symmetric surface with three-fermion topological order.  The latter surface is anomalous because, in strictly two dimensions, the three-fermion state has  gapless chiral edge modes \cite{kitaev06} and is thus incompatible with reflection symmetry.

Section~\ref{sec:tci} discusses electronic TCIs in three dimensions.  These are fermionic SPT phases protected by charge conservation and reflection symmetry [${\rm U}(1) \times \zz^P$].  In Sec.~\ref{sec:tci-classification}, we find a $\z_8 \times \zz$ classification of such phases, reproducing the $\z_8$ classification of~\cite{isobe15} obtained starting from free-fermion states, and identifying a new additional $\zz$ factor, associated with strongly correlated TCIs.  The corresponding root state can be understood as a topological paramagnet, where the spin sector is in the $E_8$ root state bosonic $\zz^P$ pgSPT phase, and is thus dubbed the $E_8$ paramagnet TCI.  Surfaces of this state are studied in Sec.~\ref{sec:tci-surface}.  In Sec.~\ref{sec:connect-bosonic}, we show that the $n=4$ state of the $\z_8$ factor (\emph{i.e.} four copies of the root state that generates the $\z_8$) can also be viewed as a different topological paramagnet, where the spin sector is in the $\zz$ root state.  This strongly interacting limit of the $n=4$ TCI is very different from the non-interacting limit of the same phase.

Section~\ref{sec:tcsc} discusses the classification of topological crystalline superconductors protected by $\zz^P$ symmetry.  There are two different cases to consider.  In the first case (Sec.~\ref{sec:tcsc-R21}), reflection squares to the identity operator, and we find a $\z_{16}$ classification.  These states can be obtained starting from free fermions, and the same classification can be obtained by a straightforward generalization of the arguments of~\cite{isobe15}, as was mentioned in~\cite{chsieh16global}.  Our analysis shows that the $\z_{16}$ classification is complete even accounting for the possibility of strongly correlated topological crystalline superconductors (barring the possibility of as yet unknown $2d$ topological phases appearing upon reduction to the mirror plane).    In the second case (Sec.~\ref{sec:tcsc-R2F}), reflection squares to the fermion parity operator, and we find a trivial classification.  These results are in agreement with~\cite{kapustin15fermionic}, which obtained the same classifications by very different methods.

In Sec.~\ref{sec:other-point-groups}, we study $3d$ bosonic pgSPT phases protected by $C_{2v}$ symmetry, which is generated by two perpendicular mirror reflections.  We find a $(\zz)^4$ classification, where two of the root states are based on $2d$ $\zz$ SPT states on the mirror planes, one is based on the $E_8$ state on the mirror planes, and one can be understood in terms of the $1d$ Haldane phase \cite{haldane83a,haldane83b} located on the line where the mirror planes intersect.  

We conclude in Sec.~\ref{sec:discussion} with a discussion of our results, and of possible directions for further work.  Appendix~\ref{app:1d} uses our approach to recover the known $\zz$ classification of bosonic $1d$ pgSPT phases with reflection symmetry \cite{gu09tensor,pollmann10,chen11a,schuch11}.  Appendix~\ref{app:toricmod} gives some technical details pertaining to a modified toric code model used in Sec.~\ref{sec:z2root} to study the gapped, topologically ordered surface of the $\zz$ root state.

Finally, we note that some of our results have appeared in the Ph.D. thesis of H.S. \cite{hsongThesis}.

\section{Bosonic point group SPT phases in three dimensions}
\label{sec:3dz2p}

\subsection{Approach: Reduction to $2d$}
\label{sec:reduction}

We illustrate our approach by considering $3d$ bosonic systems with a single mirror reflection symmetry, $\sigma : (x,y,z) \to (-x,y,z)$.  We begin with this example as it is relatively simple, leads to interesting phenomena on symmetry-preserving surfaces, and is physically relevant \emph{e.g.} for spin systems.  Moreover, the results we obtain here will be useful when we consider electronic TCIs and topological crystalline superconductors below.

In a solid, reflection symmetry would only occur as a subgroup of a larger space group including translation symmetry.  It turns out to be  important for our approach to ignore all the symmetry except for a single reflection, at least as a first step.  In Sec.~\ref{sec:3d-translation}, we will return to the role of translation symmetry.

We shall argue that a $3d$ SPT phase protected by reflection is adiabatically connected, while preserving symmetry, to an \emph{extensively trivial} state.  An extensively trivial state is a product state, except over a sub-extensive region (\emph{i.e.} one that occupies a vanishing fraction of the system in the thermodynamic limit).  In the present case, the sub-extensive region is centered on the mirror plane, and can be viewed as an effective $2d$ system, on which reflection acts as an on-site $\zz$ symmetry.  The classification of pgSPT phases protected by reflection symmetry in $3d$ then reduces to a classification of $2d$ states with $\zz$ on-site symmetry.

\begin{figure}
\includegraphics[width=0.6\columnwidth]{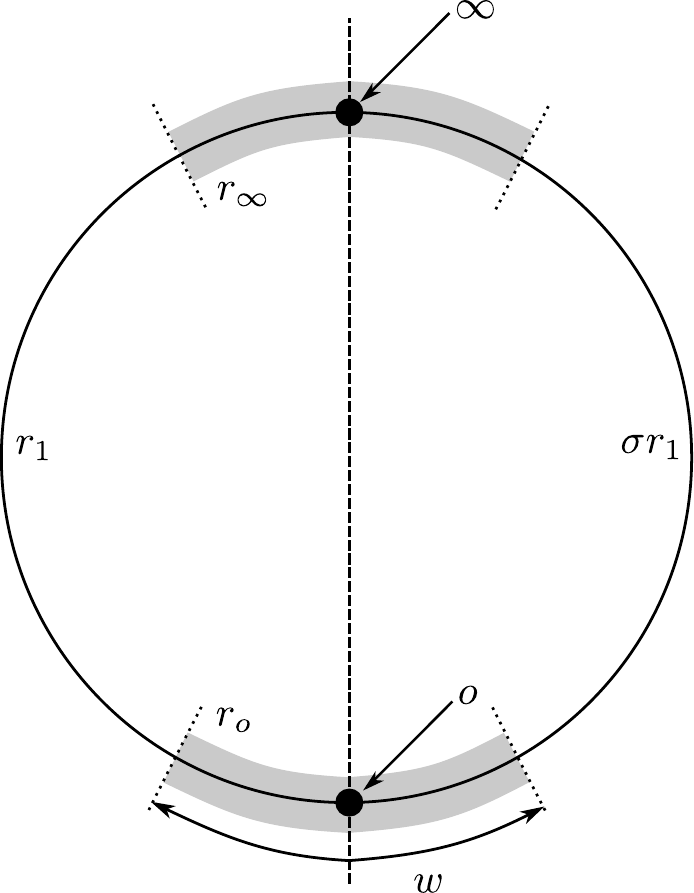}
\caption{Three-dimensional system with periodic boundary conditions and $\zz^P$ reflection symmetry.  Each point on the solid circle corresponds to a $2d$ plane with periodic boundary conditions.  The dashed line intersects the system at the two mirror planes $o$ and $\infty$, which are contained in the shaded regions $r_o$ and $r_{\infty}$, respectively.  These regions have thickness $w$.  Dotted lines indicate the boundaries of these regions with two other regions, $r_1$ and $\sigma r_1$.  The regions are chosen so that $r_o$ and $r_{\infty}$ are invariant under reflection, while $r_1$ and $\sigma r_1$ are exchanged under reflection.} 
\label{fig:3d-regions}
\end{figure}

Now, in more detail, we consider a lattice model,  refer to the degrees of freedom at each site as a spin, and refer to the symmetry group as $\zz^P = \{ 1, \sigma \}$.  The unitary operator $U_{\sigma}$ represents the action of the reflection $\sigma$ on Hilbert space.  Because $\zz^P$ has no non-trivial projective representations [formally, $H^2(\zz^P, {\rm U}(1))$ is trivial], without loss of generality we assume $U_{\sigma}^2 = 1$ acting on any individual spin, and therefore also on the entire Hilbert space.  We consider a system of linear size $L$ with periodic boundary conditions, which means there are actually two planes in the system fixed by the reflection $\sigma$, as shown in Fig.~\ref{fig:3d-regions}.  We focus on properties near one of these planes, which we refer to as $o$.  We view the other plane as spatial infinity upon taking the thermodynamic limit, and refer to it as $\infty$.  We come back to this point later in this section, where we discuss the role of boundary conditions.

\begin{figure}
\includegraphics[width=0.8\columnwidth]{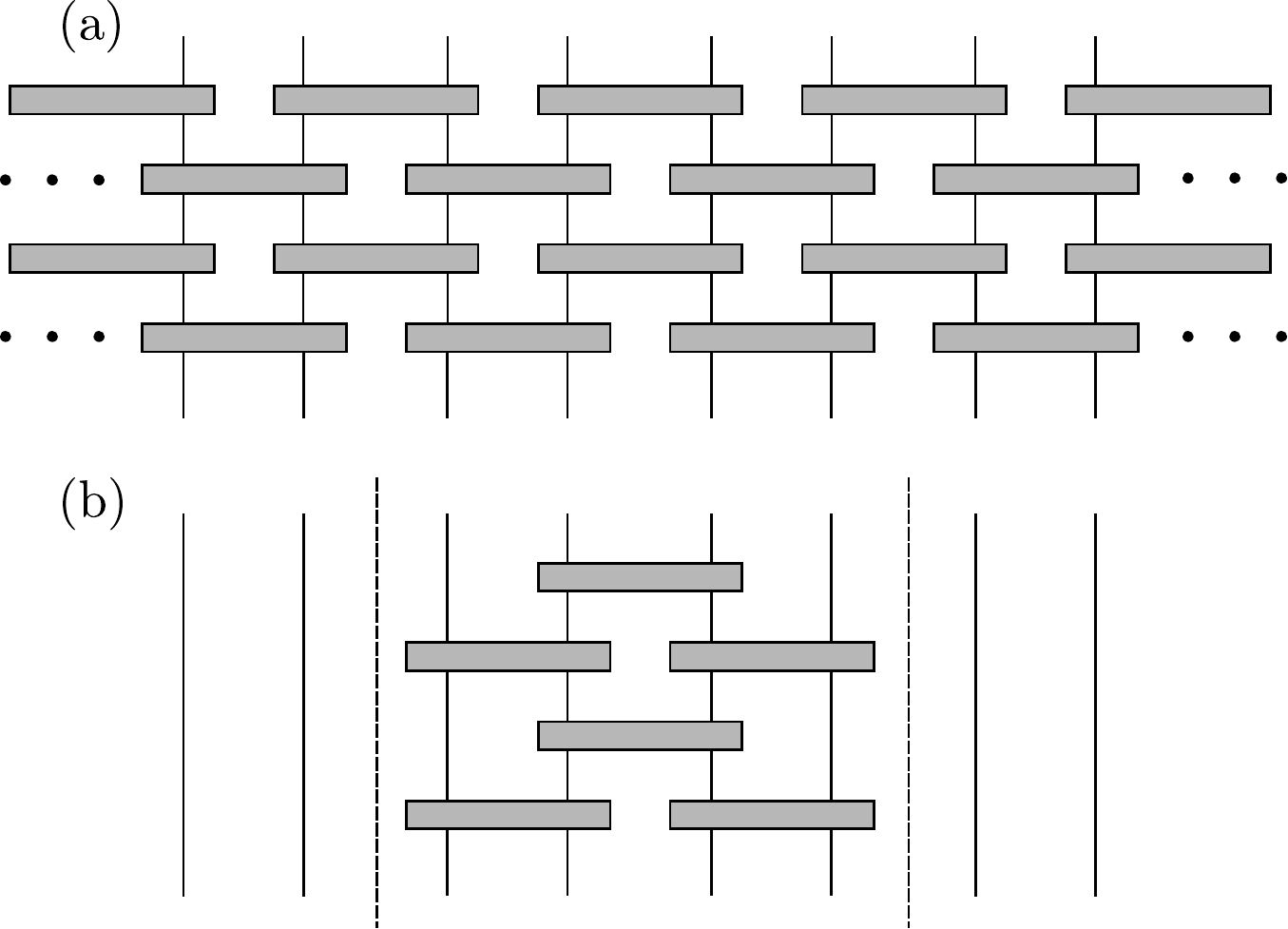}
\caption{(a) $1d$ local unitary represented as a finite-depth quantum circuit.  The vertical lines represent spins, and each shaded rectangle is a unitary operator acting on a pair of spins.  (b)  Restriction of a $1d$ local unitary to the region between the two dashed lines.  The two-spin unitary operators lying outside this region are simply omitted.  The restriction procedure is not uniquely defined near the boundaries of the region, but this freedom does not play a role in our discussion.} 
\label{fig:restrict}
\end{figure}

We suppose the ground state $|\psi \rangle$ is a SPT phase.  Precisely, we take this to mean that there is an energy gap to bulk excitations, there is no spontaneous symmetry breaking,  the ground state is unique, and, if we allow explicit breaking of symmetry, $|\psi \rangle$ is adiabatically connected to a trivial product state.  The last condition can be expressed by writing
\begin{equation}
U^{loc} | \psi \rangle = | T \rangle \text{,}  \label{eqn:trivial}
\end{equation}
where $U^{loc}$ is a local unitary described as a finite-depth quantum circuit (see Fig.~\ref{fig:restrict}), and $ | T \rangle$ is a trivial product state.

If $|\psi \rangle$ is in a non-trivial pgSPT phase, then we cannot choose $U^{loc}$ to trivialize the state while respecting the symmetry.  However, we now see that we can act with a different local unitary to \emph{extensively} trivialize $|\psi\rangle$ while preserving symmetry.  We divide the system into four regions as shown in Fig.~\ref{fig:3d-regions}.  Regions $r_o$ and $r_{\infty}$ are reflection-symmetric, while $\sigma r_1$ is the image of $r_1$ under reflection.  The thickness of $r_o$ and $r_{\infty}$ is $w$, which is held fixed in the thermodynamic limit ($L \to \infty$), so that these regions are truly two-dimensional.  An important parameter is the ratio $w / \xi$, where $\xi$ is the correlation length.  The statements we make below are expected to hold in the limit $w / \xi \gg 1$.

First, we trivialize the system in region $r_1$.  We note that a finite-depth quantum circuit can be restricted to act in a smaller region \cite{else15}, as is illustrated for $1d$ in Fig.~\ref{fig:restrict}.  We restrict $U^{loc}$ to a region $r'_1$ that contains $r_1$, and extends a small amount into regions $r_o$ and $r_{\infty}$, and denote the resulting restriction by $U^{loc}_{r_1}$.
A few correlation lengths away from the boundaries of $r'_1$, the action of $U^{loc}_{r_1}$ on $| \psi \rangle$ should be indistinguishable from that of $U^{loc}$.  Therefore, we expect
\begin{equation}
U^{loc}_{r_1} | \psi \rangle = |T_{r_1} \rangle \otimes | \psi_{\overline{r_1}}\rangle \text{,}
\end{equation}
where the system is in the product state $| T_{r_1} \rangle$ in region $r_1$, and the remainder of the system (the complement $\overline{r_1}$) is in the state $| \psi_{\overline{r_1}} \rangle$.  The intuition behind this expectation is that $| \psi \rangle$ has only short-range entanglement, so that a region can be disentangled from the rest of the system, and the entanglement within this region removed, by a finite-depth quantum circuit.

To extensively trivialize the ground state while preserving symmetry, we note that $U^{loc}_{\sigma r_1} = U_{\sigma} U^{loc}_{r_1} U^{-1}_{\sigma}$ trivializes  $\sigma r_1$, and we act on $|\psi \rangle$ with
\begin{equation}
U^{loc}_{R_1} = U^{loc}_{r_1} U^{loc}_{\sigma r_1} \text{.}
\end{equation}
This trivializes $R_1 = r_1 \cup \sigma r_1$, leaving only the two-dimensional regions $r_o$ and $r_{\infty}$ non-trivial.  Moreover, this transformation preserves symmetry; that is, $U_\sigma U^{loc}_{R_1}  =  U^{loc}_{R_1} U_\sigma$.
Acting on $|\psi\rangle$, we obtain
\begin{equation}
| \psi' \rangle \equiv U^{loc}_{R_1} | \psi \rangle = |T_{r_1} \rangle \otimes  | T_{\sigma r_1} \rangle \otimes | \psi_{o} \rangle \otimes | \psi_{\infty} \rangle \text{,}
\end{equation}
where  $r_1$ and $\sigma r_1$ are in product states $|T_{r_1} \rangle$ and $| T_{\sigma r_1} \rangle$, while  $r_o$ and $r_{\infty}$ are in the states $ | \psi_{o} \rangle$ and $| \psi_{\infty} \rangle$.  

All properties of the pgSPT phase are now encoded in the two-dimensional states $|\psi_o \rangle$ and $| \psi_{\infty} \rangle$, and we can ignore the now-trivial regions $r_1$ and $\sigma r_1$.  We shall focus on $r_o$, which we view as an effective two-dimensional system, and study its properties in Sec.~\ref{sec:3dz2p-classification} to classify $3d$ pgSPT phases.  On the other hand, we ignore $r_{\infty}$.  This is justified by the point of view that the $\infty$-plane becomes spatial infinity upon taking the thermodynamic limit, so that properties localized there are not observable.  Another point of view is that, if we include translation symmetry, the relationship between the properties of $r_o$ and $r_{\infty}$ will be determined by translation symmetry and the detailed choice of periodic boundary conditions.  Therefore, we lose nothing by ignoring $r_{\infty}$ at this stage, as long as we consider translation symmetry later on, which we do in Sec.~\ref{sec:3d-translation}.

While we have chosen to describe the reduction procedure in terms of wave functions, there is a complementary viewpoint based on Hamiltonians.  We consider the Hamiltonian density in region $r_1$.  Because there is no symmetry taking this region into itself, we expect that the Hamiltonian density can be deformed to that of a trivial state without passing through any phase transitions.  This can be done preserving reflection as long as the Hamiltonian density is changed correspondingly in $\sigma r_1$.  Therefore, we can make the Hamiltonian density trivial away from the mirror plane, leaving an effectively two-dimensional system.

\subsection{Classification}
\label{sec:3dz2p-classification}

We now use reduction to the $2d$ mirror plane to show that $3d$ pgSPT phases protected by $\zz^P$ symmetry obey a $\zz \times \zz$ classification.  This and other classifications we find obey an Abelian group structure, where two SPT states ``stacked'' on top of one another result in a third SPT phase.  We refer to this operation as addition of SPT phases.

To obtain a classification, we have to answer two questions.  First,  how do states on the $2d$ mirror plane correspond to distinct pgSPT phases?  Second, what states can the $2d$ mirror plane be in, and what is the resulting classification of pgSPT phases?  Attending to the first question, there are two kinds of operations that group states on the mirror plane into equivalence classes of pgSPT phases:
\begin{enumerate}
\item   Two $2d$ states are equivalent if they are in the same $\zz$-symmetric $2d$ phase.  That is, if they are related by a local unitary preserving the $\zz$ symmetry, and/or by adding trivial degrees of freedom.

\item  Two $2d$ states are equivalent if they are related by adjoining new degrees of freedom near the boundaries of $r_o$.    Precisely, we modify the ground state in $r_o$ by
\begin{equation}
| \psi_{r_o} \rangle \to | L \rangle \otimes | \psi_{r_o} \rangle \otimes | R \rangle \text{,}  \label{eqn:2d-adding}
\end{equation}
where $| L \rangle$ and $ | R \rangle$ each describe a $2d$ ``layer'' adjoined to $r_o$, and where reflection acts by
\begin{equation}
U_{\sigma} | L \rangle = | R \rangle \text{, } U_{\sigma} | R \rangle = | L \rangle \text{.}
\end{equation}
\end{enumerate}
The second operation may be unfamiliar, but we must allow for it; physically, it corresponds to changing the extensive trivialization by expanding the size of $r_o$.  This can also be pictured as ``bringing in'' degrees of freedom from the trivial regions $r_1$ and $\sigma r_1$.  This operation will play an important role in our analysis:  there are states on the mirror plane that are distinct as $2d$ phases, but are related by adjoining layers, and thus correspond to the same pgSPT phase.  

Moving on to the second question, it is convenient to first obtain a classification of $2d$ phases that can occur on the mirror plane.  Then, we will see how this collapses to a classification of pgSPT phases, when we allow adjoining layers.  It is clear that the $2d$ system must be gapped and must preserve $\zz$ symmetry.  Moreover, there can be no excitations with non-trivial braiding (\emph{i.e.} anyons) in the $2d$ bulk, because these excitations would then also be present in the $3d$ bulk of the original pgSPT state before reduction to $2d$.  A non-trivial possibility meeting these criteria is for the $2d$ system to be in the single non-trivial SPT phase protected by $\zz$ symmetry \cite{chen13symmetry, levin12braiding}.  We shall refer to this state as \emph{the} $\zz$ SPT state, a slight but convenient abuse of terminology (the trivial state is also a SPT state protected by $\zz$ symmetry).

Na\"{\i}vely, it might appear that the $\zz$ SPT state is the only non-trivial possibility for the $2d$ system on the mirror plane, but this is not correct. We will see below that this system can also be in an ``integer'' topological state with intrinsic topological order but no anyon excitations; that is, a state that remains non-trivial even upon breaking the $\zz$ symmetry.  In a bosonic system, the only known examples of this kind are the so-called $E_8$ state \cite{kitaev11KITP,plamadeala13shortrange}, or are obtained by taking an integer $n_{E_8}$ copies of the $E_8$ state.  The edge of the $E_8$ state supports 8 co-propagating chiral boson modes (chiral central charge $c = 8$), and thus has a quantized thermal Hall effect that is robust independent of symmetry.

We have thus identified two ``root'' states on the $2d$ mirror plane, and we now argue that all possible states are obtained as an integer number of copies of the root states.  One root state is the $\zz$ SPT phase, and the other is a single copy of the $E_8$ state ($n_{E_8} = 1$).  We take the $\zz$ symmetry to act trivially on the $E_8$ root state \footnote{By ``acting trivially,'' we mean the $\zz$ symmetry acts as the identity operator on the Hilbert space of the $E_8$ root state.}. From these root states, we obtain a $\zz \times \z$ classification of $2d$ phases on the mirror plane, where the $\zz$ factor reflects the fact that two $\zz$ SPT phases add together to a trivial phase, and the $\z$ factor is simply $n_{E_8}$.  It should be noted that we have two different $E_8$ states in the presence of $\zz$ symmetry; that is, there are two states that reduce to the usual $E_8$ state if we break the $\zz$ symmetry.  One of these is the root state, on which $\zz$ acts trivially.  The other is obtained by adding the $E_8$ and $\zz$ root states together.  These states can be distinguished by gauging the $\zz$ symmetry, and studying the braiding statistics of the resulting theory, following the analysis of Levin and Gu \cite{levin12braiding}. This analysis applies without modification because the $\zz$ symmetry acts trivially on the $E_8$ root state.

Is the $\zz \times \z$ classification complete? It would be incomplete if there exist integer topological phases that are robust in the absence of symmetry, beyond those obtained from the $E_8$ root state.  Putting this possibility aside, can there still be other states beyond the $\zz \times \z$ classification?   In particular, could there be a third distinct state that also reduces to the $E_8$ state upon breaking $\zz$ symmetry?   We argue that this is unlikely.  We expect that addition of integer topological phases occurring on the mirror plane obeys an Abelian group structure.  Making this assumption, suppose the additional state we are seeking exists.  Then we can add to it an opposite-chirality $E_8$ state, and obtain a new distinct SPT phase protected by $\zz$ symmetry.  There is compelling evidence that only one non-trivial such state exists \cite{levin12braiding}, so we believe the $\zz \times \z$ classification is most likely complete, unless there are additional states with intrinsic topological order not obtained from the $E_8$ root state.

From a certain perspective, it is surprising that the $E_8$ state can occur on the mirror plane of a $3d$ pgSPT phase.  By definition, such a phase must become trivial upon breaking $\zz^P$.  However, if the mirror plane hosts an $E_8$ state after reduction to $2d$, it seems this state remains upon breaking $\zz^P$, an apparent contradiction.

\begin{figure}
\includegraphics[width=\columnwidth]{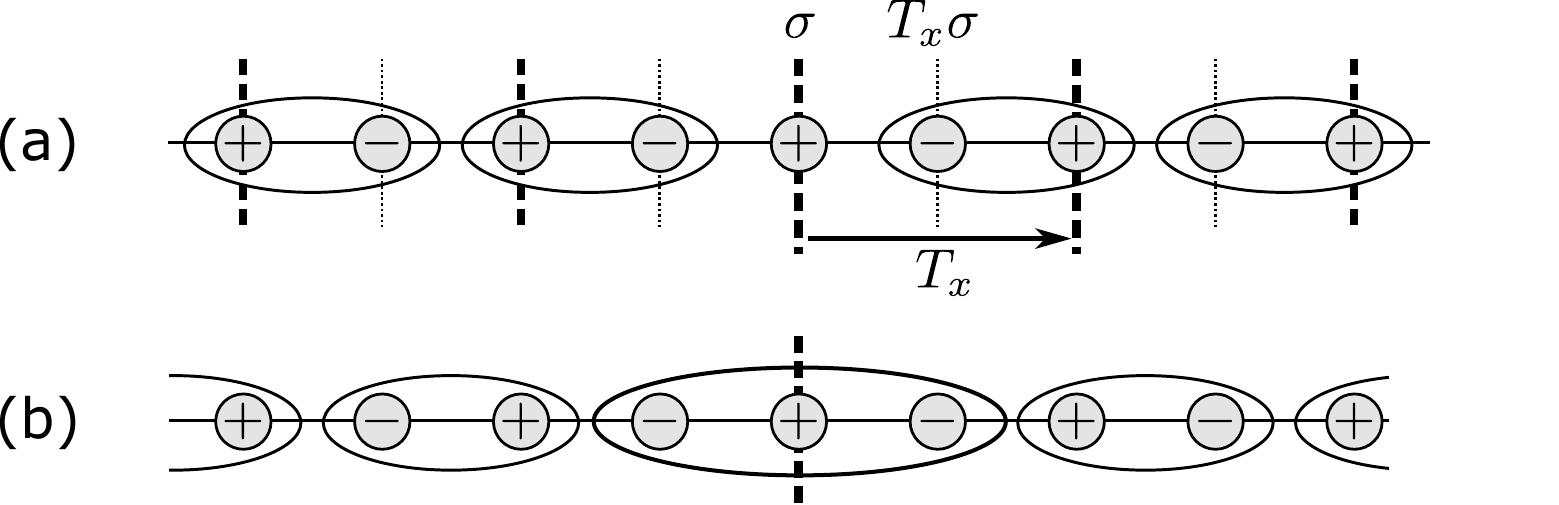}
\caption{Panel (a) depicts a system with mirror reflection $\sigma$, and discrete translation symmetry generated by $T_x$ normal to the mirror plane.  Each point on the line represents a $2d$ plane, and there are two inequivalent types of mirror planes.  One type (thick dashed lines) is obtained by translating the $\sigma$-plane, and the other type (thin dotted lines) is obtained by translating the $T_x \sigma$-plane, which is separated from the $\sigma$-plane by half a lattice.  In this setting we can have a stack of alternating-chirality $E_8$ states, where $+$ / $-$ represent $E_8$ states with $n_{E_8} = \pm 1$ on the two types of mirror planes.  Reduction to $2d$ can be visualized by pairing $E_8$ states away from the mirror plane as shown, leaving a $n_{E_8} = +1$  state on the mirror plane.  A different reduction procedure is illustrated in (b), where states are grouped to give a $n_{E_8} = -1$ state on the mirror plane.}
\label{fig:e8stack-pairing}
\end{figure}

A simple way to see there is no real contradiction is to momentarily consider adding discrete translation symmetry $T_x$ normal to the mirror plane.  This leads to two inequivalent types of mirror planes separated by half a lattice constant; as shown in Fig.~\ref{fig:e8stack-pairing}a, one type of plane is obtained by translating the $\sigma$ plane, while the other type is obtained by translating the $T_x \sigma$ plane.  We put $E_8$ states of the same chirality ($n_{E_8} = 1$) on all the $\sigma$-type planes, and $E_8$ states of the opposite chirality ($n_{E_8} = - 1$) on  the $T_x \sigma$-type planes.  
This state becomes trivial upon breaking reflection symmetry (even if translation is maintained), because adjacent pairs of opposite-chirality $E_8$ states can then be paired together and annihilated, and is thus a pgSPT phase.  Moreover, if we ignore all symmetry except for $\sigma$, our reduction procedure can lead to a single $E_8$ state ($n_{E_8} = 1$) on the $\sigma$ mirror plane, by pairing up and annihilating states away from this plane as shown in Fig.~\ref{fig:e8stack-pairing}a.

To understand how the $\zz \times \z$ gives a classification of pgSPT phases, we have to understand how the states on the mirror plane behave under adjoining layers, as in Eq.~(\ref{eqn:2d-adding}).  First of all, the same translation-symmetric example discussed above indicates that the $E_8$ index $n_{E_8}$ should only be well-defined modulo 2.  This is because we can pair up and annihilate states in a slightly different way, shown in Fig.~\ref{fig:e8stack-pairing}b, to obtain an opposite chirality ($n_{E_8} = -1$) state on the mirror plane.   The same conclusion is readily obtained from Eq.~(\ref{eqn:2d-adding}), because $|L\rangle$ and $|R \rangle$ can be $E_8$ states of the same chirality, so that adjoining layers can change the $E_8$ index of $| \psi_{r_o} \rangle$ by $\pm 2$.  Moreover, this is the only effect of adding degrees of freedom:  if the state $|L\rangle$ is not an $E_8$ state, then it should be trivial, because there is no symmetry that takes $| L \rangle$ into itself.

This discussion is not yet sufficient to completely fix the classification of pgSPT phases, but actually leaves us with two possibilities that we have to decide between.  To see why this is so, suppose we add two $E_8$ root states together, so we have $n_{E_8} = 2$ on the mirror plane.  We can then apply Eq.~(\ref{eqn:2d-adding}) to adjoin two $n_{E_8} = -1$ $E_8$ states.  The resulting state is non-chiral and has $\zz$ symmetry, so it must either be the trivial state, or the $\zz$ SPT state.  We show this state is trivial in Sec.~\ref{sec:surface}, by analyzing its surface theory.  Therefore, we obtain a $\zz \times \zz$ classification of $3d$ pgSPT phases protected by reflection symmetry.

\subsection{Role of translation symmetry}
\label{sec:3d-translation}

In crystalline solids, the reflection symmetry $\zz^P$ will always occur together with translation symmetry, which we mostly ignored in the above discussion, except in the context of the $E_8$ root state.  Here, we consider SPT phases protected by both $\zz^P$ and discrete translations normal to the mirror plane, and obtain a $(\zz)^3$ classification.  All the phases within this $(\zz)^3$ can be obtained as stacks of $2d$ topological phases.  This gives a convenient construction of $3d$ pgSPT phases that may be useful to further understand the properties of these phases in future work.

Just as for the translation-invariant stack of $E_8$ states discussed in Sec.~\ref{sec:3dz2p-classification}, we include discrete translation $T_x$ normal to the mirror plane, and ignore any translation symmetry within the mirror plane.  In the presence of both translation and reflection, there are  two types of planes of reflection symmetry, separated from one another by half a lattice constant.  The two types of planes are inequivalent in the sense that they cannot be obtained from one another by translation, or, equivalently, they are not related by conjugation in the symmetry group.   More formally, the symmetry group is generated by the reflection $\sigma$ and the elementary translation $T_x$.  These generators obey relations $\sigma^2 = 1$ and $\sigma T_x \sigma = T^{-1}_x$.  The two inequivalent reflections are $\sigma$ and $T_x \sigma$, with all other reflections related to one of these by conjugation, so we refer to $\sigma$-type and $T_x \sigma$-type reflections.

We can focus on any reflection operation, and reduce the system to a $2d$ topological phase on the corresponding plane.  Translation symmetry requires all $\sigma$-type planes to be in the same $2d$ state, and similarly for all $T_x \sigma$-type planes.  Then, from the $\zz$ root state, we can obtain two distinct root states protected by both $\zz^P$ and translation symmetry.  In one of these, the $\sigma$-type planes are in the $\zz$ SPT state while the $T_x \sigma$-type planes are trivial.  In the other state this is reversed, with the $T_x \sigma$-type planes in the non-trivial $\zz$ SPT state.  These root states generate a $\zz \times \zz$ classification.

The situation is different for the $E_8$ root state.  If we put the $\sigma$-type planes in $E_8$ states with $n_{E_8} = 1$, while keeping the $T_x \sigma$-type planes trivial, we do not have a pgSPT phase.  One way to see this is to note that any $2d$ surface cutting through the mirror planes is a chiral thermal metal that cannot be gapped out.  Therefore, if the $\sigma$-type planes have $n_{E_8} = 1$, we must put the $T_x \sigma$-type planes into $E_8$ states of opposite chirality, \emph{i.e.} $n_{E_8} = -1$.  We are thus led to the same translation-symmetric example discussed in Sec.~\ref{sec:3dz2p-classification}, and we  obtain a single translation-invariant root state from the $E_8$ root state, which generates a $\zz$ factor in the classification including translation symmetry.  Here, focusing on any particular reflection operation and ignoring other symmetries, we have the $E_8$ root state on the corresponding mirror plane.  It should be noted that we do not obtain a different phase upon reversing the overall chirality, because chirality can be reversed by adjoining layers of $E_8$ state as mentioned earlier.

Combining the three root states together, we obtain a $(\zz)^3$ classification upon including translation symmetry.
We make two comments on this result before proceeding.  First, it is \emph{not} the case that all $3d$ pgSPT phases with $\zz^P$ and translation symmetry are simply a stack of $2d$ states at the microscopic level.  However, it is true that all such phases are adiabatically connected to a $2d$ state if we focus on one specific reflection operation.  Second, there is no guarantee that we have found all SPT phases protected by both reflection and translation.  In principle, we can imagine phases with a non-trivial interplay between reflection and translation symmetries that are not captured in our approach.

\section{Surfaces of bosonic point group SPT phases}
\label{sec:surface}

We now discuss symmetry-preserving surfaces of $3d$ pgSPT phases protected by $\zz^P$ reflection symmetry.  For now, we ignore any translation symmetry.  We focus on  two types of surface states.  In the first type, the surface is gapped and trivial away from the mirror plane, and the $1d$ edge of the mirror plane is gapless.  Second, we consider gapped surfaces with topological order, in the sense that anyon quasiparticle excitations are present.  Both types of surfaces are interesting in their own right, and also allow us to establish the $\zz \times \zz$ classification, by showing that adding two $E_8$ root states results in a trivial phase.  We believe it will be interesting to study other possible surface states, a problem that we leave for future work.

The gapped surface states have the crucial property that the action of symmetry is anomalous, by which we mean it cannot be realized strictly in two dimensions.  The surface can thus be viewed as an anomalous $2d$ symmetry enriched topological (SET) phase.  When symmetry does not permute the distinct types of anyon excitations, the symmetry action fractionalizes into an action on individual anyon quasiparticles, and such anomalous SET phases are said to exhibit anomalous symmetry fractionalization.  Some SPT phases protected by internal symmetry \cite{vishwanath13, cwang13boson, chen14anomalous}, or a combination of ${\rm U}(1)$ and crystalline symmetry \cite{hermele15}, can have anomalous SET surfaces.  Apart from one recent study on electronic topological crystalline insulators \cite{yqi15b}, less is known about anomalous symmetry fractionalization at surfaces of SPT phases protected only by crystalline symmetry.

\subsection{$\zz$ root state}
\label{sec:z2root}

\begin{figure}
\includegraphics[width=0.6\columnwidth]{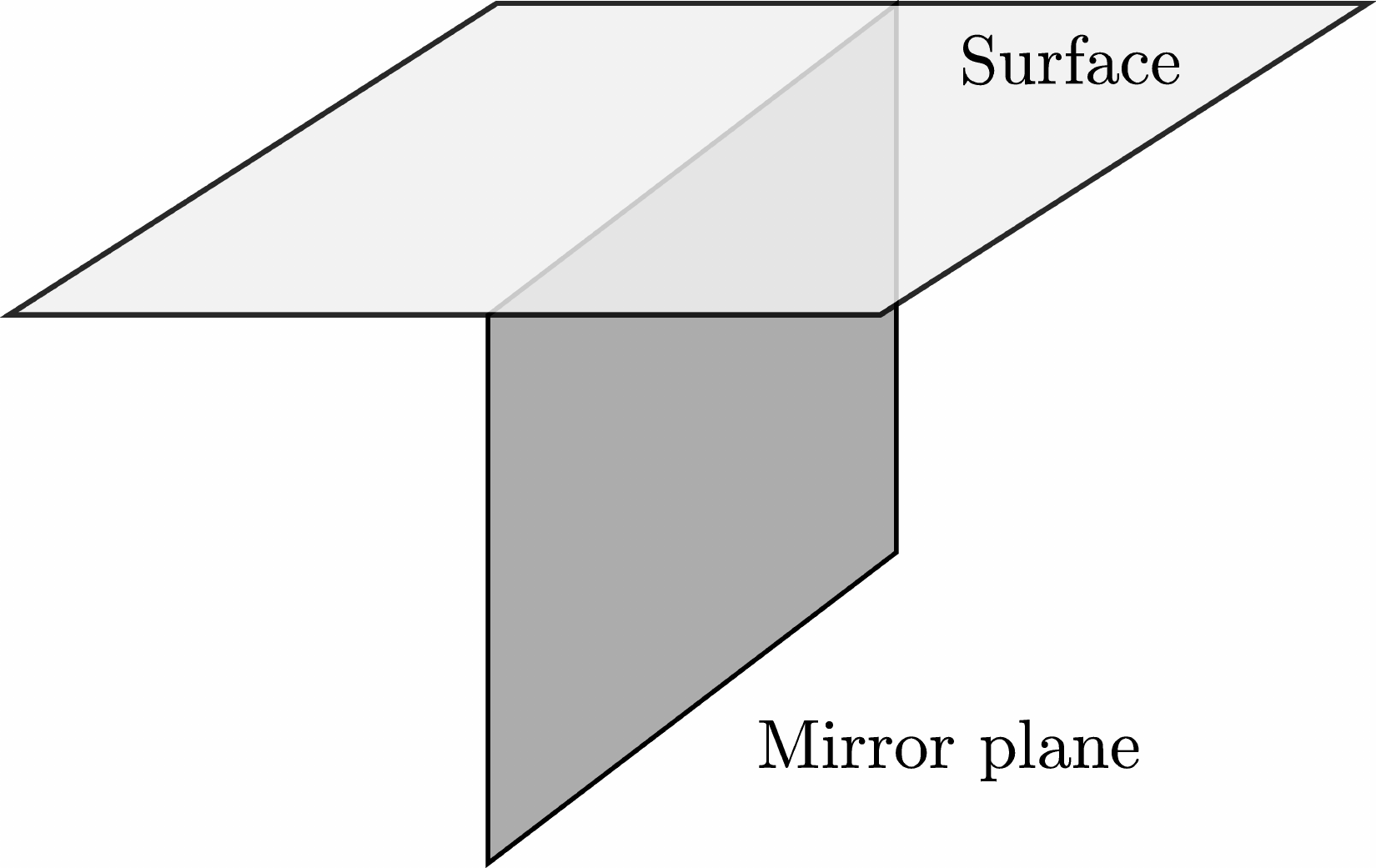}
\caption{Geometry of a symmetry preserving surface of a $3d$ pgSPT phase protected by $\zz^P$ reflection symmetry, ignoring any translation symmetry.  In the bulk, the system has been reduced to a $2d$ state lying on the mirror plane.  The edge of the mirror plane coincides with the reflection axis of the surface.} 
\label{fig:z2p-surface-geometry}
\end{figure}

We first discuss the $\zz$ root state.  Upon reduction to $2d$, there is a $\zz$ SPT state on the mirror plane.  This plane and a symmetry-preserving $2d$ surface form a ``T'' geometry, as shown in Fig.~\ref{fig:z2p-surface-geometry}.  The intersection of the mirror plane and the surface is both the edge of the SPT state on the mirror plane, and the reflection axis of the $2d$ surface.  Reflection acts on this edge as an on-site, unitary $\zz$ symmetry.  Away from the reflection axis, the surface degrees of freedom behave as in an ordinary $2d$ system. If the surface degrees of freedom away from the reflection axis are in a trivial gapped state, the surface properties are simply those of the $\zz$ SPT edge \cite{chen11b,levin12braiding}: there are either gapless reflection-protected edge modes, or the reflection symmetry is spontaneously broken.

Now we will show that there is another possibility, namely that the surface is gapped with $\zz$ topological order.  This is the topological order of the toric code model \cite{kitaev03}, or, equivalently, the deconfined phase of $2d$ Ising gauge theory.  This type of topological order has anyons $e$ and $m$, that can be thought of as bosonic $\zz$ gauge charges and bosonic $\zz$ gauge fluxes, respectively.  While these excitations have bosonic self-statistics, they should still be viewed as anyons due to their $\Theta = \pi$ mutual statistics, which is simply the Ising version of the Aharonov-Bohm effect.  This implies that the composite of $e$ and $m$, $\epsilon = e m$, is a fermion.  The fusion and braiding properties are invariant under the relabeling $e \leftrightarrow m$, so there is an arbitrary choice of which particle we call $e$ and which we call $m$.  Building on the theory of projective symmetry group for parton mean-field theories \cite{wen02}, distinct crystal symmetry fractionalization patterns have been classified \cite{essin13}, without regard to possible anomalies.  

When the surface has $\zz$ topological order, the edge of the mirror plane can be gapped out without breaking symmetry, leading to anomalous reflection symmetry fractionalization at the surface.  We establish this by constructing and solving an effective model for the surface.

% To get transparency right, first create pdf using inkscape, then rasterize it by:
%  ps2pdfwr -dCompatibilityLevel=1.4 -dHaveTransparency=false input.pdf output.pdf

Our construction is based on an effective model for the edge of the $\zz$ SPT phase, introduced in Ref.~\onlinecite{chen11b} and dubbed the CZX model; a closely related model was also introduced in Ref.~\onlinecite{levin12braiding}.  In the model of Ref.~\onlinecite{chen11b}, the edge is a $1d$ chain of Ising spins located at sites labeled by the integer $j$, with Pauli spin operators $\tau^z_j$, $\tau^x_j$.  The Ising symmetry is realized by
\begin{equation}
U_I = \prod_j \tau^x_j \prod_j CZ_{j, j+1} \text{,}
\end{equation}
where 
\begin{equation}
CZ_{i,j} =  \lvert \uparrow \uparrow \rangle \langle \uparrow \uparrow \rvert + 
 \lvert \uparrow \downarrow \rangle \langle \uparrow \downarrow \rvert + 
  \lvert \downarrow \uparrow \rangle \langle \downarrow \uparrow \rvert - 
   \lvert \downarrow \downarrow \rangle \langle \downarrow \downarrow \rvert \text{.}
\end{equation}
is the controlled-$Z$ operation acting on the pair of spins labeled by $i$ and $j$.  This acts on spin operators by
\begin{eqnarray}
U_I \tau^z_j U^{-1}_I &=& - \tau^z_j  \nonumber \\
U_I \tau^x_j U^{-1}_I &=&   \tau^z_{j-1} \tau^x_j \tau^z_{j+1} \label{eqn:czx} \\
U_I \tau^y_j U^{-1}_I &=&  - \tau^z_{j-1} \tau^y_j \tau^z_{j+1} \nonumber \text{.}
\end{eqnarray}
This ``non-on-site'' action of symmetry encodes the anomalous properties of the edge of the $2d$ $\zz$ SPT phase.

Our effective $2d$ surface model has Ising spins residing on the edges of a $2d$ square lattice.  We choose the origin so that the centers of horizontal edges have coordinates $\br = (x,y)$ with $x,y$ integers, while the centers of vertical edges have $x$ and $y$ half-odd integers.  Under reflection symmetry $U_{\sigma}$, the spins on the reflection axis at $x=0$ transform as the spins of the CZX model boundary under Ising symmetry.  Precisely, the spin at $\br = (0,j)$ transforms under $U_{\sigma}$ exactly as in Eq.~(\ref{eqn:czx}).  The remaining spins obey the ordinary transformation law
\begin{equation}
U_{\sigma} \tau^{\mu}_{(x,y)} U^{-1}_{\sigma} = \tau^{\mu}_{(-x,y)}, \qquad x \neq 0\text{,} \label{eqn:ordinary}
\end{equation}
where $\mu = x,y,z$.

\begin{figure}
\includegraphics[width=0.8\columnwidth]{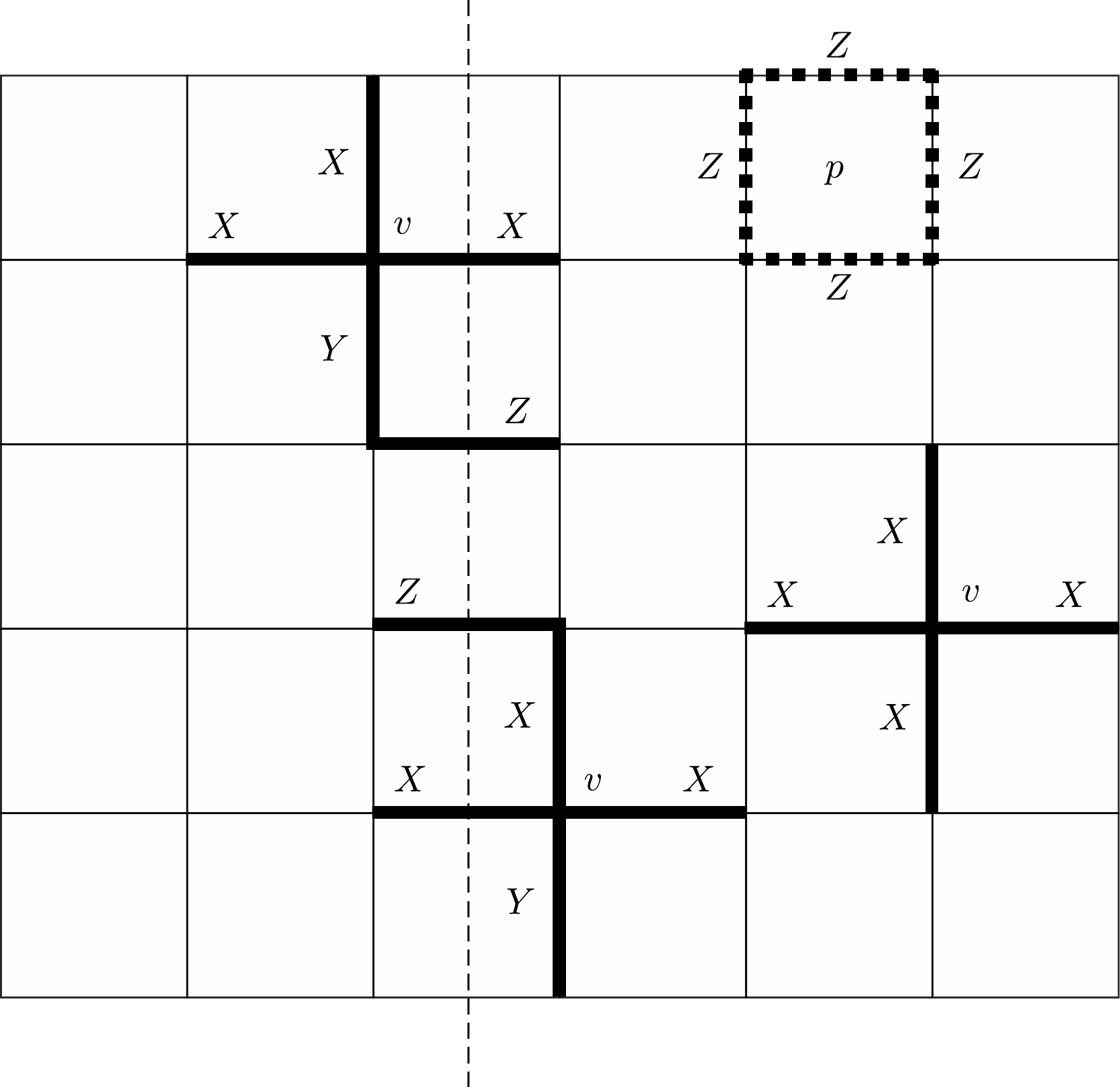}
\caption{Operators in the modified toric code model at the surface of the $\zz$ root state.  Three vertex operators $A_v$ are shown, with two adjacent to the reflection axis (dashed line), and one away from it.  Each operator is a product of Pauli spin operators on the edges marked by thick solid lines, with $X,Y,Z$ corresponding to $\tau^x, \tau^y, \tau^z$.  Plaquette operators $B_p$ are products of four $\tau^z$ operators around the perimeter of a plaquette $p$, as indicated by thick dotted lines.} 
\label{fig:modified-toric-code}
\end{figure}

The Hamiltonian is a variant of the toric code model, and can be written
\begin{equation}
H = - \sum_v A_v - \sum_p B_p \text{.}
\end{equation}
The first term is a sum over vertices $v = (v_x, v_y)$ of the square lattice, and the second term is a sum over plaquettes $p$, with operators $A_v$ and $B_p$ associated with each vertex and plaquette, respectively.  The plaquette operators are identical to those in the ordinary toric code,
\begin{equation}
B_p = \prod_{\br \sim p} \tau^z_{\br} \text{,}
\end{equation}
where the product is over the perimeter of the plaquette $p$ (Fig.~\ref{fig:modified-toric-code}).  If the vertex operators were also chosen identical to the ordinary toric code, this would not respect the anomalous action of mirror symmetry at the reflection axis.  To handle this, we modify the form of $A_v$ for vertices adjacent to the reflection axis, while, elsewhere, we choose $A_v$ as in the ordinary toric code.  For vertices away from the axis, we define
\begin{equation}
A_v = \prod_{\br \sim v} \tau^x_{\br} , \qquad v_x \neq \pm 1/2 \text{,}
\end{equation}
where the product is over the four edges touching the vertex $v$.  Then, for vertices adjacent to the axis, we choose
\begin{equation}
A_v = \left\{ \begin{array}{ll}
\left[ \prod'_{\br \sim v} \tau^x_{\br} \right] \tau^y_{v - \by/2} \tau^z_{v + \bx/2 - \by} , & v_x = -1/2 \\
- \left[ \prod'_{\br \sim v} \tau^x_{\br} \right] \tau^y_{v - \by/2} \tau^z_{v - \bx/2 + \by} , & v_x = 1/2 \text{,}
\end{array}\right.
\end{equation}
where $\bx = (1,0)$ and $\by = (0,1)$, and $\prod'_{\br \sim v}$ is a product over the edges touching $v$, excluding the  edge below.  A graphical representation of these operators is shown in Fig.~\ref{fig:modified-toric-code}.

It is straightforward to check that the Hamiltonian thus defined is invariant under the reflection symmetry, and is exactly solvable as the vertex and plaquette operators form a commuting set of observables.  It is thus not surprising that this model shares many properties with the ordinary toric code.  In particular, there is an energy gap, the mirror symmetry is unbroken in the ground state, and there is $\zz$ topological order.  $e$ particles reside at vertices where $A_v = -1$, and $m$ particles, except those at $x=0$, reside at plaquettes with $B_p = -1$.  The string operators that move $e$ and $m$ particles are products of $\tau^z$ and $\tau^x$ Pauli operators, respectively, except that  $m$ strings are decorated with a $\tau^z$ Pauli operator whenever they cross the reflection axis, as shown in Fig.~\ref{fig:strings}.  Some of the details underlying these statements are given in Appendix~\ref{app:toricmod}.

\begin{figure}
\includegraphics[width=0.7\columnwidth]{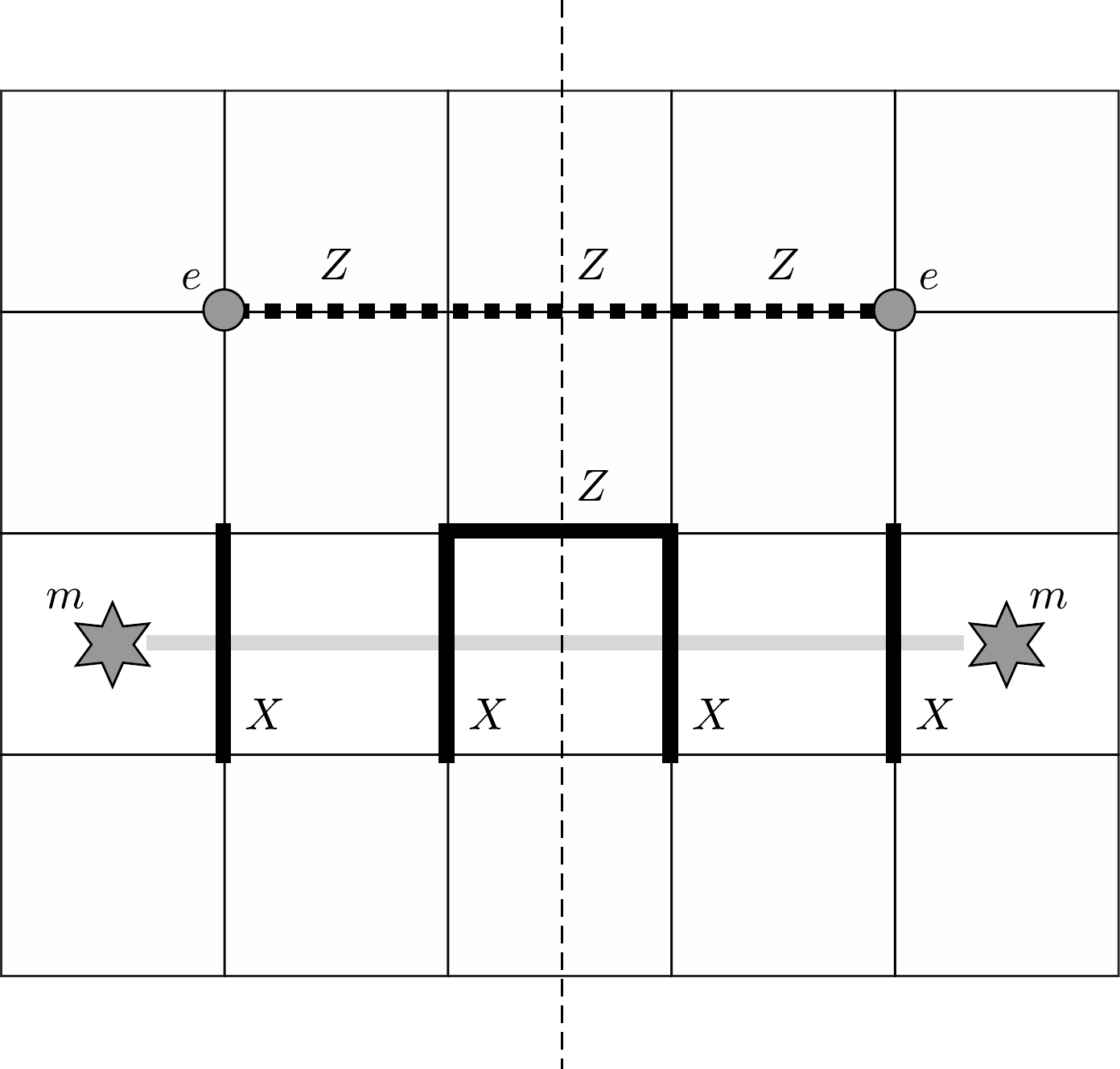}
\caption{String operators, in the modified toric code model, creating a reflection-symmetric pair of $e$ particles (top) and $m$ particles (bottom).  Operators in the $e$-string ($m$-string) are indicated by thick dotted (solid) lines.  The $m$-string, whose path is shown by the light gray line, is decorated with a single $\tau^z$ operator at the reflection axis (dashed line).}
\label{fig:strings}
\end{figure}

For a single reflection $\sigma$ that does not exchange $e \leftrightarrow m$, as is the case here, the symmetry fractionalization pattern can be described by introducing operators $U^{e,m}_\sigma$ that give the action of $\sigma$ on a single $e$ or $m$ particle, respectively \cite{essin13}.  We have
\begin{eqnarray}
( U^e_{\sigma} )^2  &=& \mu^e_{\sigma}  \\
( U^m_{\sigma} )^2  &=& \mu^m_{\sigma} \text{,}
\end{eqnarray}
where $\mu^e_{\sigma}, \mu^m_{\sigma} = \pm 1$.  This apparently gives four possible symmetry fractionalization patterns, but only three are distinct under the relabeling $e \leftrightarrow m$.  Paralleling notation introduced in Ref.~\onlinecite{cwang13boson}, we denote these by $e0m0$, $ePm0$ and $ePmP$, where $0$ ($P$) indicates $\mu_{\sigma} = 1$
($\mu_{\sigma} = -1$) for the corresponding particle.  Both $e0m0$ and $ePm0$ can be realized strictly in two dimensions \cite{hsong15}.

Reflection symmetry fractionalization can also be characterized without introducing the operators $U^{e,m}$.  We consider a string operator  $S^e$ ($S^m$) creating two $e$ ($m$) particles at positions related by reflection symmetry.
Then it can be shown \cite{yqi15detecting,zaletel15} that the reflection eigenvalue of the string operator is the same as the $\mu_{\sigma}$ parameter describing the corresponding anyon's symmetry fractionalization; that is,
\begin{equation}
U_{\sigma} S^{e,m} U^{-1}_{\sigma} = \mu^{e,m}_{\sigma} S^{e,m}  \text{.}
\end{equation}
This phenomenon, which can also be understood in terms of dimensional reduction to a $1d$ SPT phase protected by reflection symmetry \cite{zaletel15}, can in fact be viewed as an alternate definition of reflection symmetry fractionalization that does not require describing the action of symmetry on a single anyon.

Using this characterization of reflection symmetry fractionalization, it is straightforward to see that the $ePmP$ fractionalization pattern is realized in our model.  String operators creating a pair of $e$ particles and a pair of $m$ particles at positions related by reflection symmetry are shown in Fig.~\ref{fig:strings}, and both of these string operators are odd under reflection.

We have shown that $ePmP$ occurs at the surface of a non-trivial SPT phase, and it is thus natural to expect this fractionalization pattern is anomalous.  To see this is indeed the case, we will assume that there is a $2d$ system realizing  $ePmP$, and obtain a contradiction.  We add a layer of this $2d$ system to the $ePmP$ surface of the $3d$ pgSPT phase.  The resulting surface has 16 types of anyons, labeled by pairs $(a_1, a_2)$, where $a_1, a_2 = 1,e,m,\epsilon$ are particle types in the two $ePmP$ layers.  Because $e$ and $m$ particles in the two layers transform identically under reflection, the composites $(e,e)$ and $(m,m)$ transform trivially (\emph{i.e.} reflection squares to unity acting on these particles), so they can be condensed without breaking symmetry.  The resulting condensate confines all the other anyons, and we have thus obtained a gapped, symmetry-preserving surface with no topological order.  The surface can then be trivialized away from the reflection axis following the same procedure used to classify pgSPT phases, and effectively becomes a gapped, symmetric system on the reflection axis.  But this is a contradiction, because we have gapped out the edge of the non-trivial $2d$ $\zz$ SPT phase on the mirror plane, without breaking symmetry.

We note that the $ePmP$ fractionalization pattern has previously been argued to be anomalous in Ref.~\onlinecite{yqi15b}.  This was done by considering an electronic topological crystalline insulator with $\zz^P$ symmetry and $n=4$ Dirac cones, and putting the surface into the $ePmP$ state.  Our result confirms this conclusion from a different point of view.  In particular, the $ePmP$ state is a bosonic anomalous SET phase, so we should expect, as we have shown, that it can be realized at the surface of a bosonic SPT phase.

\subsection{$E_8$ root state}
\label{sec:e8root}

Here we turn to the $E_8$ root state.  Upon reduction to $2d$, $n_{E_8}$ copies of the $E_8$ state lie on the mirror plane, with $n_{E_8}$ odd.  Therefore, if the surface is gapped and trivial away from the mirror plane, the surface supports a gapless chiral $1d$ system on the reflection axis, with chiral central charge $c \operatorname{mod} 16 = 8$.

Such an effective $1d$ system cannot occur on the reflection axis of a strictly $2d$ system with $\zz^P$ symmetry, and where we assume no anyon excitations are present.  In this case, the only known non-trivial possibility is that one side of the reflection axis is in an $E_8$ state with index $n_{E_8}$.  The other side of the reflection axis then necessarily has $E_8$ index $-n_{E_8}$.  On the reflection axis, we then have gapless modes with chiral central charge $c = 16 n_{E_8}$, and $c \operatorname{mod} 16 = 0$.  

We now use this $1d$ edge theory to construct a trivial, gapped surface termination of the $E_8 \oplus E_8$ state, obtained by adding two $E_8$ root states.  This indicates that $E_8 \oplus E_8$ is a trivial pgSPT phase.  To proceed, we will need a concrete description of the $E_8$ state on the $2d$ mirror plane\cite{kitaev11KITP}.  This state can be described as a $\zz$ gauge theory; we start with a $\nu = 8$  IQH state, and couple the fermion parity to a deconfined $\zz$ gauge field.  The ${\rm U}(1)$ symmetry of the $\nu = 8$ state does not play a role, so we can  also view it as sixteen copies of a $p + i p$ topological superconductor.  The resulting state has toric code topological order, so the gauge flux is a boson and can be condensed, which results in the $E_8$ state.  We take the $\zz^P$ symmetry, which acts on the mirror plane as a $\zz$ on-site symmetry, to act trivially on the fermion and gauge field degrees of freedom.

To construct a gapped surface, we first consider a state with two different $\nu = 8$ IQH states on the mirror plane, each coupled to its own deconfined $\zz$ gauge field.  Then, we describe how to gap the edge.  Finally, we condense $\zz$ gauge fluxes in the bulk, thus producing a bulk $E_8 \oplus E_8$ state with a trivial, gapped surface.  The edge Hamiltonian density is
\begin{equation}
{\cal H} = - i v \sum_{I = 1}^8 c^\dagger_{1 I} \partial_x c^{\vphantom\dagger}_{1 I}  - i v \sum_{I = 1}^8 c^\dagger_{2 I} \partial_x c^{\vphantom\dagger}_{2 I}  \text{.}
\end{equation}
Here,  $c_{1 I}$ and $c_{2 I}$ are the chiral edge fermions of two different $\nu = 8$ IQH states.  In the bulk, the $c_{1 I}$ fermions are coupled to one $\zz$ gauge field, while the $c_{2 I}$ fermions are coupled to another.  Both gauge fields are in the deconfined phase, so we can ignore coupling between the edge fermions and the $\zz$ gauge fields.  Reflection symmetry acts trivially on the fermions; that is, $\sigma : c_{j I} \to c_{j I}$.

As it stands, we have a chiral edge that cannot be gapped.  We now adjoin two counter-propagating $E_8$ layers, resulting in a non-chiral edge.  These new layers are also described, for the moment, as $\nu = -8$ IQH states coupled to $\zz$ gauge fields, so that in total we have four different $\zz$ gauge fields, whose gauge fluxes eventually need to be condensed.  The corresponding edge fermions of the new layers are $d_{L I}$ and $d_{R I}$, obeying the Hamiltonian density
\begin{equation}
{\cal H}_{{\rm adjoined}} = i v \sum_{I = 1}^8 d^\dagger_{L I} \partial_x d^{\vphantom\dagger}_{L I}  + i v \sum_{I = 1}^8 d^\dagger_{R I} \partial_x d^{\vphantom\dagger}_{R I}  \text{.}
\end{equation}
For simplicity of notation we have taken all velocities to have the same magnitude; this assumption plays no role in our analysis.  Reflection acts on these fields by $\sigma : d_{R I} \leftrightarrow d_{L I}$.  We introduce linear combinations $d_{\pm I} = (d_{R I} \pm d_{L I} ) / \sqrt{2}$, on which reflection acts by $\sigma : d_{\pm I} \to \pm d_{\pm I}$.

We now add a mass term
\begin{equation}
\delta {\cal H} = m \sum_{I = 1}^8 (c^\dagger_{2 I} d^{\vphantom\dagger}_{+ I} + \text{H.c.} ) \text{,}
\end{equation}
which gaps out the $c_{2 I}$ and $d_{+ I}$ fermions.\footnote{This term is not gauge invariant, but it is allowed if appropriate bosonic Higgs fields are condensed at the edge.} This leaves gapless the counter-propagating $c_{1I}$ and $d_{- I}$ modes, which cannot be gapped out at the non-interacting level, because they have opposite reflection eigenvalues.  However, it has been shown that a theory of four pairs of counter-propagating fermions with opposite eigenvalues under a $\zz$ symmetry can be gapped, while preserving symmetry, by a suitable interaction term \cite{ryu12interacting,xlqi13,yao13interaction,gu14effect}.  The present theory, with eight counter-propagating pairs of modes, is just two decoupled copies of this theory, so the same conclusion holds.  Now that we have fully gapped out the edge, we can condense all four $\zz$ gauge fluxes in the bulk, obtaining a description of the $E_8 \oplus E_8$ state with a gapped, trivial surface.

Returning to the $E_8$ root state itself, we construct a different type of gapped surface, with three-fermion $\zz$ topological order, which is a variant of the toric code theory discussed above.  Here, there are three non-trivial particles $e_f, m_f$ and $\epsilon_f$, which all have fermionic self-statistics.  Any pair of distinct non-trivial particles has $\Theta = \pi$ mutual statistics.  The fusion rules are $e_f^2 = m_f^2 = \epsilon_f^2 = 1$, and $\epsilon_f = e_f m_f$.  Similarly to the $\zz$ gauge theory description of the $E_8$ state, the three-fermion state can be realized by starting with a $\nu = 4$ IQH state, and then coupling the fermion parity to a $\zz$ gauge field in its deconfined phase \cite{kitaev06}.  In the absence of the fermionic matter, this $\zz$ gauge theory would realize the toric code statistics discussed in Sec.~\ref{sec:z2root}.  Here, the topologically non-trivial fermions modify the statistics of the gauge theory, and we obtain the three-fermion state.  We  label the fermionic $\zz$ gauge charge by $e_f$, and the fermionic $\zz$ gauge flux by $m_f$.

In a strictly $2d$ system, the three-fermion state is incompatible with reflection symmetry, because it has chiral edge modes with chiral central charge $c \operatorname{mod} 8 = 4$ \cite{kitaev06}.   Therefore, any reflection-symmetric realization of the three-fermion state is anomalous.  For such a state, we can go further and study the action of reflection symmetry on the anyons.  Assuming reflection does not permute the anyons, we find there are two such actions, one of which is realized at the surface of the $E_8$ root state, while the other is realized when we add together the $E_8$ and $\zz$ root states.

\begin{figure}
\includegraphics[width=0.9\columnwidth]{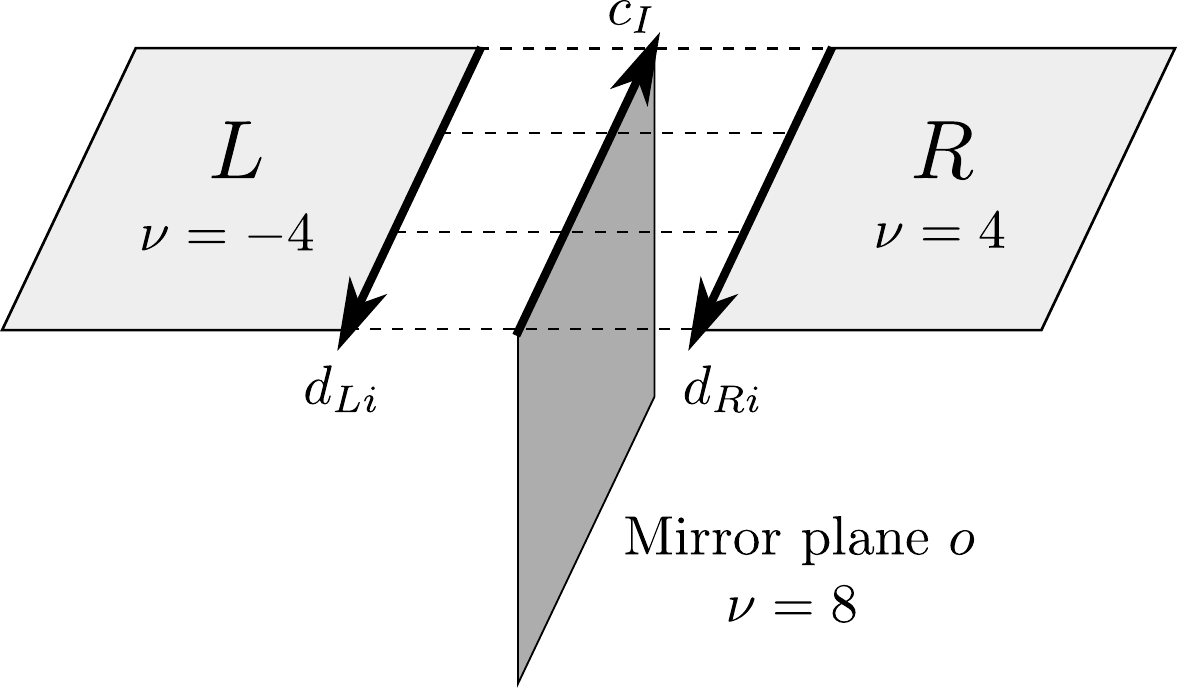}
\caption{Construction of the gapped surface for the $E_8$ root state.  A $\nu = 8$ IQH state lies on the mirror plane, while $\nu = \pm 4$ IQH states lie on the surface in regions $R$ and $L$, respectively.  Each of these three regions is a half plane, with edges supporting chiral modes indicated by the dark lines.  The $1d$ fermion fields are $c_I$, $d_{R i}$, $d_{L i}$, with chiralities as indicated.  The same $\zz$ gauge field, which resides on the ``T-shaped'' lattice formed by the union of the three regions, and connects the regions as indicated by the dashed lines, is coupled to the fermion parity.}
\label{fig:e8surface}
\end{figure}

Now we can construct the three-fermion surface of the $E_8$ root state.  Our construction is similar to the argument above that the $E_8 \oplus E_8$ state is trivial, and proceeds in a few steps.  First, we consider a theory of non-interacting fermions, where we put a $\nu = 8$ IQH state on the mirror plane, and $\nu = \pm 4$ IQH states on the surface regions $R$ and $L$, as shown in Fig.~\ref{fig:e8surface}.  These states have chiral edge modes as shown in Fig.~\ref{fig:e8surface}, all lying along the edge of the mirror plane.  The edge fermion fields for the $\nu = 8$ state are denoted $c_I$, with $I = 1,\dots,8$, and the edge fields for the $\nu = \pm 4$ states are $d_{R i}$ and $d_{L i}$, respectively, with $i = 1,\dots,4$.  Eventually, all three regions will be coupled to the same $\zz$ gauge field, and the $\zz$ flux will be condensed only on the mirror plane, so that the mirror plane is an $E_8$ state, and the surface is in the three-fermion state.  Before introducing the gauge field, we will first show that all the edge fermions can be gapped while preserving reflection symmetry.

It is sufficient for our purposes to consider any convenient edge Hamiltonian density.  We start with the simple choice
\begin{equation}
{\cal H} = - i v_c  \sum_{I = 1}^8 c^{\dagger}_I \partial_x c_I + i v_d \sum_{i = 1}^4  \big[ d^{\dagger}_{R i} \partial_x d^{\vphantom\dagger}_{R i}
+ d^{\dagger}_{L i} \partial_x d^{\vphantom\dagger}_{L i} \big] \text{,} \label{eqn:free-edge}
\end{equation}
with velocities $v_c, v_d > 0$, and will add terms as needed to open a gap.  So far, the mirror plane and the regions $R$ and $L$ are all decoupled, and there are three independent $\zz$ fermion parity symmetries.

Reflection symmetry acts on the fermion fields by  $\sigma : c_I \to c_I$ and $\sigma : d_{R i} \leftrightarrow d_{L i}$.
We introduce linear combinations $d_{\pm i} = \frac{1}{\sqrt{2}} ( d_{R i} \pm d_{L i} )$, which satisfy $\sigma : d_{\pm i} \to \pm d_{\pm i}$.
We can then gap out half of the edge modes at the non-interacting level, by adding the mass term
\begin{equation}
\delta {\cal H} = m \sum_{i = 1}^4 ( d^{\dagger}_{+ i} c^{\vphantom\dagger}_{i + 4} + \text{H.c.} ) \text{.}
\end{equation}
In addition to gapping out some of the modes, this term breaks the three fermion parity symmetries down to a single $\zz$ fermion parity, under which all the fermion fields acquire a minus sign.

This leaves four pairs of counter-propagating gapless modes.  Each pair of consists of $c_i$ and $d_{- i}$ fermions ($i = 1,\dots,4$), which are even and odd under reflection, respectively.  Again, this theory can be gapped out, preserving symmetry, by a suitable short-range interaction\cite{ryu12interacting,xlqi13,yao13interaction,gu14effect}.

Now that our theory of fermions has been gapped, we introduce a $\zz$ gauge field on the ``T-shaped'' lattice on which the fermions reside, as shown in Fig.~\ref{fig:e8surface}.  This lattice joins the mirror plane together with regions $L$ and $R$ (dashed lines in Fig.~\ref{fig:e8surface}).  The $\zz$ gauge field is minimally coupled to the fermion parity, and for the moment we suppose the gauge field is put in its deconfined phase in all three regions.  Regions $L$ and $R$ are now in the three-fermion state, while the mirror plane has toric code topological order.

\begin{figure}
\includegraphics[width=0.9\columnwidth]{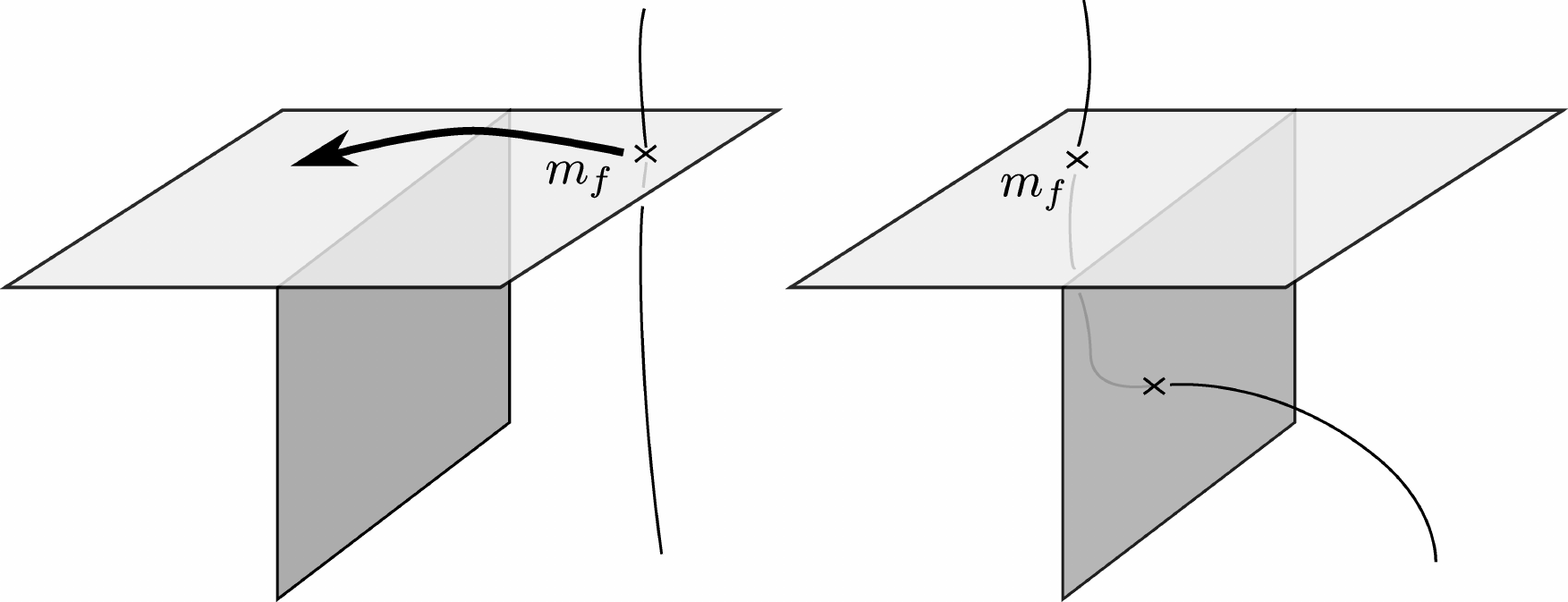}
\caption{In our construction of the three-fermion surface of the $E_8$ root state, before condensing the $\zz$ gauge flux on the mirror plane, regions $R$ and $L$ of the surface are in the three-fermion state, while the mirror plane is also in a deconfined phase of $\zz$ gauge theory, with toric code statistics.  Here, in the left panel, a $\zz$ gauge flux ($m_f$) in region $R$ is moved to region $L$, leading to the configuration shown in the right panel.  This process leaves behind a $\zz$ gauge flux excitation on the mirror plane, which can be understood by viewing the $\zz$ flux excitations as intersection points between the various regions and a flux line in three-dimensional space, as shown.  This residual excitation is eliminated upon condensing $\zz$ fluxes on the mirror plane to produce the $E_8$ root state in the bulk.}
\label{fig:moving-mf}
\end{figure}

To show we have realized the three-fermion state on the surface, it is not enough to show that each of regions $L$ and $R$ is in this state.  We also have to show that anyons in one region are free to move into the other region.  This is indeed true for the $\zz$ gauge charge $e_f$.  Before gauging, $e_f$ is nothing but a fermion excitation on the surface, and because only a single $\zz$ fermion parity is present, a fermion in $R$ can pass through the mirror plane to become a fermion in $L$, and vice versa.  What about the $\zz$ flux $m_f$?  Moving such an excitation from $R$ to $L$ leaves behind a $\zz$ gauge flux in the mirror plane, as illustrated in Fig.~\ref{fig:moving-mf}.  Fortunately, this undesired excitation is eliminated precisely by the remaining step in our construction, which is to condense the gauge flux in the mirror plane.  Upon doing this, we have an $E_8$ state on the mirror plane, and $m_f$ can move freely between $L$ and $R$ on the surface.

Now that we have obtained a reflection-symmetric realization of the three-fermion state, we can go further and characterize the action of reflection on the anyons $e_f, m_f$ and $\epsilon_f$.  As discussed in Sec.~\ref{sec:z2root}, this can be done by introducing string operators $S^a$ that create two anyons of type $a$ in positions related by reflection symmetry.  These operators transform under reflection by
\begin{equation}
U_{\sigma} S^a U^{-1}_{\sigma} = \mu^a_{\sigma} S^a \text{,}
\end{equation}
where the  $\mu^a_{\sigma} = \pm 1$ characterize the reflection symmetry fractionalization.  Because we can choose $S^{\epsilon_f} = S^{e_f} S^{m_f}$, we have $\mu^{\epsilon_f}_{\sigma} = \mu^{e_f}_{\sigma} \mu^{m_f}_{\sigma}$.  While we prefer to work with symmetry fractionalization defined in terms of string operators in this case, we note that if we also introduce operators $U^a_{\sigma}$ giving the action of $\sigma$ on a single fermionic anyon, it has been shown that\cite{yqi15detecting,zaletel15}
\begin{equation}
(U^a_{\sigma})^2 = - \mu^a_{\sigma} \text{.}
\end{equation}
This important minus sign, which was missed in Ref.~\onlinecite{essin13}, is not present for the bosonic $e$ and $m$ particles discussed in Sec.~\ref{sec:z2root}.

There are two distinct patterns of reflection symmetry fractionalization possible for the three-fermion state.  One of these has $\mu^{e_f}_{\sigma} = \mu^{m_f}_{\sigma} = 1$, and we refer to this as $e_f 0 m_f 0$.  The other has $\mu^{e_f}_{\sigma} = \mu^{m_f}_{\sigma} = -1$, and is referred to as $e_f P m_f P$.  While two other choices of $\mu^{e_f}_\sigma, \mu^{m_f}_{\sigma}$ are possible, these are equivalent to $e_f P m_f P$ under a relabeling of anyons.  It is important to note that any permutation of $e_f, m_f, \epsilon_f$ is a legitimate relabeling in the three-fermion state.  Therefore, unlike for the toric code, $e_f P m_f 0$ is not distinct from $e_f P m_f P$.

Which symmetry fractionalization pattern is realized at the surface of the $E_8$ root state?  We can answer this question by explicitly constructing the $S^{e_f}$ string operator.  It is enough to construct this operator for our $1d$ theory describing the edge of the mirror plane, where, for instance, we can choose $S^{e_f} =  d^\dagger_{L i} d^\dagger_{R i}$.  This operator creates one fermion in $L$ and one in $R$, thus creating a single $e_f$ excitation in each region after gauging.  To make $S^{e_f}$ gauge-invariant, we should include a Wilson line built from the $\zz$ vector potential, joining the insertion points of the two fermions.  However, the Wilson line can be chosen as a product of $\zz$ vector potential operators on two of the dashed edges in Fig.~\ref{fig:e8surface}, and thus does not contribute the the transformation of $S^{e_f}$ under reflection.  Therefore, due to the fermion anticommutation relations, we have $\mu^{e_f} = -1$.  This immediately implies that the surface of the $E_8$ root state realizes the $e_f P m_f P$ fractionalization pattern;  in the other pattern, all the string operators are even under reflection.

This surface theory provides an alternative demonstration that $E_8 \oplus E_8$ is trivial.
Upon adding two $E_8$ root states with three-fermion surfaces, we have a surface theory of two decoupled ``layers'' of the three-fermion state.  There are 16 types of particles that are composites of anyons in the two layers, which are labeled by ordered pairs such as $(e_f, e_f)$, $(e_f, m_f)$, $(1, e_f)$, and so on.  We can trivialize the surface by condensing both $(e_f, e_f)$ and $(m_f, m_f)$.  Both these particles are bosonic, and they have trivial mutual statistics so that they can be simultaneously condensed.  Moreover, since $e_f$ and $m_f$ transform identically under reflection in both layers, these composites transform trivially under reflection and can be condensed while preserving symmetry.  It is then straightforward to see that all non-trivial particles are either confined by this condensate, or have condensed, and we have obtained a trivial gapped surface, so the bulk pgSPT phase is also trivial.  

\subsection{Adding the root states}

Now we discuss the surface of the $E_8 \oplus \zz$ state, obtained by adding the two root states.  Again, we can start from a surface theory  comprised of two decoupled layers.  One is a three-fermion state with fractionalization pattern $e_f P m_f P$, while the other has toric code topological order and $e P m P$ fractionalization.  We can obtain a simpler theory by condensing the bosonic particle $(\epsilon_f, \epsilon)$, which transforms trivially under reflection, because both $\epsilon_f$ and $\epsilon$ transform in the same way.  The remaining deconfined particles form a three-fermion state, and are generated by fusing $e'_f = (e_f, e)$ and $m'_f = (m_f,e)$.  The resulting symmetry fractionalization pattern is $e_f 0 m_f 0$.  Therefore we see that both the $E_8$ root state, and the $E_8 \oplus \zz$ state, have three-fermion surfaces, with these two states realizing the two different possible patterns of reflection symmetry fractionalization.

\section{Electronic topological crystalline insulators in three dimensions}
\label{sec:tci}

Here we consider $3d$ electronic topological crystalline insulators (TCIs) with $\zz^P$ reflection symmetry.  These states are insulating SPT phases of electrons, with  symmetry group ${\rm U}(1) \times \zz^P$, where the ${\rm U}(1)$ is charge conservation.  Because there are no $3d$ topological insulators protected by ${\rm U}(1)$ symmetry alone \cite{cwang14classification}, without loss of generality we consider SPT phases that become trivial if ${\rm U}(1) \times \zz^P$ is broken down to ${\rm U}(1)$.  At the level of non-interacting electrons, it is known that there is a $\z$ classification of such TCIs, which is reduced to $\z_8$ by interactions \cite{isobe15}.

Below in \ref{sec:tci-classification}, via reduction to the $2d$ mirror plane, we obtain a larger $\z_8 \times \zz$ classification.  The corresponding phases are thus labeled by the ordered pair $(n,m)$, with $n$ defined modulo 8 and $m$ defined modulo $2$.  The additional $\zz$ factor arises from a TCI that requires finite-strength bulk interactions, and is thus inaccessible to previous approaches.  This state, labeled by $(0,1)$, can be understood as a topological paramagnet, were the spin sector is a bosonic pgSPT state with $\zz^P$ symmetry, and more specifically is in the $E_8$ root state described in Sec.~\ref{sec:3dz2p}.  This TCI, dubbed the $E_8$ paramagnet, is analogous to fermionic SPT phases protected by internal symmetry that have been studied previously, where a bosonic sector is put into a bosonic SPT phase.

In Sec.~\ref{sec:tci-surface}, we consider the surface properties of the $E_8$ paramagnet TCI; the results bolster the conclusion that 
this state remains non-trivial and distinct from the $(n,0)$ TCIs in the presence of electron excitations.  Then, in Sec.~\ref{sec:connect-bosonic}, we  show that the $(4,0)$ TCI can be also viewed as a different topological paramagnet, where the spin sector is in the bosonic $\zz$ root state.  This gives a physical picture of the $(4,0)$ TCI very different than that provided by the limit of weakly interacting electrons.

Before proceeding, a word is in order about how to describe the action of symmetry in fermionic systems \cite{wen12noninteracting}.  We let $G$ be the symmetry group acting on bosonic operators, and $G_f$ the symmetry group acting on all operators, including fermionic operators.  We view fermion parity $\zz^f$ as a symmetry, and it is a subgroup of $G_f$.  Then these groups are related by $G = G_f / \zz^f$.  This means that we can view $G_f$ as a group extension of $G$ with coefficients in $\zz^f$.  For given $G$ acting on bosonic operators, distinct actions of symmetry on fermions then correspond to elements of $H^2(G, \zz^f)$, which classify the different possible group extensions.

More physically, this discussion implies that we can view the symmetry acting on fermions in terms of symmetry fractionalization of the bosonic symmetry $G$.  For example, in the present case, $G = {\rm U}(1) \times \zz^P$, and we view bosonic Cooper pairs as unit charges.  Electrons then carry half-charge, and it follows that a $2\pi$ ${\rm U}(1)$ rotation is equal to $(-1)^F$, the fermion parity operator.  In fact, this fully characterizes the symmetry action on electrons in the present case.  Suppose that reflection squares to fermion parity, $U_{\sigma}^2 = (-1)^{F}$.  Then we can redefine the reflection by $U'_{\sigma} = R(\pi) U_{\sigma}$, where $R(\pi)$ is a $\pi$ ${\rm U}(1)$ rotation.  Since $R(\pi)^2 = (-1)^{F}$, we have $(U'_{\sigma})^2 = 1$, so that we can always choose the reflection to square to the identity operator, a choice we make below.

\subsection{Classification}
\label{sec:tci-classification}

Because we consider states that are trivial under the protection of ${\rm U}(1)$ symmetry alone, we can apply the same reduction procedure to obtain a $2d$ system on the mirror plane.  Here, the $\zz^P$ acts as a $\zz$ on-site, unitary symmetry, and the full internal symmetry group of the $2d$ system is $G = {\rm U}(1) \times \zz$.   We find a $\z_4 \times \z \times \z$ classification of $2d$ phases on the mirror plane.

The $\z_4$ factor labels electronic SPT phases, which were studied in Ref.~\onlinecite{isobe15}, where it was shown that the non-interacting $\z$ classification reduces to $\z_4$ in the presence of interactions.  In principle, there could be other such SPT phases not obtainable starting from a non-interacting limit, but we do not consider this possibility here.  We refer to the root state associated with the $\z_4$ factor as the SPT root state.

The two $\z$ factors correspond to integer quantum Hall (IQH) and $E_8$ states.  One of these factors is generated by the IQH root state, which is simply a $\nu = 1$ IQH state.  The other factor is generated by the $E_8$ root state.  This state is a topological paramagnet; 
 we start with a charge-neutral bosonic $E_8$ state (which can be thought of as describing the spin sector), and take a product of this state with a trivial electronic insulator.  This is distinct from a $\nu = 8$ IQH state (8 copies of the IQH root state), because the Hall conductance is different, so that the IQH and $E_8$ root states indeed generate independent $\z$ factors in the classification.  We might also consider an $E_8$ state built from charge-2 Cooper pairs, but this state is identical to 8 copies of the IQH root state, and does not need to be considered separately \cite{cano14bulkedge,chengPC}.  For both the IQH and $E_8$ root states, we choose the $\zz$ symmetry to act trivially.

As in the bosonic case, we now ask how the $\z_4 \times \z \times \z$ classification collapses under adjoining reflection-symmetric layers to give a classification of pgSPT phases.  Here, we can choose the added layers $| L \rangle$ and $| R \rangle$ to be some combination of IQH and (charge-neutral) $E_8$ states.  The crucial issue is to understand the effect of adjoining layers when we add two IQH root states or two $E_8$ root states.  Because the $E_8$ root state is a product of a bosonic pgSPT phase with a trivial electronic insulator, it follows immediately from the discussion of Sec.~\ref{sec:3dz2p} that adding two $E_8$ root states produces a trivial state.

Upon adding two IQH root states, we have fermions $c_1$ and $c_2$ each forming a $\nu = 1$ IQH state on the mirror plane.  The $\zz^P$ symmetry acts trivially, that is $\sigma : c_{1,2} \to c_{1,2}$.  Now we adjoin layers, so that each of $| L \rangle , | R \rangle$ is a $\nu = -1$ IQH state with fermions $d_{L, R}$, where $\zz^P$ acts by
\begin{equation}
\sigma : d_{L} \leftrightarrow d_{R} \text{.}
\end{equation}
We take linear combinations $d_{\pm} = d_R \pm d_L$ with eigenvalue $\pm 1$ under the action of $\sigma$.  We can  combine the $d_{+}$ and $c_2$ IQH states and gap them out while preserving $\zz^P$ symmetry; this is easily seen via the edge theory, similar to the discussion of Sec.~\ref{sec:e8root}.  This leaves a non-chiral state, where the $c_1$ and $d_{-}$ fermions have opposite eigenvalues under $\zz^P$ and form IQH states of opposite chirality.  This is precisely the  SPT root state \cite{isobe15}.

We have thus shown that adding two IQH root states does \emph{not} give a trivial pgSPT phase, but instead is equivalent to the SPT root state.  Therefore, the IQH and SPT root states combine together to give a $\z_8$ factor in the classification of pgSPT phases, and the full $\z_8$ is generated by the IQH root state, even though half of the corresponding pgSPT phases are related to $2d$ SPT phases.  The $E_8$ root state generates a separate $\zz$ factor, and the full classification we find is $\z_8 \times \zz$.

\subsection{Surfaces of the $E_8$ paramagnet TCI}
\label{sec:tci-surface}

In some cases, it is known that taking a product of a non-trivial bosonic SPT phase with a trivial fermionic insulator does not produce a new distinct fermionic SPT phase \cite{cwang14classification,cwang14interacting}.  While our approach of reduction to $2d$ already shows the $E_8$ paramagnet TCI is non-trivial and distinct from the TCIs in the $\z_8$ classification, it is desirable to confirm this from other points of view.  Here, we do this by studying two different surface states.

First, we consider a surface which is gapped and trivial away from the reflection axis.  The mirror plane then supports gapless chiral modes at its $1d$ edge, characterized by a chiral central charge $c=8$ and vanishing Hall conductivity.  Similarly to the bosonic $E_8$ root state discussed in Sec.~\ref{sec:e8root}, this situation cannot occur in a strictly $2d$ electron system with ${\rm U}(1) \times \zz^P$ symmetry, with no anyons away from the reflection axis.  To see this, in such a $2d$ system the most general possibility on one side of the reflection axis is to have $n_{E_8}$ copies of a charge-neutral $E_8$ state, and $n_{I}$ copies of a $\nu = 1$ IQH state.  On the other side of the reflection axis there are then $-n_{E_8}$ and $-n_I$ copies of the corresponding states, respectively.  This leads to gapless modes on the reflection axis characterized by chiral central charge $c = 16 n_{E_8} + 2 n_I$ and Hall conductivity $2 n_I e^2 / h$, which cannot reproduce the surface of the $E_8$ paramagnet TCI.

It should be noted that if ${\rm U}(1)$ symmetry is broken, the Hall conductivity is not meaningful, and we can achieve $c = 8$ by choosing $n_I = 4$ and $n_{E_8} = 0$.  The resulting $1d$ theory is the same as the edge of a $\nu = 8$ IQH state.  This is equivalent to the edge of the $E_8$ state, in the sense that adding appropriate perturbations localized to the edge can drive the theory across a quantum phase transition and into an $E_8$ edge \cite{cano14bulkedge}.  Therefore, the $E_8$ paramagnet TCI requires both ${\rm U}(1)$ charge conservation and $\zz^P$ symmetry for its protection.

Next, we consider a gapped, topologically ordered surface, building on the three-fermion state surface of the bosonic $E_8$ root state.  Because the $E_8$ paramagnet TCI is a product of the bosonic $E_8$ root state and a trivial fermionic insulator, the particle types at this surface are also products
\begin{equation}
\{ 1, e_f, m_f, \epsilon_f \} \times \{ 1 , c \} \text{,}
\end{equation}
where $c$ represents the electron.  It is important to note that $e_f, m_f$ and $\epsilon_f$ are all charge-neutral, where $c$ carries unit charge. In order for this surface to be non-trivial, it should be impossible to rewrite it as a different product, where one factor of the product can occur in a strictly $2d$ bosonic system, and the other factor is again a trivial fermionic insulator.  

Here, we can also view the particle types as a product
\begin{equation}
\{ 1, e_f c, m_f c, \epsilon_f \} \times \{ 1, c \} \text{,}
\end{equation}
where the choice of attaching $c$ to $e_f$ and $m_f$ is arbitrary; we could choose any two anyons of the three-fermion state, and the discussion below applies.
The first factor has the  topological order of the toric code, with $e = e_f c, m = m_f c$, and $\epsilon = \epsilon_f$.  While this topological order certainly can occur in a strictly $2d$ bosonic system with reflection symmetry, we also have to consider the role of ${\rm U}(1)$ symmetry.  For a bosonic sector of the underlying electronic system, $e$ and $m$ must be viewed as carrying half-charge; they carry the charge of the electron, which is half the elementary charge of bosonic particles (Cooper pairs).  Within the simplest possible description in terms of Abelian Chern-Simons theory (using a $2 \times 2$ $K$-matrix), such a fractionalization pattern, where both $e$ and $m$ carry half charge, leads to a non-zero quantized Hall conductivity \cite{cwang13boson}.  This is incompatible with reflection symmetry, and suggests that any reflection-symmetric realization of this fractionalization pattern is anomalous.  Indeed, this can be shown by generalizing the flux fusion approach of Ref.~\onlinecite{hermele15} \cite{sjhuangUNPUB}.  This conclusion is consistent with the non-triviality of the $E_8$ paramagnet TCI.

\subsection{Connection to bosonic pgSPT phases}
\label{sec:connect-bosonic}

Here, we consider the $(4,0)$ TCI, and show that it can be viewed as a topological paramagnet, where the spin sector is in the bosonic $\zz$ root state.  This result is closely related to, and indeed can be understood to follow from, prior work showing that a certain two-dimensional SPT phase of fermions with $\zz$ symmetry is related to the non-trivial bosonic $\zz$ SPT phase.\cite{you15bridging, you16quantum}

We proceed by considering a surface that is trivial away from the reflection axis.  On the axis, there are two counter-propagating pairs of chiral fermions.  We denote electron creation operators by $\psi^\dagger_{p, R}$ and $\psi^\dagger_{p, L}$, with $p = 1,2$, for right and left movers, respectively.  Reflection symmetry acts by
\begin{eqnarray}
\sigma : \psi^\dagger_{p, R} &\to& \psi^\dagger_{p, R} \\
\sigma : \psi^\dagger_{p, L}  &\to& - \psi^\dagger_{p, L} \text{.}
\end{eqnarray}

The effect of interactions in this theory was analyzed in Ref.~\onlinecite{isobe15} using a bosonized description, and we adopt the same approach here.  We introduce bosonic fields $\phi_i$ ($i = 1,\dots,4$), related to electron operators by
\begin{eqnarray}
\psi^\dagger_{p,R} &\sim& e^{i \phi_p} \\
\psi^{\dagger}_{p,L} &\sim& e^{- i \phi_{p + 2} } \text{.}
\end{eqnarray}
The Lagrangian is
\begin{eqnarray}
\mathcal{L}=\frac{1}{4\pi}(K_{ij}\partial_{x}\phi_{i}\partial_{t}\phi_{j}-V_{ij}\partial_{x}\phi_{i}\partial_{x}\phi_{j}),
\label{L}
\end{eqnarray}
where 
\begin{eqnarray}
K= \left( \begin{array}{ccc}
1_{2 \times 2} & 0  \\
0 & -1_{2 \times 2} \\
\end{array} \right),
\end{eqnarray}
and $V$ is a $4 \times 4$ velocity matrix. Since $V$ is not universal, the exact form is not important here.
The ${\rm U}(1)$ symmetry, acting on electron operators by $\psi^\dagger \to e^{i \alpha} \psi^\dagger$, acts on the bosonic fields by
\begin{equation}
\vec{\phi} \to \vec{\phi} + \alpha (1,1,-1,-1)^T \text{,}
\end{equation}
while reflection acts by
\begin{equation}
\sigma : \vec{\phi} \to \vec{\phi} + \pi (0,0,1,1)^T \text{.}
\end{equation}

To proceed, we make a change of variables $\vec{\phi} = W \vec{\phi}'$, where $W$ is a $GL(4, \z)$ matrix\begin{equation}
W = \left( \begin{array}{cccc}
1 & 0 & 0 & -1 \\
1 & 0 & -1 & 0 \\
-1 & 0 & 1 & 1 \\
0 & -1 & 0 & 0 
\end{array}\right) \text{.}
\end{equation}
The $K$-matrix in the new basis is
\begin{equation}
K' = W^T K W = \left( \begin{array}{cccc}
1 & 0 & 0 & 0 \\
0 & -1 & 0 & 0 \\
0 & 0 & 0 & -1 \\
0 & 0 & -1 & 0 \end{array}\right) \text{.}
\end{equation}
The block form of $K'$ implies that we can decouple the edge modes into a fermionic sector with
\begin{equation}
K_f = \left( \begin{array}{cc}
1 & 0 \\
0 & -1 \end{array}\right) \text{,}
\end{equation}
and a bosonic sector with
\begin{equation}
K_b = - \left( \begin{array}{cc}
0 & 1 \\
1 & 0 \end{array}\right) \text{.}
\end{equation}
This is possible because the velocity matrix is non-universal, and can be tuned to achieve such a decoupling.

The fields in the fermionic sector transform under ${\rm U}(1)$ by  $\phi'_{1,2} \to \phi'_{1,2} + \alpha$, and under reflection by
\begin{eqnarray}
\sigma : \phi'_1 &\to& \phi'_1 + \pi \\
\sigma : \phi'_2 &\to& \phi'_2 - \pi \text{.}
\end{eqnarray} 
We can refermionize this sector by defining $\Psi^\dagger_i = e^{i \phi'_i}$ for $i=1,2$.  The mass term
\begin{equation}
\delta {\cal H} = m (\Psi^\dagger_1 \Psi_2 + \text{H.c.} ) 
\end{equation}
is clearly allowed by symmetry, and trivially gaps out the fermionic sector, which thus describes the edge of a trivial electronic insulator.

The fields of the bosonic sector are neutral under ${\rm U}(1)$, so we can interpret this as the spin sector of the $(4,0)$ TCI.  The transformations
under reflection are
\begin{equation}
\sigma : \phi'_{3,4} \to \phi'_{3,4} + \pi \text{.}
\end{equation}
This is precisely the edge of the $\zz$ bosonic SPT phase \cite{levin12braiding}, and we can identify the neutral bosonic sector with the bosonic $\zz$ root state.

\section{Topological crystalline superconductors}
\label{sec:tcsc}

Here, we consider electronic SPT phases in three dimensions with only $\zz^P$ reflection symmetry.  These states are referred to as topological crystalline superconductors (TCSCs), because they lack ${\rm U}(1)$ charge conservation symmetry.  In accord with the discussion of Sec.~\ref{sec:tci}, we need to describe in more detail how symmetry acts on electrons, and here there are two possibilities.  One is that reflection squares to the identity, $\sigma^2 = 1$, in which case we find a $\z_{16}$ classification.  The other is reflection squaring to fermion parity, $\sigma^2 = (-1)^F$, in which case we find a trivial classification.

The \emph{same} classifications can be obtained by following the approach of Ref.~\onlinecite{isobe15}, where one starts with non-interacting TCSCs, and then asks how the non-interacting classification collapses in the presence of interactions.  Our treatment allows for the possibility of TCSCs that require strong interactions in the bulk, and we find no such states for the symmetries considered.  

We note that these results agree with those obtained in Ref.~\onlinecite{kapustin15fermionic}, where some fermionic SPT phases were classified based on a cobordism approach.  

\subsection{$\sigma^2 = 1$: $\z_{16}$ classification}
\label{sec:tcsc-R21}

As in the cases analyzed above, the first step is to analyze the possible $2d$ states on the mirror plane.  The $\zz^P$ reflection acts as an on-site $\zz$ symmetry where $\sigma^2 = 1$.  One possibility is to have a SPT phase on the mirror plane, and such phases were studied in Refs.~\onlinecite{ryu12interacting,xlqi13,yao13interaction,gu14effect}, where a $\z_8$ classification was found.  All the SPT phases are obtained by starting with a free-fermion state with $n_{SPT}$ pairs of counter-propagating Majorana modes at its edge, where  the right-moving (left-moving) Majoranas are even (odd) under $\sigma$.  This state is trivial in the presence of interactions for $n_{SPT}  = 8$, but the states with $1 \leq n_{SPT} \leq 7$ are non-trivial, leading to a $\z_8$ classification.  We refer to the corresponding root state (with $n_{SPT} = 1$) as the SPT root state.

Another possible state for the mirror plane is to have $n_p$ copies of a topological $p+ip$ superconductor, with $n_p$ chiral Majorana fermions at the edge.
We take the reflection symmetry to act trivially on these states.  We have a $\z$ classification generated by the state with $n_p = 1$, dubbed the $p+ip$ root state.  We can also consider the possibility of an $E_8$ state on the mirror plane, but in a fermionic system, this state is not distinct from $n_p = 16$ copies of the $p+ip$ root state \cite{cano14bulkedge}.

We thus obtain a $\z_8 \times \z$ classification of $2d$ states on the mirror plane, and we need to ask how this collapses to a classification of pgSPT phases.  We can adjoin pairs of $p+ip$ superconductors that go into one another under reflection, which changes $n_p \to n_p \pm 2$, so that, similar to previous cases, $n_p$ is only well-defined modulo two.  This means that the $(p + ip) \oplus (p + ip)$ state, obtained by adding two $p + ip$ root states, can either be equivalent to a non-trivial SPT phase on the mirror plane, or it can be trivial.  In fact, $(p + ip) \oplus (p + ip)$ is equivalent to the SPT root state, so that the $p + ip$ root state generates a $\z_{16}$ classification, which includes all the SPT states on the mirror plane.

To establish this result, let $\gamma_1, \gamma_2$ represent Majorana fermions making up the two copies of the $p+ip$ root state in $(p + ip) \oplus (p + ip)$.  Reflection acts trivially on $\gamma_i$.  Then we adjoin two $p - i p$ layers with Majorana fermions $\delta_R, \delta_L$, that are exchanged by reflection.  We take linear combinations $\delta_{\pm}  = \delta_R \pm \delta_L$.  At the non-interacting level, the $p-ip$ state with $\delta_{+}$ fermions and the $p+ip$ state with $\gamma_2$ fermions can be combined and gapped out.  This leaves $\gamma_1$ and $\delta_{-}$ fermions gapless, which is precisely the SPT root state.

\subsection{$\sigma^2 = (-1)^F$: Trivial classification}
\label{sec:tcsc-R2F}

For the case where reflection squares to fermion parity, we first consider possible non-interacting states on the mirror plane.  We refer to the on-site symmetry in this case as $\z_4^f$, because $\sigma$, which squares to fermion parity, generates a $\z_4$ group.  We can always choose a basis of fermion operators in which $\sigma$ is diagonal, so that
\begin{equation}
\sigma : \psi \to i \psi \text{.}
\end{equation}
This immediately implies that $2d$ quadratic fermion Hamiltonians are identical to those with ${\rm U}(1)$ symmetry, and with no other symmetries.  There is a $\z$ classification of such Hamiltonians, where the integer index $n_{IQH}$ gives the integer quantized Hall conductivity associated with the effective ${\rm U}(1)$ symmetry, or, equivalently, the number of chiral (Dirac) edge modes.  This exhausts the possibilities for free-fermion states on the mirror plane; in particular, there are no free-fermion SPT phases in this case.

We also need to consider the possibility of interacting states on the mirror plane.  Recent works have found that there are no non-trivial $2d$ SPT phases with $\z_4^f$ symmetry \cite{chenjiewang16braiding, lan16classification}.  For instance, one possibility to consider is a product of bosonic $\zz$ SPT phase with a trivial fermionic state, but Ref.~\onlinecite{chenjiewang16braiding} showed the edge of this state can be trivially gapped out.  Ref.~\onlinecite{chenjiewang16braiding} also found a $\z$ classification of integer topological phases, allowing for strong interactions, indicating the $\z$ classification of free-fermion states is complete for interacting systems.  For example, we can consider the possibility of a bosonic $E_8$ state with $n_{E_8} = 1$ on the mirror plane, with some action of $\zz = \z_4^f / \zz^f$ symmetry.  We can add to this a state with $n_{IQH} = -8$, to produce a non-chiral state that should thus be a $\z_4^f$ symmetric SPT phase.  But since such a state is trivial, the $n_{E_8} = 1$ state is equivalent to the free-fermion  $n_{IQH} = 8$ state, and we do not obtain any new states in this manner.

Now, we show that the $\z$ classification of $2d$ phases on the mirror plane collapses to a trivial classification of $3d$ pgSPT phases.  We start with the $n_{IQH} = 1$ root state, which is built from a single species of fermion $\psi$, with $\sigma : \psi \to i \psi$.  Then we adjoin layers of $(p - ip)$ superconductors, whose edge modes propagate in the opposite direction to that of the root state.  Denoting  Majorana fermions making up these two states by $\gamma_L, \gamma_R$, reflection acts by
\begin{eqnarray}
\sigma : \gamma_L &\to& \gamma_R \\
\sigma: \gamma_R &\to& - \gamma_L \text{,}
\end{eqnarray}
where the minus sign is present because $\sigma^2 = (-1)^F$.  We introduce a Dirac fermion $c = \gamma_R + i \gamma_L$ on which reflection acts by
$\sigma : c \to i c$.  The adjoined layers are this equivalent to a $n_{IQH} = -1$ state, so that combined with the $n_{IQH} = 1$ root state, we are left with a trivial state.  This $\z$ classification in $2d$ thus becomes trivial upon passing to a classification of $3d$ pgSPT phases.

\section{Beyond reflection: bosonic SPT phases with $C_{2v}$ symmetry}
\label{sec:other-point-groups}

So far, we have only considered $\zz^P$ reflection symmetry.  Our approach can be applied for any point group, and we illustrate this here by considering $3d$ bosonic pgSPT phases protected by  $C_{2v}$ symmetry.  Other cases are left for future work.  The $C_{2v}$ point group is generated by two reflections, $\sigma_a$ and $\sigma_b$, whose mirror planes are perpendicular; as an abstract group, $C_{2v} \simeq \zz \times \zz$.  We take all spins to transform as linear (\emph{i.e.} not projective) representations of $C_{2v}$.

\begin{figure}
\includegraphics[width=0.6\columnwidth]{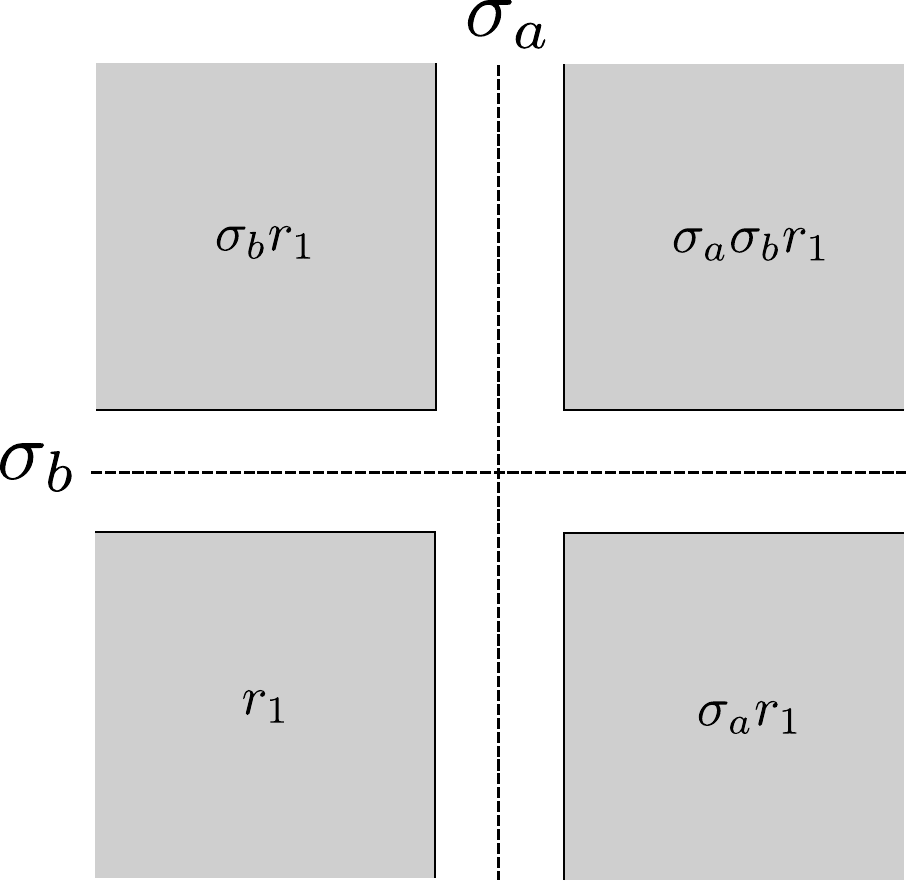}
\caption{Cross-section of a $3d$ system with $C_{2v}$ symmetry, which is generated by the two mirror reflections $\sigma_a$ and $\sigma_b$.  The dashed lines are mirror planes.  The shaded regions can be trivialized by applying a symmetry-preserving local unitary, reducing the system to the cross-shaped region near the mirror planes.}
\label{fig:c2v}
\end{figure}

Figure~\ref{fig:c2v} shows a cross-section of a system with $C_{2v}$ symmetry.  As before, there is a local unitary $U^{loc}$ that trivializes the SPT ground state, and we can trivialize region $r_1$ by restricting $U^{loc}$.  Then we can copy the restricted local unitary to the regions $\sigma_a r_1$, $\sigma_b r_1$ and $\sigma_a \sigma_b r_1$, to extensively trivialize the ground state while respecting symmetry.  We are left with a system composed of intersecting slabs centered on the two mirror planes.

The reduced system can be viewed as four half-planes, each with on-site $\zz$ symmetry, joined together in a $1d$ region with $\zz \times \zz$ on-site symmetry.  Following the same approach laid out above, we first obtain a $(\zz)^3 \times \z$ classification of states in the reduced system, then ask how it collapses to a classification of pgSPT phases.  

The three $\zz$ factors in the dimensionally reduced classification come from different SPT root states.  In two of these, the $\z_{2a}$ and $\z_{2b}$ root states, we put a $\zz$ SPT state on the $\sigma_a$ or $\sigma_b$ mirror plane.  It should be noted that, for instance, $\sigma_b$ acts as a $2d$ reflection symmetry on the $\sigma_a$ mirror plane.  The two-dimensional $\zz$ SPT state is compatible with reflection symmetry; for example, the model of Ref.~\onlinecite{levin12braiding} is manifestly reflection-invariant.  The third SPT root state is the unique non-trivial $1d$ SPT phase with $\zz \times \zz$ symmetry (the Haldane phase) \cite{haldane83a,haldane83b, gu09tensor,pollmann10,fidkowski11,turner11,chen11a,schuch11}, placed on the axis where the planes intersect.

\begin{figure}
\includegraphics[width=\columnwidth]{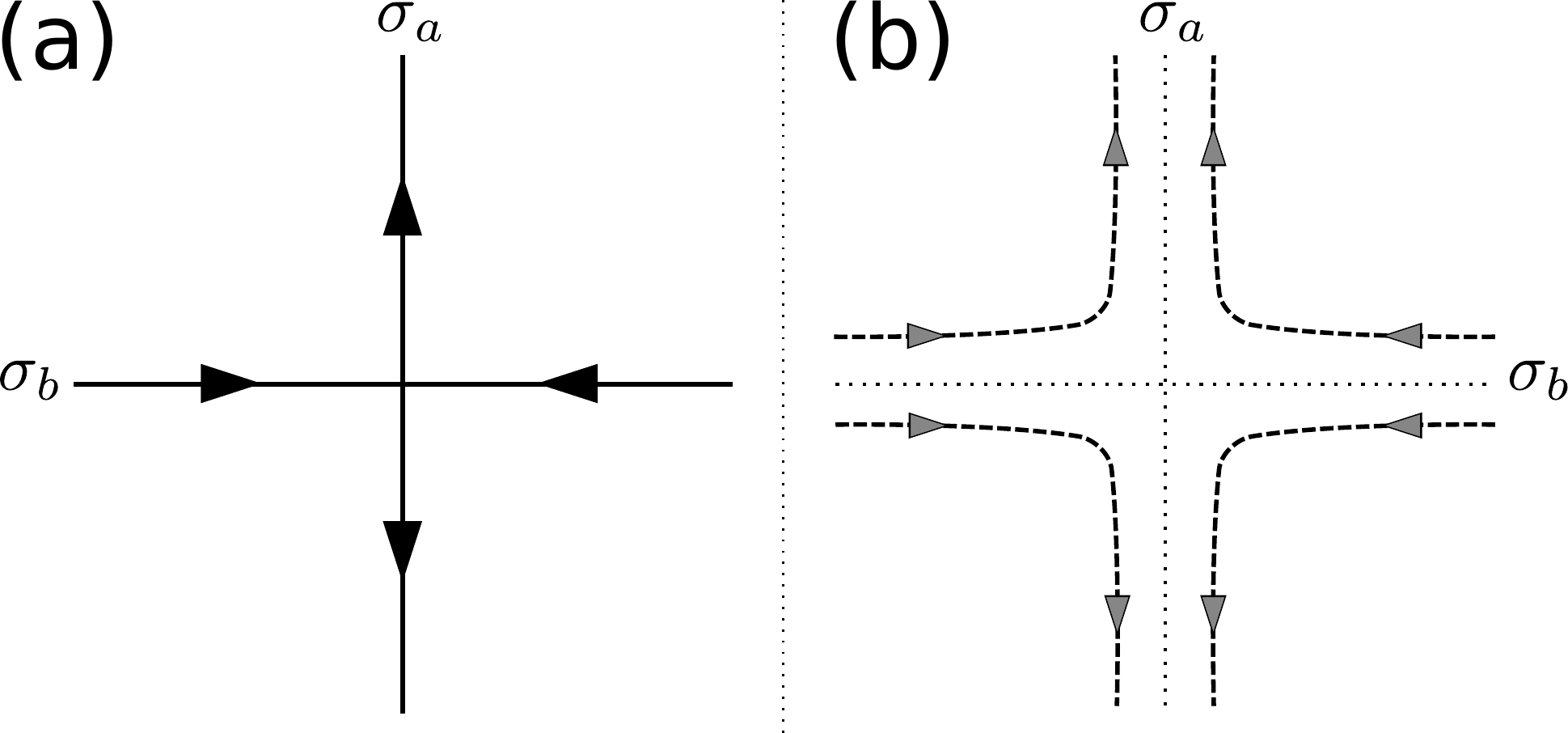}
\caption{(a) $E_8$ root state for $C_{2v}$ symmetry.  Solid lines represent an $E_8$ state on each half-plane, with edge chiralities indicated by the arrows.  (b) Microscopic construction of the $E_8$ root state in terms of sheets of $\nu = 4$ IQH states (dashed lines), with edge chiralities as indicated by the arrows.}
\label{fig:e8-c2v}
\end{figure}

There is also an $E_8$ root state, with edge chiralities arranged as shown in Fig.~\ref{fig:e8-c2v}a to respect the $C_{2v}$ symmetry.  To show this arrangement of $E_8$ states is actually possible, and compatible with an energy gap everywhere in the bulk, we give a more microscopic construction, which is illustrated in Fig.~\ref{fig:e8-c2v}b.  The construction starts with four sheets of $\nu = 4$ IQH state, with edge chiralities as shown, so that each plane hosts two sheets making up a $\nu = 8$ IQH state.  The fermion parity is coupled to a single $\zz$ gauge field residing on the cross-shaped lattice of the reduced system, and the $\zz$ gauge flux is condensed everywhere, resulting in $E_8$ states on each half-plane with chiralities as shown.  This state generates the $\z$ factor in the reduced classification.

Considering the $E_8$ root state on one of the half planes (say, the lower $\sigma_a$ half-plane), we see from our construction that the on-site $\zz$ symmetry (coming from $\sigma_a$) acts non-trivially on its degrees of freedom.  However, we now show that this $E_8$ state on the half-plane is in the same phase as an $E_8$ state with trivial action of $\zz$ symmetry.  To see this, we go back to our construction before introducing the $\zz$ gauge field, and let $c_{1i}$ and $c_{2i}$ ($i = 1,\dots,4$) denote the fermions making up the two $\nu = 4$ IQH sheets.  The $\zz$ symmetry acts by $\sigma_a : c_{1 i} \leftrightarrow c_{2i}$, and we introduce linear combinations $c_{\pm i} = c_{1i} \pm c_{2i}$ satisfying $\sigma_a : c_{\pm i} \to \pm c_{\pm i}$.

Now, we are free to add a trivial fermionic state, because this will not affect the phase that results upon gauging fermion parity and condensing $\zz$ gauge fluxes.  Therefore, we add $\nu = 4$ and $\nu = -4$ IQH layers, with fermions $a_i$ and $b_i$ respectively, where the symmetry acts by $\sigma_a : a_i \to a_i$ and $\sigma_a : b_i \to - b_i$.  This is precisely the state shown to be trivial in Refs.~\cite{ryu12interacting,xlqi13,yao13interaction,gu14effect}, so we are free to add it.  Then, we combine the counter-propagating $c_{-i}$ and $b_i$ fermions into a trivial state, leaving a $\nu = 8$ IQH state (with $c_{+i}$ and $a_i$ fermions) on which $\sigma_a$ acts trivially.  Upon gauging fermion parity and condensing the $\zz$ gauge flux, we get an $E_8$ state with trivial action of $\sigma_a$.

\begin{figure}
\includegraphics[width=0.8\columnwidth]{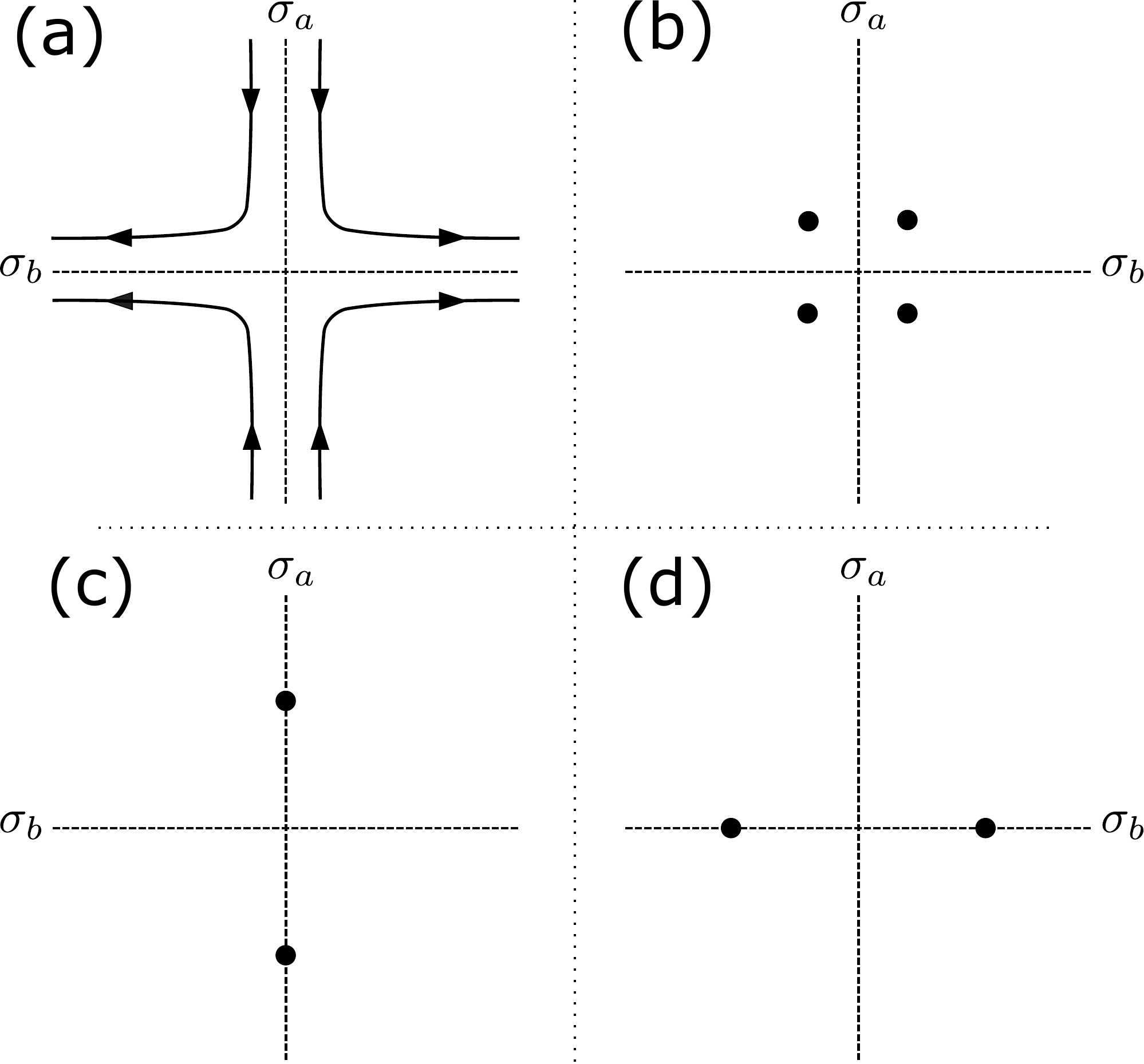}
\caption{(a) Adjoining $2d$ sheets (solid lines) to a reduced system with $C_{2v}$ symmetry defined on the mirror planes (dashed lines).  The arrows indicate one choice of edge chiralities when the adjoined sheets are $E_8$ states.  (b), (c) and (d) Adjoining $1d$ systems, shown as filled circles.}
\label{fig:c2v-adjoined}
\end{figure}

To determine how the $(\zz)^3 \times \z$ classification for the reduced system collapses to a classification of pgSPT phases, we need to understand the analog of adjoining layers for $C_{2v}$ symmetry.  There are a few operations that need to be considered.  First, we can adjoin four $2d$ sheets as shown in Fig.~\ref{fig:c2v-adjoined}a.  Second, we can adjoin four $1d$ systems away from the mirror planes, as in Fig.~\ref{fig:c2v-adjoined}b, or a pair of $1d$ systems lying on one of the mirror planes (Fig.~\ref{fig:c2v-adjoined}c and~\ref{fig:c2v-adjoined}d).  Each adjoined $1d$ system either has no symmetry taking it into itself (in Fig.~\ref{fig:c2v-adjoined}b), or has a $\zz$ on-site symmetry (in Fig.~\ref{fig:c2v-adjoined}c and Fig.~\ref{fig:c2v-adjoined}d).  In either case, these $1d$ systems must be trivial, and adjoining them has no effect.  Each adjoined $2d$ sheet also has no symmetry taking it into itself, but the sheets can be $E_8$ states, with chiralities arranged as shown in Fig.~\ref{fig:c2v-adjoined}a, or reversed from the chiralities shown in the figure.

Just as for the $E_8$ root state of the bosonic $\zz^P$ pgSPT phase, the chirality of the $E_8$ root state here can be reversed by adjoining sheets of $E_8$ state.  Therefore, the integer index of the reduced classification is only well-defined modulo two when passing to a classification of pgSPT phases.

We now show that the $E_8 \oplus E_8$ state is trivial, leading to a $(\zz)^4$ classification of pgSPT phases.  The argument parallels that given in Sec.~\ref{sec:e8root} for the case of $\zz^P$ symmetry.  We first consider two copies of the $E_8$ root state, constructed in terms of sheets of $\nu = 4$ IQH state (Fig.~\ref{fig:e8-c2v}b) coupled to two different $\zz$ gauge fields.  Ignoring coupling to the gauge fields for the moment, we can combine the $\nu = 4$ IQH sheets together, and think of this state in terms of four sheets of $\nu = 8$ IQH state.  Then, we can adjoin sheets of $E_8$ state, with chiralities opposite to the $\nu = 8$ sheets.  Representing the adjoined $E_8$ states as $\nu = 8$ IQH sheets coupled to $\zz$ gauge fields, we now have in each quadrant two $\nu = 8$ IQH sheets with opposite chiralities, which can be trivially gapped at the surface while preserving symmetry.  We can then condense the $\zz$ gauge fields in the bulk, to obtain the $E_8 \oplus E_8$ state with a trivial gapped surface, indicating that $E_8 \oplus E_8$ is trivial.  We note there are six different $\zz$ gauge fields in this construction -- one for each of the two $E_8$ states we started with, and one for each of the four adjoined sheets of $E_8$ state.

Finally, we note that all the states in the $(\zz)^4$ classification can be constructed as arrays of lower-dimensional topological phases.  This can be done by adding two-dimensional translation symmetry, with elementary translations normal to the mirror planes, and periodically repeating the extensively trivialized states.

\section{Discussion}
\label{sec:discussion}

Via consideration of a few examples in three dimensions, we developed a general framework to classify, characterize and construct pgSPT phases in terms of lower-dimensional topological phases with on-site symmetry.  Our framework applies to bosonic and fermionic pgSPT phases in any spatial dimension.  The classifications we find are given in Table~\ref{tab}.  We also showed that some of the pgSPT phases we identified admit gapped, topologically ordered surface states, where symmetry is realized in an anomalous fashion.  

We would like to note a striking correspondence between pgSPT phases protected by $\zz^P$ reflection symmetry and SPT phases protected by $\zz^T$ time reversal.  For bosonic systems with $\zz^P$ only or $\zz^T$ only, the classification of SPT phases is $\zz$ in $d=1$ \cite{gu09tensor,pollmann10,chen11a,schuch11}, and $\zz \times \zz$ in $d=3$ \cite{vishwanath13,cwang13boson,burnell14exactly,kapustin14symmetry}.  For fermions in $d=3$ with $\zz^P$ and $\sigma^2 = 1$, or with $\zz^T$ and $T^2 = (-1)^F$, the classification is $\z_{16}$ \cite{fidkowski13nonabelian,cwang14interacting,metlitski14interaction,kapustin15fermionic}.  Similarly, for fermions in $d=3$ with $\zz^P$ and $\sigma^2 = (-1)^F$, or with $\zz^T$ and $T^2 = 1$, there is a trivial classification \cite{kapustin15fermionic}.  Finally, there is a $\z_8 \times \zz$ classification for SPT phases of electrons in $d=3$ with either ${\rm U}(1) \times \zz^P$ or ${\rm U}(1) \times \zz^T$ symmetry \cite{cwang14interacting}.

This correspondence follows in general if we make the assumption that all the relevant SPT phases admit a description in terms of Lorentz-invariant field theory.  Then, reflection of one space time coordinate in Euclidean space time, \emph{e.g.} $x_0 \to -x_0$, can be analytically continued to Minkowski space time either as a spatial reflection, or as a time reversal transformation.  In fermionic theories, due to CPT symmetry, one of these operations squares to $1$, while the other squares to $(-1)^F$.  Of course, the classifications we quote for reflection and time reversal SPT phases are obtained without assuming Lorentz invariance.  This suggests that there may be a way to argue for the correspondence more directly, without invoking Lorentz-invariant field theory.  We believe this is an interesting problem for future work, that could shed new light on the physics of both reflection and time reversal SPT phases.

We now discuss the outlook for further developments building on the results presented here.  Clearly, the examples considered do not exhaust the possibilities for physically interesting pgSPT phases.  One-dimensional bosonic pgSPT phases protected by $\zz^P$ are discussed in Appendix~\ref{app:1d}, where the $\zz$ classification obtained in  Refs.~\onlinecite{gu09tensor,pollmann10,chen11a,schuch11} is recovered.  For crystallographic point groups in two dimensions, straightforward application of our approach shows that, while there are non-trivial bosonic pgSPT phases, none of them have protected edge modes~\cite{2dedgeUNPUB}.  However, there are interesting possibilities for fermions in $2d$; for example, one can obtain a non-trivial fermionic topological crystalline superconductor with reflection symmetry by making a stack of topological $p$-wave superconducting chains \cite{kitaev01unpaired}.  A symmetry-preserving edge is then a chain of end-state Majorana fermions.  If both reflection and translation symmetry are present at the edge, the quadratic part of the edge Hamiltonian vanishes, leading to an unusual interaction-dominated system of Majorana fermions that merits further study.  Along similar lines, we have not exhausted the possibilities for three-dimensional pgSPT phases with symmetry that can be preserved at a clean surface.

Our approach can be directly applied to SPT phases protected by a combination of point group and internal symmetries, as for the electronic TCIs we studied, protected by ${\rm U}(1) \times \zz^P$.  In that case, there were no non-trivial SPT phases protected by internal symmetry alone, but the situation is different for other symmetries, \emph{e.g.} for time reversal symmetry.  We can still apply our approach in such a case, by first adding a layer to cancel any SPT phase non-trivial under the internal symmetry, and then applying the extensive trivialization procedure.

A different direction for future work is to use  constructions of pgSPT phases as stacks and arrays as a starting point to study physical properties, especially surface properties.  In particular, various physical properties could be studied using coupled wire constructions, which might also be useful to obtain continuum field theory descriptions of pgSPT surfaces.  It would also be desirable to obtain bulk field theories, and to find realizations of pgSPT phases in models that are not simply stacks and arrays at the microscopic level.

Finally, we believe the approach developed here can form the basis for an approach to symmetry enriched topological (SET) phases with point group symmetry.  To illustrate the basic idea, we consider a specific example of a bosonic system in $2d$ with toric code topological order and $\zz^P$ reflection symmetry.  In the absence of any symmetry, perhaps after adding some trivial degrees of freedom, this system is adiabatically connected to the exactly solvable toric code Hamiltonian.  This adiabatic continuity may fail in the presence of the symmetry, but we can still reduce the system to a solvable toric code away from the reflection axis.  This is just like extensive trivialization for pgSPT phases, except now we are not trivializing the system, but rather reducing it to a non-trivial but simple reference state away from the reflection axis.  SET phases will then be distinguished by properties of the $1d$ reflection axis, where $\zz^P$ acts as a $\zz$ on-site symmetry.  To proceed, it will be necessary to classify $1d$ systems with $\zz$ on-site symmetry, embedded in a reflection-symmetric toric code medium.

\acknowledgments{M.H. is grateful for discussions with Maissam Barkeshli and Ari Turner during the early stages of this work, and for useful correspondence with Xie Chen, Meng Cheng,  Anton Kapustin, and Michael Levin.  This research is supported by the U.S. Department of Energy, Office of Science, Basic Energy Sciences, Division of Materials Sciences and Engineering, under Award numbers DE-SC0014415 (H.S., S.-J. H., and M.H.) and DE-SC0010526 (L.F.).  H.S. acknowledges financial support from the Spanish MINECO grants FIS2012-33152, FIS2015-67411, and the CAM research consortium QUITEMAD+ Grant No. S2013/ICE-2801 during the final process of preparing the manuscript.}

\appendix

\section{One dimension}
\label{app:1d}

We consider pgSPT phases in $1d$, where the only non-trivial point group is $\zz^P$ reflection generated by $\sigma : x \to -x$. Such phases are  known to obey a $\zz$ classification \cite{gu09tensor,pollmann10,chen11a,schuch11}, a conclusion reproduced by our analysis here.

As for $3d$ pgSPT phases with reflection symmetry, we can extensively trivialize the system away from the center of reflection symmetry $o$, which is now a point.  Therefore, we can reduce a SPT ground state to a zero-dimensional state $|\psi_o \rangle$.  This effectively $0d$ system must be gapped, with $|\psi_o\rangle$ a unique symmetry-preserving ground state.  Such a state transforms as a one-dimensional representation of $\zz^P$, which acts on the zero-dimensional region as a $\zz$ on-site symmetry.  We have
\begin{equation}
U_{\sigma} | \psi_{o} \rangle = \lambda_{\sigma} | \psi_{o} \rangle  \text{,}
\end{equation}
where $\lambda = \pm 1$ labels the two representations.  The operation of adjoining layers has no effect on $\lambda_{\sigma}$, so the two values of $\lambda_{\sigma}$ correspond to a $\zz$ classification of pgSPT phases.  In a system with open boundaries, there are no protected boundary states, and $\lambda_{\sigma}$ is simply the reflection eigenvalue of the ground-state; it was pointed out in~\cite{zaletel15} that this labels distinct reflection-symmetric SPT phases.

There are some subtleties with the $\zz$ classification that do not arise in the three-dimensional examples we focused on.  To expose one of these subtleties, consider a $1d$ system with open boundary conditions, where $\lambda_{\sigma}$ is the ground-state reflection eigenvalue.  If the reflection is site-centered, there are an odd number of lattice sites, and we are free to redefine the unitary realizing reflection symmetry by $U_{\sigma} \to -U_{\sigma}$, by adding a minus sign to the action of reflection on each lattice site.  This reverses the sign of $\lambda_{\sigma}$, which means there is not an invariant notion of which pgSPT phase is trivial, and which is non-trivial, although there are still two distinct phases.  Therefore, in this case, we should say the classification is a $\zz$ torsor rather than a $\zz$ group.  Such pgSPT phases with site-centered reflection were referred to as ``symmetry protected trivial'' states in Ref.~\onlinecite{fuji15distinct}.  

This subtlety does not arise for bond-centered reflections, because in that case the overall sign of $U_{\sigma}$ cannot be changed while maintaining the site structure of reflection symmetry.  This is consistent with earlier work showing that, for bond-centered reflections, the non-trivial SPT phase can be identified by symmetry protected multiplets in the entanglement spectrum, while the trivial phase lacks these multiplets \cite{pollmann10}.

There is a further subtlety if we consider stable equivalence, \emph{i.e.} if we allow for adding trivial bulk degrees of freedom.  In that case, the $\zz$ classification is always a $\zz$ torsor, with no invariant notion of which phase is trivial and which non-trivial.  To see this, begin with a system with bond-centered reflection, in the non-trivial pgSPT phase.  We then add trivial degrees of freedom (\emph{e.g.}, polarized spins) at bond centers, so that we have a site-centered reflection acting on these degrees of freedom.  Now, with reflection-symmetric open boundaries, we can reverse the sign of $U_{\sigma}$ as in the site-centered reflection case above.  This conclusion is consistent with the fact that the non-trivial phase for bond-centered reflection can be robustly identified via the entanglement spectrum, because it is no longer possible to make a symmetric entanglement cut after adding the new degrees of freedom.  We emphasize that it is natural to add degrees of freedom at different locations from those already included in some model.  For example, in a tight-binding model where the electron orbitals included lie at ionic positions, there can be tightly bound bonding or anti-bonding orbitals lying in between ions, that can be added to the model as trivial degrees of freedom.

\section{Properties of the modified toric code model}
\label{app:toricmod}

Here we study some basic properties of the toric code model introduced in Sec.~\ref{sec:z2root}.  We consider a $L \times L$ square lattice, where for technical convenience we take $L$ to be a multiple of 4.  As described in Sec.~\ref{sec:z2root}, the spins on the reflection axis transform as the boundary spins of the CZX model [Eq.~(\ref{eqn:czx})], while spins away from the reflection axis transform in an ordinary manner  [Eq.~(\ref{eqn:ordinary})].  In a finite system with periodic boundary conditions there is a second reflection axis ``at infinity,''  just as there are two mirror planes in Fig.~\ref{fig:3d-regions}.  We take the spins on the axis at infinity to transform in the ordinary manner.  

The model has $2 L^2$ Ising spins, and because there are $L^2$ each of vertices $v$ and plaquettes $p$, there are a total of $2 L^2$ commuting $A_v$ and $B_p$ operators.  Just as in the ordinary toric code, these operators are not all independent and obey the constraints
\begin{eqnarray}
\prod_p B_p &=& 1 \\
\prod_v A_v &=& \prod_{p \in \text{axis}} B_p \text{,}
\end{eqnarray}
where the second constraint differs from that in the ordinary toric code.  The products on the left-hand sides are over all plaquettes and vertices in the system, respectively.  The product on the right-hand side of the second equation is over those plaquettes $p$ intersected by the reflection axis.

Due to the constraints, specifying simultaneous eigenvalues for the $A_v$ and $B_p$ operators fixes $2L^2 - 2$ Ising degrees of freedom, leaving 2 degrees of freedom that give rise to a four-fold topological degeneracy.  We now show that it is possible to construct a complete basis of eigenstates of $A_v$ and $B_p$.  We let $a_v = \pm 1$ and $b_p = \pm 1$ be sets of eigenvalues of $A_v$ and $B_p$, respectively, satisfying the constraints above.  It is possible to find a product state in the $\tau^z$ basis, $| \psi_{{\rm ref}} \rangle$, realizing any desired choice of $b_p$.  In addition, the same state can be chosen to fix the eigenvalues of products of $\tau^z$ along non-contractible loops, thus fixing the degrees of freedom associated with the topological degeneracy.  Then we consider the state
\begin{equation}
| \psi \rangle = C \prod_v \Big[ \frac{1}{2} \big( 1 + a_v A_v \big) \Big] | \psi_{{\rm ref}} \rangle \text{,}
\end{equation}
where $C$ is a normalization constant.  It is straightforward to check that this state has the desired eigenvalues of $A_v$ and $B_p$.  Moreover, it can be checked this state is non-zero, by computing its norm, as long as $a_v$ and $b_p$ satisfy the constraints.

It follows from the above discussion that the modified toric code model has a four-fold topological degeneracy, and a gap to local excitations.  Moreover, away from the reflection axis the model is identical to the ordinary toric code,
so it has the same topological order (same theory of anyons)  as the ordinary toric code.  We note that anyons can be transported across the reflection axis by the string operators described in Sec.~\ref{sec:z2root}.

We now show that the model has a ground state respecting the reflection symmetry.  We start with the ground state
\begin{equation}
| \psi_0^a \rangle = C  \prod_v \Big[ \frac{1}{2} \big( 1 + A_v \big) \Big] | \{ \tau^z = 1\} \rangle \text{,}
\end{equation}
where we have chosen a reference state with all spins polarized to $\tau^z = 1$, which has $b_p = 1$ for all $p$, as required for a ground state.  Similarly, we choose $a_v = 1$ for all vertices $v$.  This state on its own is not invariant under reflection, but instead can be seen to transform as
\begin{equation}
U_{\sigma} | \psi^a_0 \rangle = \Big[ \prod_{\ell \in \text{axis} } \tau_{\ell}^y  \Big]  | \psi^a_0 \rangle \equiv | \psi^b_0 \rangle \text{.}
\end{equation}
The operator $[ \prod_{\ell \in \text{axis} } \tau_{\ell}^y ]$ is a string operator on the reflection axis, that commutes with all the $A_v$ and $B_p$ operators, and thus acts only within the space of topologically degenerate ground states.  It follows that $|\psi^a_0 \rangle$ and $| \psi^b_0 \rangle$ cannot be distinguished by local measurements, and thus cannot correspond to degenerate ground states associated with spontaneous symmetry breaking.  Therefore we obtain the reflection-invariant ground state
\begin{equation}
| \psi_0 \rangle = \frac{1}{\sqrt{2}} \big[ | \psi^a_0 \rangle + | \psi^b_0 \rangle \big]  \text{.}
\end{equation}

\bibliography{pgSPT}

\begin{thebibliography}{78}%
\makeatletter
\providecommand \@ifxundefined [1]{%
 \@ifx{#1\undefined}
}%
\providecommand \@ifnum [1]{%
 \ifnum #1\expandafter \@firstoftwo
 \else \expandafter \@secondoftwo
 \fi
}%
\providecommand \@ifx [1]{%
 \ifx #1\expandafter \@firstoftwo
 \else \expandafter \@secondoftwo
 \fi
}%
\providecommand \natexlab [1]{#1}%
\providecommand \enquote  [1]{``#1''}%
\providecommand \bibnamefont  [1]{#1}%
\providecommand \bibfnamefont [1]{#1}%
\providecommand \citenamefont [1]{#1}%
\providecommand \href@noop [0]{\@secondoftwo}%
\providecommand \href [0]{\begingroup \@sanitize@url \@href}%
\providecommand \@href[1]{\@@startlink{#1}\@@href}%
\providecommand \@@href[1]{\endgroup#1\@@endlink}%
\providecommand \@sanitize@url [0]{\catcode `\\12\catcode `\$12\catcode
  `\&12\catcode `\#12\catcode `\^12\catcode `\_12\catcode `\%12\relax}%
\providecommand \@@startlink[1]{}%
\providecommand \@@endlink[0]{}%
\providecommand \url  [0]{\begingroup\@sanitize@url \@url }%
\providecommand \@url [1]{\endgroup\@href {#1}{\urlprefix }}%
\providecommand \urlprefix  [0]{URL }%
\providecommand \Eprint [0]{\href }%
\providecommand \doibase [0]{http://dx.doi.org/}%
\providecommand \selectlanguage [0]{\@gobble}%
\providecommand \bibinfo  [0]{\@secondoftwo}%
\providecommand \bibfield  [0]{\@secondoftwo}%
\providecommand \translation [1]{[#1]}%
\providecommand \BibitemOpen [0]{}%
\providecommand \bibitemStop [0]{}%
\providecommand \bibitemNoStop [0]{.\EOS\space}%
\providecommand \EOS [0]{\spacefactor3000\relax}%
\providecommand \BibitemShut  [1]{\csname bibitem#1\endcsname}%
\let\auto@bib@innerbib\@empty
%</preamble>
\bibitem [{\citenamefont {Hasan}\ and\ \citenamefont {Kane}(2010)}]{hasan10}%
  \BibitemOpen
  \bibfield  {author} {\bibinfo {author} {\bibfnamefont {M.~Z.}\ \bibnamefont
  {Hasan}}\ and\ \bibinfo {author} {\bibfnamefont {C.~L.}\ \bibnamefont
  {Kane}},\ }\href {\doibase 10.1103/RevModPhys.82.3045} {\bibfield  {journal}
  {\bibinfo  {journal} {Rev. Mod. Phys.}\ }\textbf {\bibinfo {volume} {82}},\
  \bibinfo {pages} {3045} (\bibinfo {year} {2010})}\BibitemShut {NoStop}%
\bibitem [{\citenamefont {Hasan}\ and\ \citenamefont {Moore}(2011)}]{hasan11}%
  \BibitemOpen
  \bibfield  {author} {\bibinfo {author} {\bibfnamefont {M.~Z.}\ \bibnamefont
  {Hasan}}\ and\ \bibinfo {author} {\bibfnamefont {J.~E.}\ \bibnamefont
  {Moore}},\ }\href {\doibase 10.1146/annurev-conmatphys-062910-140432}
  {\bibfield  {journal} {\bibinfo  {journal} {Annual Review of Condensed Matter
  Physics}\ }\textbf {\bibinfo {volume} {2}},\ \bibinfo {pages} {55} (\bibinfo
  {year} {2011})}\BibitemShut {NoStop}%
\bibitem [{\citenamefont {Qi}\ and\ \citenamefont {Zhang}(2011)}]{qi11}%
  \BibitemOpen
  \bibfield  {author} {\bibinfo {author} {\bibfnamefont {X.-L.}\ \bibnamefont
  {Qi}}\ and\ \bibinfo {author} {\bibfnamefont {S.-C.}\ \bibnamefont {Zhang}},\
  }\href {\doibase 10.1103/RevModPhys.83.1057} {\bibfield  {journal} {\bibinfo
  {journal} {Rev. Mod. Phys.}\ }\textbf {\bibinfo {volume} {83}},\ \bibinfo
  {pages} {1057} (\bibinfo {year} {2011})}\BibitemShut {NoStop}%
\bibitem [{\citenamefont {Kitaev}(2009)}]{kitaev09}%
  \BibitemOpen
  \bibfield  {author} {\bibinfo {author} {\bibfnamefont {A.}~\bibnamefont
  {Kitaev}},\ }\href {http://arxiv.org/abs/0901.2686} {\enquote {\bibinfo
  {title} {Periodic table for topological insulators and superconductors},}\ }
  (\bibinfo {year} {2009}),\ \Eprint {http://arxiv.org/abs/arXiv:0901.2686}
  {arXiv:0901.2686} \BibitemShut {NoStop}%
\bibitem [{\citenamefont {Ryu}\ \emph {et~al.}(2010)\citenamefont {Ryu},
  \citenamefont {Schnyder}, \citenamefont {Furusaki},\ and\ \citenamefont
  {Ludwig}}]{ryu10}%
  \BibitemOpen
  \bibfield  {author} {\bibinfo {author} {\bibfnamefont {S.}~\bibnamefont
  {Ryu}}, \bibinfo {author} {\bibfnamefont {A.~P.}\ \bibnamefont {Schnyder}},
  \bibinfo {author} {\bibfnamefont {A.}~\bibnamefont {Furusaki}}, \ and\
  \bibinfo {author} {\bibfnamefont {A.~W.~W.}\ \bibnamefont {Ludwig}},\ }\href
  {http://stacks.iop.org/1367-2630/12/i=6/a=065010} {\bibfield  {journal}
  {\bibinfo  {journal} {New Journal of Physics}\ }\textbf {\bibinfo {volume}
  {12}},\ \bibinfo {pages} {065010} (\bibinfo {year} {2010})}\BibitemShut
  {NoStop}%
\bibitem [{\citenamefont {Gu}\ and\ \citenamefont {Wen}(2009)}]{gu09tensor}%
  \BibitemOpen
  \bibfield  {author} {\bibinfo {author} {\bibfnamefont {Z.-C.}\ \bibnamefont
  {Gu}}\ and\ \bibinfo {author} {\bibfnamefont {X.-G.}\ \bibnamefont {Wen}},\
  }\href {\doibase 10.1103/PhysRevB.80.155131} {\bibfield  {journal} {\bibinfo
  {journal} {Phys. Rev. B}\ }\textbf {\bibinfo {volume} {80}},\ \bibinfo
  {pages} {155131} (\bibinfo {year} {2009})}\BibitemShut {NoStop}%
\bibitem [{\citenamefont {Pollmann}\ \emph {et~al.}(2010)\citenamefont
  {Pollmann}, \citenamefont {Turner}, \citenamefont {Berg},\ and\ \citenamefont
  {Oshikawa}}]{pollmann10}%
  \BibitemOpen
  \bibfield  {author} {\bibinfo {author} {\bibfnamefont {F.}~\bibnamefont
  {Pollmann}}, \bibinfo {author} {\bibfnamefont {A.~M.}\ \bibnamefont
  {Turner}}, \bibinfo {author} {\bibfnamefont {E.}~\bibnamefont {Berg}}, \ and\
  \bibinfo {author} {\bibfnamefont {M.}~\bibnamefont {Oshikawa}},\ }\href
  {\doibase 10.1103/PhysRevB.81.064439} {\bibfield  {journal} {\bibinfo
  {journal} {Phys. Rev. B}\ }\textbf {\bibinfo {volume} {81}},\ \bibinfo
  {pages} {064439} (\bibinfo {year} {2010})}\BibitemShut {NoStop}%
\bibitem [{\citenamefont {Fidkowski}\ and\ \citenamefont
  {Kitaev}(2011)}]{fidkowski11}%
  \BibitemOpen
  \bibfield  {author} {\bibinfo {author} {\bibfnamefont {L.}~\bibnamefont
  {Fidkowski}}\ and\ \bibinfo {author} {\bibfnamefont {A.}~\bibnamefont
  {Kitaev}},\ }\href {\doibase 10.1103/PhysRevB.83.075103} {\bibfield
  {journal} {\bibinfo  {journal} {Phys. Rev. B}\ }\textbf {\bibinfo {volume}
  {83}},\ \bibinfo {pages} {075103} (\bibinfo {year} {2011})}\BibitemShut
  {NoStop}%
\bibitem [{\citenamefont {Turner}\ \emph {et~al.}(2011)\citenamefont {Turner},
  \citenamefont {Pollmann},\ and\ \citenamefont {Berg}}]{turner11}%
  \BibitemOpen
  \bibfield  {author} {\bibinfo {author} {\bibfnamefont {A.~M.}\ \bibnamefont
  {Turner}}, \bibinfo {author} {\bibfnamefont {F.}~\bibnamefont {Pollmann}}, \
  and\ \bibinfo {author} {\bibfnamefont {E.}~\bibnamefont {Berg}},\ }\href
  {\doibase 10.1103/PhysRevB.83.075102} {\bibfield  {journal} {\bibinfo
  {journal} {Phys. Rev. B}\ }\textbf {\bibinfo {volume} {83}},\ \bibinfo
  {pages} {075102} (\bibinfo {year} {2011})}\BibitemShut {NoStop}%
\bibitem [{\citenamefont {Chen}\ \emph
  {et~al.}(2011{\natexlab{a}})\citenamefont {Chen}, \citenamefont {Gu},\ and\
  \citenamefont {Wen}}]{chen11a}%
  \BibitemOpen
  \bibfield  {author} {\bibinfo {author} {\bibfnamefont {X.}~\bibnamefont
  {Chen}}, \bibinfo {author} {\bibfnamefont {Z.-C.}\ \bibnamefont {Gu}}, \ and\
  \bibinfo {author} {\bibfnamefont {X.-G.}\ \bibnamefont {Wen}},\ }\href
  {\doibase 10.1103/PhysRevB.83.035107} {\bibfield  {journal} {\bibinfo
  {journal} {Phys. Rev. B}\ }\textbf {\bibinfo {volume} {83}},\ \bibinfo
  {pages} {035107} (\bibinfo {year} {2011}{\natexlab{a}})}\BibitemShut
  {NoStop}%
\bibitem [{\citenamefont {Schuch}\ \emph {et~al.}(2011)\citenamefont {Schuch},
  \citenamefont {P{\'e}rez-Garc{\'\i}a},\ and\ \citenamefont
  {Cirac}}]{schuch11}%
  \BibitemOpen
  \bibfield  {author} {\bibinfo {author} {\bibfnamefont {N.}~\bibnamefont
  {Schuch}}, \bibinfo {author} {\bibfnamefont {D.}~\bibnamefont
  {P{\'e}rez-Garc{\'\i}a}}, \ and\ \bibinfo {author} {\bibfnamefont
  {I.}~\bibnamefont {Cirac}},\ }\href {\doibase 10.1103/PhysRevB.84.165139}
  {\bibfield  {journal} {\bibinfo  {journal} {Phys. Rev. B}\ }\textbf {\bibinfo
  {volume} {84}},\ \bibinfo {pages} {165139} (\bibinfo {year}
  {2011})}\BibitemShut {NoStop}%
\bibitem [{\citenamefont {Chen}\ \emph {et~al.}(2013)\citenamefont {Chen},
  \citenamefont {Gu}, \citenamefont {Liu},\ and\ \citenamefont
  {Wen}}]{chen13symmetry}%
  \BibitemOpen
  \bibfield  {author} {\bibinfo {author} {\bibfnamefont {X.}~\bibnamefont
  {Chen}}, \bibinfo {author} {\bibfnamefont {Z.-C.}\ \bibnamefont {Gu}},
  \bibinfo {author} {\bibfnamefont {Z.-X.}\ \bibnamefont {Liu}}, \ and\
  \bibinfo {author} {\bibfnamefont {X.-G.}\ \bibnamefont {Wen}},\ }\href
  {\doibase 10.1103/PhysRevB.87.155114} {\bibfield  {journal} {\bibinfo
  {journal} {Phys. Rev. B}\ }\textbf {\bibinfo {volume} {87}},\ \bibinfo
  {pages} {155114} (\bibinfo {year} {2013})}\BibitemShut {NoStop}%
\bibitem [{\citenamefont {Levin}\ and\ \citenamefont
  {Gu}(2012)}]{levin12braiding}%
  \BibitemOpen
  \bibfield  {author} {\bibinfo {author} {\bibfnamefont {M.}~\bibnamefont
  {Levin}}\ and\ \bibinfo {author} {\bibfnamefont {Z.-C.}\ \bibnamefont {Gu}},\
  }\href {\doibase 10.1103/PhysRevB.86.115109} {\bibfield  {journal} {\bibinfo
  {journal} {Phys. Rev. B}\ }\textbf {\bibinfo {volume} {86}},\ \bibinfo
  {pages} {115109} (\bibinfo {year} {2012})}\BibitemShut {NoStop}%
\bibitem [{\citenamefont {Dzero}\ \emph {et~al.}(2016)\citenamefont {Dzero},
  \citenamefont {Xia}, \citenamefont {Galitski},\ and\ \citenamefont
  {Coleman}}]{dzero16topological}%
  \BibitemOpen
  \bibfield  {author} {\bibinfo {author} {\bibfnamefont {M.}~\bibnamefont
  {Dzero}}, \bibinfo {author} {\bibfnamefont {J.}~\bibnamefont {Xia}}, \bibinfo
  {author} {\bibfnamefont {V.}~\bibnamefont {Galitski}}, \ and\ \bibinfo
  {author} {\bibfnamefont {P.}~\bibnamefont {Coleman}},\ }\href {\doibase
  10.1146/annurev-conmatphys-031214-014749} {\bibfield  {journal} {\bibinfo
  {journal} {Annual Review of Condensed Matter Physics}\ }\textbf {\bibinfo
  {volume} {7}},\ \bibinfo {pages} {249} (\bibinfo {year} {2016})}\BibitemShut
  {NoStop}%
\bibitem [{\citenamefont {Ando}\ and\ \citenamefont
  {Fu}(2015)}]{ando15topological}%
  \BibitemOpen
  \bibfield  {author} {\bibinfo {author} {\bibfnamefont {Y.}~\bibnamefont
  {Ando}}\ and\ \bibinfo {author} {\bibfnamefont {L.}~\bibnamefont {Fu}},\
  }\href {\doibase 10.1146/annurev-conmatphys-031214-014501} {\bibfield
  {journal} {\bibinfo  {journal} {Annual Review of Condensed Matter Physics}\
  }\textbf {\bibinfo {volume} {6}},\ \bibinfo {pages} {361} (\bibinfo {year}
  {2015})}\BibitemShut {NoStop}%
\bibitem [{\citenamefont {Isobe}\ and\ \citenamefont {Fu}(2015)}]{isobe15}%
  \BibitemOpen
  \bibfield  {author} {\bibinfo {author} {\bibfnamefont {H.}~\bibnamefont
  {Isobe}}\ and\ \bibinfo {author} {\bibfnamefont {L.}~\bibnamefont {Fu}},\
  }\href {\doibase 10.1103/PhysRevB.92.081304} {\bibfield  {journal} {\bibinfo
  {journal} {Phys. Rev. B}\ }\textbf {\bibinfo {volume} {92}},\ \bibinfo
  {pages} {081304} (\bibinfo {year} {2015})}\BibitemShut {NoStop}%
\bibitem [{\citenamefont {Qi}\ and\ \citenamefont
  {Fu}(2015{\natexlab{a}})}]{yqi15anomalous}%
  \BibitemOpen
  \bibfield  {author} {\bibinfo {author} {\bibfnamefont {Y.}~\bibnamefont
  {Qi}}\ and\ \bibinfo {author} {\bibfnamefont {L.}~\bibnamefont {Fu}},\ }\href
  {\doibase 10.1103/PhysRevLett.115.236801} {\bibfield  {journal} {\bibinfo
  {journal} {Phys. Rev. Lett.}\ }\textbf {\bibinfo {volume} {115}},\ \bibinfo
  {pages} {236801} (\bibinfo {year} {2015}{\natexlab{a}})}\BibitemShut
  {NoStop}%
\bibitem [{\citenamefont {Yoshida}\ and\ \citenamefont
  {Furusaki}(2015)}]{yoshida15correlation}%
  \BibitemOpen
  \bibfield  {author} {\bibinfo {author} {\bibfnamefont {T.}~\bibnamefont
  {Yoshida}}\ and\ \bibinfo {author} {\bibfnamefont {A.}~\bibnamefont
  {Furusaki}},\ }\href {\doibase 10.1103/PhysRevB.92.085114} {\bibfield
  {journal} {\bibinfo  {journal} {Phys. Rev. B}\ }\textbf {\bibinfo {volume}
  {92}},\ \bibinfo {pages} {085114} (\bibinfo {year} {2015})}\BibitemShut
  {NoStop}%
\bibitem [{\citenamefont {Morimoto}\ \emph {et~al.}(2015)\citenamefont
  {Morimoto}, \citenamefont {Furusaki},\ and\ \citenamefont
  {Mudry}}]{morimoto15breakdown}%
  \BibitemOpen
  \bibfield  {author} {\bibinfo {author} {\bibfnamefont {T.}~\bibnamefont
  {Morimoto}}, \bibinfo {author} {\bibfnamefont {A.}~\bibnamefont {Furusaki}},
  \ and\ \bibinfo {author} {\bibfnamefont {C.}~\bibnamefont {Mudry}},\ }\href
  {\doibase 10.1103/PhysRevB.92.125104} {\bibfield  {journal} {\bibinfo
  {journal} {Phys. Rev. B}\ }\textbf {\bibinfo {volume} {92}},\ \bibinfo
  {pages} {125104} (\bibinfo {year} {2015})}\BibitemShut {NoStop}%
\bibitem [{\citenamefont {Hsieh}\ \emph {et~al.}(2016)\citenamefont {Hsieh},
  \citenamefont {Cho},\ and\ \citenamefont {Ryu}}]{chsieh16global}%
  \BibitemOpen
  \bibfield  {author} {\bibinfo {author} {\bibfnamefont {C.-T.}\ \bibnamefont
  {Hsieh}}, \bibinfo {author} {\bibfnamefont {G.~Y.}\ \bibnamefont {Cho}}, \
  and\ \bibinfo {author} {\bibfnamefont {S.}~\bibnamefont {Ryu}},\ }\href
  {\doibase 10.1103/PhysRevB.93.075135} {\bibfield  {journal} {\bibinfo
  {journal} {Phys. Rev. B}\ }\textbf {\bibinfo {volume} {93}},\ \bibinfo
  {pages} {075135} (\bibinfo {year} {2016})}\BibitemShut {NoStop}%
\bibitem [{\citenamefont {Hsieh}\ \emph
  {et~al.}(2014{\natexlab{a}})\citenamefont {Hsieh}, \citenamefont {Sule},
  \citenamefont {Cho}, \citenamefont {Ryu},\ and\ \citenamefont
  {Leigh}}]{chsieh14symmetry}%
  \BibitemOpen
  \bibfield  {author} {\bibinfo {author} {\bibfnamefont {C.-T.}\ \bibnamefont
  {Hsieh}}, \bibinfo {author} {\bibfnamefont {O.~M.}\ \bibnamefont {Sule}},
  \bibinfo {author} {\bibfnamefont {G.~Y.}\ \bibnamefont {Cho}}, \bibinfo
  {author} {\bibfnamefont {S.}~\bibnamefont {Ryu}}, \ and\ \bibinfo {author}
  {\bibfnamefont {R.~G.}\ \bibnamefont {Leigh}},\ }\href {\doibase
  10.1103/PhysRevB.90.165134} {\bibfield  {journal} {\bibinfo  {journal} {Phys.
  Rev. B}\ }\textbf {\bibinfo {volume} {90}},\ \bibinfo {pages} {165134}
  (\bibinfo {year} {2014}{\natexlab{a}})}\BibitemShut {NoStop}%
\bibitem [{\citenamefont {Hsieh}\ \emph
  {et~al.}(2014{\natexlab{b}})\citenamefont {Hsieh}, \citenamefont {Morimoto},\
  and\ \citenamefont {Ryu}}]{chsieh14CPT}%
  \BibitemOpen
  \bibfield  {author} {\bibinfo {author} {\bibfnamefont {C.-T.}\ \bibnamefont
  {Hsieh}}, \bibinfo {author} {\bibfnamefont {T.}~\bibnamefont {Morimoto}}, \
  and\ \bibinfo {author} {\bibfnamefont {S.}~\bibnamefont {Ryu}},\ }\href
  {\doibase 10.1103/PhysRevB.90.245111} {\bibfield  {journal} {\bibinfo
  {journal} {Phys. Rev. B}\ }\textbf {\bibinfo {volume} {90}},\ \bibinfo
  {pages} {245111} (\bibinfo {year} {2014}{\natexlab{b}})}\BibitemShut
  {NoStop}%
\bibitem [{\citenamefont {Cho}\ \emph {et~al.}(2015)\citenamefont {Cho},
  \citenamefont {Hsieh}, \citenamefont {Morimoto},\ and\ \citenamefont
  {Ryu}}]{cho15}%
  \BibitemOpen
  \bibfield  {author} {\bibinfo {author} {\bibfnamefont {G.~Y.}\ \bibnamefont
  {Cho}}, \bibinfo {author} {\bibfnamefont {C.-T.}\ \bibnamefont {Hsieh}},
  \bibinfo {author} {\bibfnamefont {T.}~\bibnamefont {Morimoto}}, \ and\
  \bibinfo {author} {\bibfnamefont {S.}~\bibnamefont {Ryu}},\ }\href {\doibase
  10.1103/PhysRevB.91.195142} {\bibfield  {journal} {\bibinfo  {journal} {Phys.
  Rev. B}\ }\textbf {\bibinfo {volume} {91}},\ \bibinfo {pages} {195142}
  (\bibinfo {year} {2015})}\BibitemShut {NoStop}%
\bibitem [{\citenamefont {Yoshida}\ \emph {et~al.}(2015)\citenamefont
  {Yoshida}, \citenamefont {Morimoto},\ and\ \citenamefont
  {Furusaki}}]{yoshida15bosonic}%
  \BibitemOpen
  \bibfield  {author} {\bibinfo {author} {\bibfnamefont {T.}~\bibnamefont
  {Yoshida}}, \bibinfo {author} {\bibfnamefont {T.}~\bibnamefont {Morimoto}}, \
  and\ \bibinfo {author} {\bibfnamefont {A.}~\bibnamefont {Furusaki}},\ }\href
  {\doibase 10.1103/PhysRevB.92.245122} {\bibfield  {journal} {\bibinfo
  {journal} {Phys. Rev. B}\ }\textbf {\bibinfo {volume} {92}},\ \bibinfo
  {pages} {245122} (\bibinfo {year} {2015})}\BibitemShut {NoStop}%
\bibitem [{\citenamefont {Ware}\ \emph {et~al.}(2015)\citenamefont {Ware},
  \citenamefont {Kimchi}, \citenamefont {Parameswaran},\ and\ \citenamefont
  {Bauer}}]{ware15topological}%
  \BibitemOpen
  \bibfield  {author} {\bibinfo {author} {\bibfnamefont {B.}~\bibnamefont
  {Ware}}, \bibinfo {author} {\bibfnamefont {I.}~\bibnamefont {Kimchi}},
  \bibinfo {author} {\bibfnamefont {S.~A.}\ \bibnamefont {Parameswaran}}, \
  and\ \bibinfo {author} {\bibfnamefont {B.}~\bibnamefont {Bauer}},\ }\href
  {\doibase 10.1103/PhysRevB.92.195105} {\bibfield  {journal} {\bibinfo
  {journal} {Phys. Rev. B}\ }\textbf {\bibinfo {volume} {92}},\ \bibinfo
  {pages} {195105} (\bibinfo {year} {2015})}\BibitemShut {NoStop}%
\bibitem [{\citenamefont {Lapa}\ \emph {et~al.}(2016)\citenamefont {Lapa},
  \citenamefont {Teo},\ and\ \citenamefont {Hughes}}]{lapa16interaction}%
  \BibitemOpen
  \bibfield  {author} {\bibinfo {author} {\bibfnamefont {M.~F.}\ \bibnamefont
  {Lapa}}, \bibinfo {author} {\bibfnamefont {J.~C.~Y.}\ \bibnamefont {Teo}}, \
  and\ \bibinfo {author} {\bibfnamefont {T.~L.}\ \bibnamefont {Hughes}},\
  }\href {\doibase 10.1103/PhysRevB.93.115131} {\bibfield  {journal} {\bibinfo
  {journal} {Phys. Rev. B}\ }\textbf {\bibinfo {volume} {93}},\ \bibinfo
  {pages} {115131} (\bibinfo {year} {2016})}\BibitemShut {NoStop}%
\bibitem [{\citenamefont {You}\ and\ \citenamefont {Xu}(2014)}]{you14symmetry}%
  \BibitemOpen
  \bibfield  {author} {\bibinfo {author} {\bibfnamefont {Y.-Z.}\ \bibnamefont
  {You}}\ and\ \bibinfo {author} {\bibfnamefont {C.}~\bibnamefont {Xu}},\
  }\href {\doibase 10.1103/PhysRevB.90.245120} {\bibfield  {journal} {\bibinfo
  {journal} {Phys. Rev. B}\ }\textbf {\bibinfo {volume} {90}},\ \bibinfo
  {pages} {245120} (\bibinfo {year} {2014})}\BibitemShut {NoStop}%
\bibitem [{\citenamefont {Kapustin}\ \emph {et~al.}(2015)\citenamefont
  {Kapustin}, \citenamefont {Thorngren}, \citenamefont {Turzillo},\ and\
  \citenamefont {Wang}}]{kapustin15fermionic}%
  \BibitemOpen
  \bibfield  {author} {\bibinfo {author} {\bibfnamefont {A.}~\bibnamefont
  {Kapustin}}, \bibinfo {author} {\bibfnamefont {R.}~\bibnamefont {Thorngren}},
  \bibinfo {author} {\bibfnamefont {A.}~\bibnamefont {Turzillo}}, \ and\
  \bibinfo {author} {\bibfnamefont {Z.}~\bibnamefont {Wang}},\ }\href
  {http://dx.doi.org/10.1007/JHEP12(2015)052} {\bibfield  {journal} {\bibinfo
  {journal} {Journal of High Energy Physics}\ }\textbf {\bibinfo {volume}
  {2015}} (\bibinfo {year} {2015})}\BibitemShut {NoStop}%
\bibitem [{\citenamefont {Hermele}\ and\ \citenamefont
  {Chen}(2016)}]{hermele15}%
  \BibitemOpen
  \bibfield  {author} {\bibinfo {author} {\bibfnamefont {M.}~\bibnamefont
  {Hermele}}\ and\ \bibinfo {author} {\bibfnamefont {X.}~\bibnamefont {Chen}},\
  }\href {\doibase 10.1103/PhysRevX.6.041006} {\bibfield  {journal} {\bibinfo
  {journal} {Phys. Rev. X}\ }\textbf {\bibinfo {volume} {6}},\ \bibinfo {pages}
  {041006} (\bibinfo {year} {2016})}\BibitemShut {NoStop}%
\bibitem [{\citenamefont {Cheng}\ \emph {et~al.}(2016)\citenamefont {Cheng},
  \citenamefont {Zaletel}, \citenamefont {Barkeshli}, \citenamefont
  {Vishwanath},\ and\ \citenamefont {Bonderson}}]{cheng15translational}%
  \BibitemOpen
  \bibfield  {author} {\bibinfo {author} {\bibfnamefont {M.}~\bibnamefont
  {Cheng}}, \bibinfo {author} {\bibfnamefont {M.}~\bibnamefont {Zaletel}},
  \bibinfo {author} {\bibfnamefont {M.}~\bibnamefont {Barkeshli}}, \bibinfo
  {author} {\bibfnamefont {A.}~\bibnamefont {Vishwanath}}, \ and\ \bibinfo
  {author} {\bibfnamefont {P.}~\bibnamefont {Bonderson}},\ }\href {\doibase
  10.1103/PhysRevX.6.041068} {\bibfield  {journal} {\bibinfo  {journal} {Phys.
  Rev. X}\ }\textbf {\bibinfo {volume} {6}},\ \bibinfo {pages} {041068}
  (\bibinfo {year} {2016})}\BibitemShut {NoStop}%
\bibitem [{Note1()}]{Note1}%
  \BibitemOpen
  \bibinfo {note} {The symmetries we focus on are more properly called site
  symmetries; in crystallography, the point group is defined as the quotient of
  the space group by its translation subgroup. We abuse terminology slightly
  and refer to site symmetries as point group symmetries, as this is a more
  evocative and commonly understood term, and because site symmetry groups are
  always themselves crystallographic point groups.}\BibitemShut {Stop}%
\bibitem [{\citenamefont {Hsieh}\ \emph {et~al.}(2012)\citenamefont {Hsieh},
  \citenamefont {Lin}, \citenamefont {Liu}, \citenamefont {Duan}, \citenamefont
  {Bansi},\ and\ \citenamefont {Fu}}]{thsieh12topological}%
  \BibitemOpen
  \bibfield  {author} {\bibinfo {author} {\bibfnamefont {T.~H.}\ \bibnamefont
  {Hsieh}}, \bibinfo {author} {\bibfnamefont {H.}~\bibnamefont {Lin}}, \bibinfo
  {author} {\bibfnamefont {J.}~\bibnamefont {Liu}}, \bibinfo {author}
  {\bibfnamefont {W.}~\bibnamefont {Duan}}, \bibinfo {author} {\bibfnamefont
  {A.}~\bibnamefont {Bansi}}, \ and\ \bibinfo {author} {\bibfnamefont
  {L.}~\bibnamefont {Fu}},\ }\href@noop {} {\bibfield  {journal} {\bibinfo
  {journal} {Nature Communications}\ }\textbf {\bibinfo {volume} {3}},\
  \bibinfo {pages} {982} (\bibinfo {year} {2012})}\BibitemShut {NoStop}%
\bibitem [{\citenamefont {Tanaka}\ \emph {et~al.}(2012)\citenamefont {Tanaka},
  \citenamefont {Ren}, \citenamefont {Sato}, \citenamefont {Nakayama},
  \citenamefont {Souma}, \citenamefont {Takahashi}, \citenamefont {Segawa},\
  and\ \citenamefont {Ando}}]{tanaka12experimental}%
  \BibitemOpen
  \bibfield  {author} {\bibinfo {author} {\bibfnamefont {Y.}~\bibnamefont
  {Tanaka}}, \bibinfo {author} {\bibfnamefont {Z.}~\bibnamefont {Ren}},
  \bibinfo {author} {\bibfnamefont {T.}~\bibnamefont {Sato}}, \bibinfo {author}
  {\bibfnamefont {K.}~\bibnamefont {Nakayama}}, \bibinfo {author}
  {\bibfnamefont {S.}~\bibnamefont {Souma}}, \bibinfo {author} {\bibfnamefont
  {T.}~\bibnamefont {Takahashi}}, \bibinfo {author} {\bibfnamefont
  {K.}~\bibnamefont {Segawa}}, \ and\ \bibinfo {author} {\bibfnamefont
  {Y.}~\bibnamefont {Ando}},\ }\href@noop {} {\bibfield  {journal} {\bibinfo
  {journal} {Nature Physics}\ }\textbf {\bibinfo {volume} {8}},\ \bibinfo
  {pages} {800} (\bibinfo {year} {2012})}\BibitemShut {NoStop}%
\bibitem [{\citenamefont {Dziawa}\ \emph {et~al.}(2012)\citenamefont {Dziawa},
  \citenamefont {Kowalski}, \citenamefont {Dybko}, \citenamefont {Buczko},
  \citenamefont {Szczerbakow}, \citenamefont {Szot}, \citenamefont
  {{\L}usakowska}, \citenamefont {Balasubramanian}, \citenamefont {Wojek},
  \citenamefont {Berntsen}, \citenamefont {Tjernberg},\ and\ \citenamefont
  {Story}}]{dziawa12topological}%
  \BibitemOpen
  \bibfield  {author} {\bibinfo {author} {\bibfnamefont {P.}~\bibnamefont
  {Dziawa}}, \bibinfo {author} {\bibfnamefont {B.~J.}\ \bibnamefont
  {Kowalski}}, \bibinfo {author} {\bibfnamefont {K.}~\bibnamefont {Dybko}},
  \bibinfo {author} {\bibfnamefont {R.}~\bibnamefont {Buczko}}, \bibinfo
  {author} {\bibfnamefont {A.}~\bibnamefont {Szczerbakow}}, \bibinfo {author}
  {\bibfnamefont {M.}~\bibnamefont {Szot}}, \bibinfo {author} {\bibfnamefont
  {E.}~\bibnamefont {{\L}usakowska}}, \bibinfo {author} {\bibfnamefont
  {T.}~\bibnamefont {Balasubramanian}}, \bibinfo {author} {\bibfnamefont
  {B.~M.}\ \bibnamefont {Wojek}}, \bibinfo {author} {\bibfnamefont {M.~H.}\
  \bibnamefont {Berntsen}}, \bibinfo {author} {\bibfnamefont {O.}~\bibnamefont
  {Tjernberg}}, \ and\ \bibinfo {author} {\bibfnamefont {T.}~\bibnamefont
  {Story}},\ }\href@noop {} {\bibfield  {journal} {\bibinfo  {journal} {Nature
  Materials}\ }\textbf {\bibinfo {volume} {11}},\ \bibinfo {pages} {1023}
  (\bibinfo {year} {2012})}\BibitemShut {NoStop}%
\bibitem [{\citenamefont {Xu}\ \emph {et~al.}(2012)\citenamefont {Xu},
  \citenamefont {Liu}, \citenamefont {Alidoust}, \citenamefont {Neupane},
  \citenamefont {Qian}, \citenamefont {Belopolski}, \citenamefont {Denlinger},
  \citenamefont {Wang}, \citenamefont {Lin}, \citenamefont {Wray},
  \citenamefont {Landolt}, \citenamefont {Slomski}, \citenamefont {Dil},
  \citenamefont {Marcinkova}, \citenamefont {Morosan}, \citenamefont {Gibson},
  \citenamefont {Sankar}, \citenamefont {Chou}, \citenamefont {Cava},
  \citenamefont {Bansil},\ and\ \citenamefont {Hasan}}]{syxu12observation}%
  \BibitemOpen
  \bibfield  {author} {\bibinfo {author} {\bibfnamefont {S.-Y.}\ \bibnamefont
  {Xu}}, \bibinfo {author} {\bibfnamefont {C.}~\bibnamefont {Liu}}, \bibinfo
  {author} {\bibfnamefont {N.}~\bibnamefont {Alidoust}}, \bibinfo {author}
  {\bibfnamefont {M.}~\bibnamefont {Neupane}}, \bibinfo {author} {\bibfnamefont
  {D.}~\bibnamefont {Qian}}, \bibinfo {author} {\bibfnamefont {I.}~\bibnamefont
  {Belopolski}}, \bibinfo {author} {\bibfnamefont {J.~D.}\ \bibnamefont
  {Denlinger}}, \bibinfo {author} {\bibfnamefont {Y.~J.}\ \bibnamefont {Wang}},
  \bibinfo {author} {\bibfnamefont {H.}~\bibnamefont {Lin}}, \bibinfo {author}
  {\bibfnamefont {L.~A.}\ \bibnamefont {Wray}}, \bibinfo {author}
  {\bibfnamefont {G.}~\bibnamefont {Landolt}}, \bibinfo {author} {\bibfnamefont
  {B.}~\bibnamefont {Slomski}}, \bibinfo {author} {\bibfnamefont {J.~H.}\
  \bibnamefont {Dil}}, \bibinfo {author} {\bibfnamefont {A.}~\bibnamefont
  {Marcinkova}}, \bibinfo {author} {\bibfnamefont {E.}~\bibnamefont {Morosan}},
  \bibinfo {author} {\bibfnamefont {Q.}~\bibnamefont {Gibson}}, \bibinfo
  {author} {\bibfnamefont {R.}~\bibnamefont {Sankar}}, \bibinfo {author}
  {\bibfnamefont {F.~C.}\ \bibnamefont {Chou}}, \bibinfo {author}
  {\bibfnamefont {R.~J.}\ \bibnamefont {Cava}}, \bibinfo {author}
  {\bibfnamefont {A.}~\bibnamefont {Bansil}}, \ and\ \bibinfo {author}
  {\bibfnamefont {M.~Z.}\ \bibnamefont {Hasan}},\ }\href@noop {} {\bibfield
  {journal} {\bibinfo  {journal} {Nature Communications}\ }\textbf {\bibinfo
  {volume} {3}},\ \bibinfo {pages} {1192} (\bibinfo {year} {2012})}\BibitemShut
  {NoStop}%
\bibitem [{\citenamefont {Wang}\ \emph {et~al.}(2014)\citenamefont {Wang},
  \citenamefont {Potter},\ and\ \citenamefont
  {Senthil}}]{cwang14classification}%
  \BibitemOpen
  \bibfield  {author} {\bibinfo {author} {\bibfnamefont {C.}~\bibnamefont
  {Wang}}, \bibinfo {author} {\bibfnamefont {A.~C.}\ \bibnamefont {Potter}}, \
  and\ \bibinfo {author} {\bibfnamefont {T.}~\bibnamefont {Senthil}},\ }\href
  {\doibase 10.1126/science.1243326} {\bibfield  {journal} {\bibinfo  {journal}
  {Science}\ }\textbf {\bibinfo {volume} {343}},\ \bibinfo {pages} {629}
  (\bibinfo {year} {2014})},\ \Eprint
  {http://arxiv.org/abs/http://www.sciencemag.org/content/343/6171/629.full.pdf}
  {http://www.sciencemag.org/content/343/6171/629.full.pdf} \BibitemShut
  {NoStop}%
\bibitem [{\citenamefont {Wen}(2002)}]{wen02}%
  \BibitemOpen
  \bibfield  {author} {\bibinfo {author} {\bibfnamefont {X.-G.}\ \bibnamefont
  {Wen}},\ }\href {\doibase 10.1103/PhysRevB.65.165113} {\bibfield  {journal}
  {\bibinfo  {journal} {Phys. Rev. B}\ }\textbf {\bibinfo {volume} {65}},\
  \bibinfo {pages} {165113} (\bibinfo {year} {2002})},\ \Eprint
  {http://arxiv.org/abs/cond-mat/0107071} {arXiv:cond-mat/0107071} \BibitemShut
  {NoStop}%
\bibitem [{\citenamefont {Essin}\ and\ \citenamefont
  {Hermele}(2013)}]{essin13}%
  \BibitemOpen
  \bibfield  {author} {\bibinfo {author} {\bibfnamefont {A.~M.}\ \bibnamefont
  {Essin}}\ and\ \bibinfo {author} {\bibfnamefont {M.}~\bibnamefont
  {Hermele}},\ }\href {\doibase 10.1103/PhysRevB.87.104406} {\bibfield
  {journal} {\bibinfo  {journal} {Phys. Rev. B}\ }\textbf {\bibinfo {volume}
  {87}},\ \bibinfo {pages} {104406} (\bibinfo {year} {2013})},\ \Eprint
  {http://arxiv.org/abs/1212.0593} {arXiv:1212.0593} \BibitemShut {NoStop}%
\bibitem [{\citenamefont {Barkeshli}\ \emph {et~al.}(2014)\citenamefont
  {Barkeshli}, \citenamefont {Bonderson}, \citenamefont {Cheng},\ and\
  \citenamefont {Wang}}]{barkeshli14}%
  \BibitemOpen
  \bibfield  {author} {\bibinfo {author} {\bibfnamefont {M.}~\bibnamefont
  {Barkeshli}}, \bibinfo {author} {\bibfnamefont {P.}~\bibnamefont
  {Bonderson}}, \bibinfo {author} {\bibfnamefont {M.}~\bibnamefont {Cheng}}, \
  and\ \bibinfo {author} {\bibfnamefont {Z.}~\bibnamefont {Wang}},\ }\href@noop
  {} {\enquote {\bibinfo {title} {Symmetry, defects, and gauging of topological
  phases},}\ } (\bibinfo {year} {2014}),\ \Eprint
  {http://arxiv.org/abs/arXiv:1410.4540} {arXiv:1410.4540} \BibitemShut
  {NoStop}%
\bibitem [{\citenamefont {Qi}\ and\ \citenamefont
  {Fu}(2015{\natexlab{b}})}]{yqi15detecting}%
  \BibitemOpen
  \bibfield  {author} {\bibinfo {author} {\bibfnamefont {Y.}~\bibnamefont
  {Qi}}\ and\ \bibinfo {author} {\bibfnamefont {L.}~\bibnamefont {Fu}},\ }\href
  {\doibase 10.1103/PhysRevB.91.100401} {\bibfield  {journal} {\bibinfo
  {journal} {Phys. Rev. B}\ }\textbf {\bibinfo {volume} {91}},\ \bibinfo
  {pages} {100401} (\bibinfo {year} {2015}{\natexlab{b}})}\BibitemShut
  {NoStop}%
\bibitem [{\citenamefont {Zaletel}\ \emph {et~al.}(2015)\citenamefont
  {Zaletel}, \citenamefont {Lu},\ and\ \citenamefont {Vishwanath}}]{zaletel15}%
  \BibitemOpen
  \bibfield  {author} {\bibinfo {author} {\bibfnamefont {M.}~\bibnamefont
  {Zaletel}}, \bibinfo {author} {\bibfnamefont {Y.-M.}\ \bibnamefont {Lu}}, \
  and\ \bibinfo {author} {\bibfnamefont {A.}~\bibnamefont {Vishwanath}},\
  }\href@noop {} {\enquote {\bibinfo {title} {Measuring space-group symmetry
  fractionalization in z$_2$ spin liquids},}\ } (\bibinfo {year} {2015}),\
  \Eprint {http://arxiv.org/abs/arXiv:1501.01395} {arXiv:1501.01395}
  \BibitemShut {NoStop}%
\bibitem [{\citenamefont {Kapustin}(2014{\natexlab{a}})}]{kapustin14symmetry}%
  \BibitemOpen
  \bibfield  {author} {\bibinfo {author} {\bibfnamefont {A.}~\bibnamefont
  {Kapustin}},\ }\href@noop {} {\enquote {\bibinfo {title} {Symmetry protected
  topological phases, anomalies, and cobordisms: Beyond group cohomology},}\ }
  (\bibinfo {year} {2014}{\natexlab{a}}),\ \Eprint
  {http://arxiv.org/abs/arXiv:1403.1467} {arXiv:1403.1467} \BibitemShut
  {NoStop}%
\bibitem [{\citenamefont {Kapustin}(2014{\natexlab{b}})}]{kapustin14bosonic}%
  \BibitemOpen
  \bibfield  {author} {\bibinfo {author} {\bibfnamefont {A.}~\bibnamefont
  {Kapustin}},\ }\href@noop {} {\enquote {\bibinfo {title} {Bosonic topological
  insulators and paramagnets: a view from cobordisms},}\ } (\bibinfo {year}
  {2014}{\natexlab{b}}),\ \Eprint {http://arxiv.org/abs/arXiv:1404.6659}
  {arXiv:1404.6659} \BibitemShut {NoStop}%
\bibitem [{\citenamefont {Kitaev}()}]{kitaev11KITP}%
  \BibitemOpen
  \bibfield  {author} {\bibinfo {author} {\bibfnamefont {A.}~\bibnamefont
  {Kitaev}},\ }\href@noop {} {}\bibinfo {note} {\begin{scriptsize}
  {h}ttp://online.kitp.ucsb.edu/online/topomat11/kitaev\end{scriptsize}}\BibitemShut
  {NoStop}%
\bibitem [{\citenamefont {Kitaev}(2006)}]{kitaev06}%
  \BibitemOpen
  \bibfield  {author} {\bibinfo {author} {\bibfnamefont {A.}~\bibnamefont
  {Kitaev}},\ }\href {\doibase 10.1016/j.aop.2005.10.005} {\bibfield  {journal}
  {\bibinfo  {journal} {Annals of Physics}\ }\textbf {\bibinfo {volume}
  {321}},\ \bibinfo {pages} {2 } (\bibinfo {year} {2006})}\BibitemShut
  {NoStop}%
\bibitem [{\citenamefont {Haldane}(1983{\natexlab{a}})}]{haldane83a}%
  \BibitemOpen
  \bibfield  {author} {\bibinfo {author} {\bibfnamefont {F.}~\bibnamefont
  {Haldane}},\ }\href {\doibase http://dx.doi.org/10.1016/0375-9601(83)90631-X}
  {\bibfield  {journal} {\bibinfo  {journal} {Physics Letters A}\ }\textbf
  {\bibinfo {volume} {93}},\ \bibinfo {pages} {464 } (\bibinfo {year}
  {1983}{\natexlab{a}})}\BibitemShut {NoStop}%
\bibitem [{\citenamefont {Haldane}(1983{\natexlab{b}})}]{haldane83b}%
  \BibitemOpen
  \bibfield  {author} {\bibinfo {author} {\bibfnamefont {F.~D.~M.}\
  \bibnamefont {Haldane}},\ }\href {\doibase 10.1103/PhysRevLett.50.1153}
  {\bibfield  {journal} {\bibinfo  {journal} {Phys. Rev. Lett.}\ }\textbf
  {\bibinfo {volume} {50}},\ \bibinfo {pages} {1153} (\bibinfo {year}
  {1983}{\natexlab{b}})}\BibitemShut {NoStop}%
\bibitem [{\citenamefont {Song}(2015)}]{hsongThesis}%
  \BibitemOpen
  \bibfield  {author} {\bibinfo {author} {\bibfnamefont {H.}~\bibnamefont
  {Song}},\ }\emph {\bibinfo {title}
  {\href{http://gradworks.umi.com/37/43/3743730.html}{Interplay between
  Symmetry and Topological Order in Quantum Spin Systems}}},\ \href@noop {}
  {Ph.D. thesis},\ \bibinfo  {school} {University of Colorado Boulder}
  (\bibinfo {year} {2015})\BibitemShut {NoStop}%
\bibitem [{\citenamefont {Else}\ and\ \citenamefont {Nayak}(2014)}]{else15}%
  \BibitemOpen
  \bibfield  {author} {\bibinfo {author} {\bibfnamefont {D.~V.}\ \bibnamefont
  {Else}}\ and\ \bibinfo {author} {\bibfnamefont {C.}~\bibnamefont {Nayak}},\
  }\href {\doibase 10.1103/PhysRevB.90.235137} {\bibfield  {journal} {\bibinfo
  {journal} {Phys. Rev. B}\ }\textbf {\bibinfo {volume} {90}},\ \bibinfo
  {pages} {235137} (\bibinfo {year} {2014})}\BibitemShut {NoStop}%
\bibitem [{\citenamefont {Plamadeala}\ \emph {et~al.}(2013)\citenamefont
  {Plamadeala}, \citenamefont {Mulligan},\ and\ \citenamefont
  {Nayak}}]{plamadeala13shortrange}%
  \BibitemOpen
  \bibfield  {author} {\bibinfo {author} {\bibfnamefont {E.}~\bibnamefont
  {Plamadeala}}, \bibinfo {author} {\bibfnamefont {M.}~\bibnamefont
  {Mulligan}}, \ and\ \bibinfo {author} {\bibfnamefont {C.}~\bibnamefont
  {Nayak}},\ }\href {\doibase 10.1103/PhysRevB.88.045131} {\bibfield  {journal}
  {\bibinfo  {journal} {Phys. Rev. B}\ }\textbf {\bibinfo {volume} {88}},\
  \bibinfo {pages} {045131} (\bibinfo {year} {2013})}\BibitemShut {NoStop}%
\bibitem [{Note2()}]{Note2}%
  \BibitemOpen
  \bibinfo {note} {By ``acting trivially,'' we mean the $\protect \mathbb
  {Z}_2$ symmetry acts as the identity operator on the Hilbert space of the
  $E_8$ root state.}\BibitemShut {Stop}%
\bibitem [{\citenamefont {Vishwanath}\ and\ \citenamefont
  {Senthil}(2013)}]{vishwanath13}%
  \BibitemOpen
  \bibfield  {author} {\bibinfo {author} {\bibfnamefont {A.}~\bibnamefont
  {Vishwanath}}\ and\ \bibinfo {author} {\bibfnamefont {T.}~\bibnamefont
  {Senthil}},\ }\href {\doibase 10.1103/PhysRevX.3.011016} {\bibfield
  {journal} {\bibinfo  {journal} {Phys. Rev. X}\ }\textbf {\bibinfo {volume}
  {3}},\ \bibinfo {pages} {011016} (\bibinfo {year} {2013})}\BibitemShut
  {NoStop}%
\bibitem [{\citenamefont {Wang}\ and\ \citenamefont
  {Senthil}(2013)}]{cwang13boson}%
  \BibitemOpen
  \bibfield  {author} {\bibinfo {author} {\bibfnamefont {C.}~\bibnamefont
  {Wang}}\ and\ \bibinfo {author} {\bibfnamefont {T.}~\bibnamefont {Senthil}},\
  }\href {\doibase 10.1103/PhysRevB.87.235122} {\bibfield  {journal} {\bibinfo
  {journal} {Phys. Rev. B}\ }\textbf {\bibinfo {volume} {87}},\ \bibinfo
  {pages} {235122} (\bibinfo {year} {2013})}\BibitemShut {NoStop}%
\bibitem [{\citenamefont {Chen}\ \emph {et~al.}(2014)\citenamefont {Chen},
  \citenamefont {Burnell}, \citenamefont {Vishwanath},\ and\ \citenamefont
  {Fidkowski}}]{chen14anomalous}%
  \BibitemOpen
  \bibfield  {author} {\bibinfo {author} {\bibfnamefont {X.}~\bibnamefont
  {Chen}}, \bibinfo {author} {\bibfnamefont {F.~J.}\ \bibnamefont {Burnell}},
  \bibinfo {author} {\bibfnamefont {A.}~\bibnamefont {Vishwanath}}, \ and\
  \bibinfo {author} {\bibfnamefont {L.}~\bibnamefont {Fidkowski}},\ }\href
  {http://arxiv.org/abs/1403.6491} {\enquote {\bibinfo {title} {Anomalous
  symmetry fractionalization and surface topological order},}\ } (\bibinfo
  {year} {2014}),\ \Eprint {http://arxiv.org/abs/arXiv:1403.6491}
  {arXiv:1403.6491} \BibitemShut {NoStop}%
\bibitem [{\citenamefont {Qi}\ and\ \citenamefont
  {Fu}(2015{\natexlab{c}})}]{yqi15b}%
  \BibitemOpen
  \bibfield  {author} {\bibinfo {author} {\bibfnamefont {Y.}~\bibnamefont
  {Qi}}\ and\ \bibinfo {author} {\bibfnamefont {L.}~\bibnamefont {Fu}},\ }\href
  {\doibase 10.1103/PhysRevLett.115.236801} {\bibfield  {journal} {\bibinfo
  {journal} {Phys. Rev. Lett.}\ }\textbf {\bibinfo {volume} {115}},\ \bibinfo
  {pages} {236801} (\bibinfo {year} {2015}{\natexlab{c}})}\BibitemShut
  {NoStop}%
\bibitem [{\citenamefont {Chen}\ \emph
  {et~al.}(2011{\natexlab{b}})\citenamefont {Chen}, \citenamefont {Liu},\ and\
  \citenamefont {Wen}}]{chen11b}%
  \BibitemOpen
  \bibfield  {author} {\bibinfo {author} {\bibfnamefont {X.}~\bibnamefont
  {Chen}}, \bibinfo {author} {\bibfnamefont {Z.-X.}\ \bibnamefont {Liu}}, \
  and\ \bibinfo {author} {\bibfnamefont {X.-G.}\ \bibnamefont {Wen}},\ }\href
  {\doibase 10.1103/PhysRevB.84.235141} {\bibfield  {journal} {\bibinfo
  {journal} {Phys. Rev. B}\ }\textbf {\bibinfo {volume} {84}},\ \bibinfo
  {pages} {235141} (\bibinfo {year} {2011}{\natexlab{b}})}\BibitemShut
  {NoStop}%
\bibitem [{\citenamefont {Kitaev}(2003)}]{kitaev03}%
  \BibitemOpen
  \bibfield  {author} {\bibinfo {author} {\bibfnamefont {A.~{\relax Yu}.}\
  \bibnamefont {Kitaev}},\ }\href {\doibase 10.1016/S0003-4916(02)00018-0}
  {\bibfield  {journal} {\bibinfo  {journal} {Ann. Phys.}\ }\textbf {\bibinfo
  {volume} {303}},\ \bibinfo {pages} {2} (\bibinfo {year} {2003})},\ \Eprint
  {http://arxiv.org/abs/quant-ph/9707021} {arXiv:quant-ph/9707021} \BibitemShut
  {NoStop}%
\bibitem [{\citenamefont {Song}\ and\ \citenamefont {Hermele}(2015)}]{hsong15}%
  \BibitemOpen
  \bibfield  {author} {\bibinfo {author} {\bibfnamefont {H.}~\bibnamefont
  {Song}}\ and\ \bibinfo {author} {\bibfnamefont {M.}~\bibnamefont {Hermele}},\
  }\href {\doibase 10.1103/PhysRevB.91.014405} {\bibfield  {journal} {\bibinfo
  {journal} {Phys. Rev. B}\ }\textbf {\bibinfo {volume} {91}},\ \bibinfo
  {pages} {014405} (\bibinfo {year} {2015})}\BibitemShut {NoStop}%
\bibitem [{Note3()}]{Note3}%
  \BibitemOpen
  \bibinfo {note} {This term is not gauge invariant, but it is allowed if
  appropriate bosonic Higgs fields are condensed at the edge.}\BibitemShut
  {Stop}%
\bibitem [{\citenamefont {Ryu}\ and\ \citenamefont
  {Zhang}(2012)}]{ryu12interacting}%
  \BibitemOpen
  \bibfield  {author} {\bibinfo {author} {\bibfnamefont {S.}~\bibnamefont
  {Ryu}}\ and\ \bibinfo {author} {\bibfnamefont {S.-C.}\ \bibnamefont
  {Zhang}},\ }\href {\doibase 10.1103/PhysRevB.85.245132} {\bibfield  {journal}
  {\bibinfo  {journal} {Phys. Rev. B}\ }\textbf {\bibinfo {volume} {85}},\
  \bibinfo {pages} {245132} (\bibinfo {year} {2012})}\BibitemShut {NoStop}%
\bibitem [{\citenamefont {Qi}(2013)}]{xlqi13}%
  \BibitemOpen
  \bibfield  {author} {\bibinfo {author} {\bibfnamefont {X.-L.}\ \bibnamefont
  {Qi}},\ }\href {http://stacks.iop.org/1367-2630/15/i=6/a=065002} {\bibfield
  {journal} {\bibinfo  {journal} {New Journal of Physics}\ }\textbf {\bibinfo
  {volume} {15}},\ \bibinfo {pages} {065002} (\bibinfo {year}
  {2013})}\BibitemShut {NoStop}%
\bibitem [{\citenamefont {Yao}\ and\ \citenamefont
  {Ryu}(2013)}]{yao13interaction}%
  \BibitemOpen
  \bibfield  {author} {\bibinfo {author} {\bibfnamefont {H.}~\bibnamefont
  {Yao}}\ and\ \bibinfo {author} {\bibfnamefont {S.}~\bibnamefont {Ryu}},\
  }\href {\doibase 10.1103/PhysRevB.88.064507} {\bibfield  {journal} {\bibinfo
  {journal} {Phys. Rev. B}\ }\textbf {\bibinfo {volume} {88}},\ \bibinfo
  {pages} {064507} (\bibinfo {year} {2013})}\BibitemShut {NoStop}%
\bibitem [{\citenamefont {Gu}\ and\ \citenamefont {Levin}(2014)}]{gu14effect}%
  \BibitemOpen
  \bibfield  {author} {\bibinfo {author} {\bibfnamefont {Z.-C.}\ \bibnamefont
  {Gu}}\ and\ \bibinfo {author} {\bibfnamefont {M.}~\bibnamefont {Levin}},\
  }\href {\doibase 10.1103/PhysRevB.89.201113} {\bibfield  {journal} {\bibinfo
  {journal} {Phys. Rev. B}\ }\textbf {\bibinfo {volume} {89}},\ \bibinfo
  {pages} {201113} (\bibinfo {year} {2014})}\BibitemShut {NoStop}%
\bibitem [{\citenamefont {Wen}(2012)}]{wen12noninteracting}%
  \BibitemOpen
  \bibfield  {author} {\bibinfo {author} {\bibfnamefont {X.-G.}\ \bibnamefont
  {Wen}},\ }\href {\doibase 10.1103/PhysRevB.85.085103} {\bibfield  {journal}
  {\bibinfo  {journal} {Phys. Rev. B}\ }\textbf {\bibinfo {volume} {85}},\
  \bibinfo {pages} {085103} (\bibinfo {year} {2012})}\BibitemShut {NoStop}%
\bibitem [{\citenamefont {Cano}\ \emph {et~al.}(2014)\citenamefont {Cano},
  \citenamefont {Cheng}, \citenamefont {Mulligan}, \citenamefont {Nayak},
  \citenamefont {Plamadeala},\ and\ \citenamefont {Yard}}]{cano14bulkedge}%
  \BibitemOpen
  \bibfield  {author} {\bibinfo {author} {\bibfnamefont {J.}~\bibnamefont
  {Cano}}, \bibinfo {author} {\bibfnamefont {M.}~\bibnamefont {Cheng}},
  \bibinfo {author} {\bibfnamefont {M.}~\bibnamefont {Mulligan}}, \bibinfo
  {author} {\bibfnamefont {C.}~\bibnamefont {Nayak}}, \bibinfo {author}
  {\bibfnamefont {E.}~\bibnamefont {Plamadeala}}, \ and\ \bibinfo {author}
  {\bibfnamefont {J.}~\bibnamefont {Yard}},\ }\href {\doibase
  10.1103/PhysRevB.89.115116} {\bibfield  {journal} {\bibinfo  {journal} {Phys.
  Rev. B}\ }\textbf {\bibinfo {volume} {89}},\ \bibinfo {pages} {115116}
  (\bibinfo {year} {2014})}\BibitemShut {NoStop}%
\bibitem [{\citenamefont {Cheng}()}]{chengPC}%
  \BibitemOpen
  \bibfield  {author} {\bibinfo {author} {\bibfnamefont {M.}~\bibnamefont
  {Cheng}},\ }\href@noop {} {}\bibinfo {note} {{p}rivate
  communication}\BibitemShut {NoStop}%
\bibitem [{\citenamefont {Wang}\ and\ \citenamefont
  {Senthil}(2014)}]{cwang14interacting}%
  \BibitemOpen
  \bibfield  {author} {\bibinfo {author} {\bibfnamefont {C.}~\bibnamefont
  {Wang}}\ and\ \bibinfo {author} {\bibfnamefont {T.}~\bibnamefont {Senthil}},\
  }\href {\doibase 10.1103/PhysRevB.89.195124} {\bibfield  {journal} {\bibinfo
  {journal} {Phys. Rev. B}\ }\textbf {\bibinfo {volume} {89}},\ \bibinfo
  {pages} {195124} (\bibinfo {year} {2014})}\BibitemShut {NoStop}%
\bibitem [{\citenamefont {Huang}()}]{sjhuangUNPUB}%
  \BibitemOpen
  \bibfield  {author} {\bibinfo {author} {\bibfnamefont {S.~J.}\ \bibnamefont
  {Huang}},\ }\href@noop {} {}\bibinfo {note} {{u}npublished}\BibitemShut
  {NoStop}%
\bibitem [{\citenamefont {You}\ \emph {et~al.}(2015)\citenamefont {You},
  \citenamefont {Bi}, \citenamefont {Rasmussen}, \citenamefont {Cheng},\ and\
  \citenamefont {Xu}}]{you15bridging}%
  \BibitemOpen
  \bibfield  {author} {\bibinfo {author} {\bibfnamefont {Y.-Z.}\ \bibnamefont
  {You}}, \bibinfo {author} {\bibfnamefont {Z.}~\bibnamefont {Bi}}, \bibinfo
  {author} {\bibfnamefont {A.}~\bibnamefont {Rasmussen}}, \bibinfo {author}
  {\bibfnamefont {M.}~\bibnamefont {Cheng}}, \ and\ \bibinfo {author}
  {\bibfnamefont {C.}~\bibnamefont {Xu}},\ }\href
  {http://stacks.iop.org/1367-2630/17/i=7/a=075010} {\bibfield  {journal}
  {\bibinfo  {journal} {New Journal of Physics}\ }\textbf {\bibinfo {volume}
  {17}},\ \bibinfo {pages} {075010} (\bibinfo {year} {2015})}\BibitemShut
  {NoStop}%
\bibitem [{\citenamefont {You}\ \emph {et~al.}(2016)\citenamefont {You},
  \citenamefont {Bi}, \citenamefont {Mao},\ and\ \citenamefont
  {Xu}}]{you16quantum}%
  \BibitemOpen
  \bibfield  {author} {\bibinfo {author} {\bibfnamefont {Y.-Z.}\ \bibnamefont
  {You}}, \bibinfo {author} {\bibfnamefont {Z.}~\bibnamefont {Bi}}, \bibinfo
  {author} {\bibfnamefont {D.}~\bibnamefont {Mao}}, \ and\ \bibinfo {author}
  {\bibfnamefont {C.}~\bibnamefont {Xu}},\ }\href {\doibase
  10.1103/PhysRevB.93.125101} {\bibfield  {journal} {\bibinfo  {journal} {Phys.
  Rev. B}\ }\textbf {\bibinfo {volume} {93}},\ \bibinfo {pages} {125101}
  (\bibinfo {year} {2016})}\BibitemShut {NoStop}%
\bibitem [{\citenamefont {Wang}(2016)}]{chenjiewang16braiding}%
  \BibitemOpen
  \bibfield  {author} {\bibinfo {author} {\bibfnamefont {C.}~\bibnamefont
  {Wang}},\ }\href@noop {} {\enquote {\bibinfo {title} {Braiding statistics and
  classification of two-dimensional charge-$2m$ superconductors},}\ } (\bibinfo
  {year} {2016}),\ \Eprint {http://arxiv.org/abs/arXiv:1601.02028}
  {arXiv:1601.02028} \BibitemShut {NoStop}%
\bibitem [{\citenamefont {Lan}\ \emph {et~al.}(2016)\citenamefont {Lan},
  \citenamefont {Kong},\ and\ \citenamefont {Wen}}]{lan16classification}%
  \BibitemOpen
  \bibfield  {author} {\bibinfo {author} {\bibfnamefont {T.}~\bibnamefont
  {Lan}}, \bibinfo {author} {\bibfnamefont {L.}~\bibnamefont {Kong}}, \ and\
  \bibinfo {author} {\bibfnamefont {X.-G.}\ \bibnamefont {Wen}},\ }\href@noop
  {} {\enquote {\bibinfo {title} {Classification of 2+1d topological orders and
  spt orders for bosonic and fermionic systems with on-site symmetries},}\ }
  (\bibinfo {year} {2016}),\ \Eprint {http://arxiv.org/abs/arXiv:1602.05946}
  {arXiv:1602.05946} \BibitemShut {NoStop}%
\bibitem [{\citenamefont {Burnell}\ \emph {et~al.}(2014)\citenamefont
  {Burnell}, \citenamefont {Chen}, \citenamefont {Fidkowski},\ and\
  \citenamefont {Vishwanath}}]{burnell14exactly}%
  \BibitemOpen
  \bibfield  {author} {\bibinfo {author} {\bibfnamefont {F.~J.}\ \bibnamefont
  {Burnell}}, \bibinfo {author} {\bibfnamefont {X.}~\bibnamefont {Chen}},
  \bibinfo {author} {\bibfnamefont {L.}~\bibnamefont {Fidkowski}}, \ and\
  \bibinfo {author} {\bibfnamefont {A.}~\bibnamefont {Vishwanath}},\ }\href
  {\doibase 10.1103/PhysRevB.90.245122} {\bibfield  {journal} {\bibinfo
  {journal} {Phys. Rev. B}\ }\textbf {\bibinfo {volume} {90}},\ \bibinfo
  {pages} {245122} (\bibinfo {year} {2014})}\BibitemShut {NoStop}%
\bibitem [{\citenamefont {Fidkowski}\ \emph {et~al.}(2013)\citenamefont
  {Fidkowski}, \citenamefont {Chen},\ and\ \citenamefont
  {Vishwanath}}]{fidkowski13nonabelian}%
  \BibitemOpen
  \bibfield  {author} {\bibinfo {author} {\bibfnamefont {L.}~\bibnamefont
  {Fidkowski}}, \bibinfo {author} {\bibfnamefont {X.}~\bibnamefont {Chen}}, \
  and\ \bibinfo {author} {\bibfnamefont {A.}~\bibnamefont {Vishwanath}},\
  }\href {\doibase 10.1103/PhysRevX.3.041016} {\bibfield  {journal} {\bibinfo
  {journal} {Phys. Rev. X}\ }\textbf {\bibinfo {volume} {3}},\ \bibinfo {pages}
  {041016} (\bibinfo {year} {2013})}\BibitemShut {NoStop}%
\bibitem [{\citenamefont {Metlitski}\ \emph {et~al.}(2014)\citenamefont
  {Metlitski}, \citenamefont {Fidkowski}, \citenamefont {Chen},\ and\
  \citenamefont {Vishwanath}}]{metlitski14interaction}%
  \BibitemOpen
  \bibfield  {author} {\bibinfo {author} {\bibfnamefont {M.~A.}\ \bibnamefont
  {Metlitski}}, \bibinfo {author} {\bibfnamefont {L.}~\bibnamefont
  {Fidkowski}}, \bibinfo {author} {\bibfnamefont {X.}~\bibnamefont {Chen}}, \
  and\ \bibinfo {author} {\bibfnamefont {A.}~\bibnamefont {Vishwanath}},\
  }\href@noop {} {\enquote {\bibinfo {title} {Interaction effects on 3d
  topological superconductors: surface topological order from vortex
  condensation, the 16 fold way and fermionic kramers doublets},}\ } (\bibinfo
  {year} {2014}),\ \Eprint {http://arxiv.org/abs/arXiv:1406.3032}
  {arXiv:1406.3032} \BibitemShut {NoStop}%
\bibitem [{\citenamefont {Hermele}()}]{2dedgeUNPUB}%
  \BibitemOpen
  \bibfield  {author} {\bibinfo {author} {\bibfnamefont {M.}~\bibnamefont
  {Hermele}},\ }\href@noop {} {}\bibinfo {note} {{u}npublished}\BibitemShut
  {NoStop}%
\bibitem [{\citenamefont {Kitaev}(2001)}]{kitaev01unpaired}%
  \BibitemOpen
  \bibfield  {author} {\bibinfo {author} {\bibfnamefont {A.~Y.}\ \bibnamefont
  {Kitaev}},\ }\href {http://stacks.iop.org/1063-7869/44/i=10S/a=S29}
  {\bibfield  {journal} {\bibinfo  {journal} {Physics-Uspekhi}\ }\textbf
  {\bibinfo {volume} {44}},\ \bibinfo {pages} {131} (\bibinfo {year}
  {2001})}\BibitemShut {NoStop}%
\bibitem [{\citenamefont {Fuji}\ \emph {et~al.}(2015)\citenamefont {Fuji},
  \citenamefont {Pollmann},\ and\ \citenamefont {Oshikawa}}]{fuji15distinct}%
  \BibitemOpen
  \bibfield  {author} {\bibinfo {author} {\bibfnamefont {Y.}~\bibnamefont
  {Fuji}}, \bibinfo {author} {\bibfnamefont {F.}~\bibnamefont {Pollmann}}, \
  and\ \bibinfo {author} {\bibfnamefont {M.}~\bibnamefont {Oshikawa}},\ }\href
  {\doibase 10.1103/PhysRevLett.114.177204} {\bibfield  {journal} {\bibinfo
  {journal} {Phys. Rev. Lett.}\ }\textbf {\bibinfo {volume} {114}},\ \bibinfo
  {pages} {177204} (\bibinfo {year} {2015})}\BibitemShut {NoStop}%
\end{thebibliography}%

\end{document}